\documentclass[12pt,letterpaper,usenames,dvipsnames]{article}
\usepackage{jheppub}
\usepackage{scalerel}
\usepackage{mathtools}
\usepackage{dsfont}
\usepackage{amssymb}
\usepackage{latexsym}
\usepackage{braket}
\usepackage{graphicx}
\usepackage{subcaption}
\usepackage[utf8]{inputenc}
\usepackage{multirow}
\usepackage{graphicx,tikz}
\usepackage[framemethod=TikZ]{mdframed} 
\usepackage{tikz-cd} 
\usetikzlibrary{arrows,snakes,shapes.arrows,decorations.markings}
     \tikzset{>=triangle 90}
     \tikzstyle{bbc}=[draw,circle,fill=black,scale=.75]
     \tikzstyle{rc}=[circle,fill=red,scale=.6]
     \tikzstyle{wc}=[draw,circle,scale=.75]

\usepackage{hyperref}
\definecolor{darkred}{rgb}{0.8,0.1,0.1}
\hypersetup{colorlinks=true, linkcolor=darkred, citecolor=blue, linktoc=page}

\renewcommand{\thanks}[1]{\footnote{#1}}

\newcommand{\bea}{\begin{eqnarray}}
\newcommand{\eea}{\end{eqnarray}}
\newcommand{\ee}{\end{equation}}
\newcommand{\be}{\begin{equation}}

\usepackage{adjustbox}
\usepackage{multirow}
\usepackage{hhline}
\usepackage{arydshln}

\def\a{\alpha}
\def\m{\mu}

\def\l{\lambda}

\def\La{\Lambda}

\definecolor{Cyan}{cmyk}{1.,0,0,0}
\definecolor{Magenta}{cmyk}{0,1.,0,0}
\definecolor{Yellow}{cmyk}{0,0,1.,0}
\definecolor{White}{cmyk}{0,0,0,0}
\definecolor{Orange}{cmyk}{0,0.61,0.87,0}
\definecolor{RedOrange}{cmyk}{0,0.77,0.87,0}
\definecolor{Red}{cmyk}{0,1.,1.,0}
\definecolor{Purple}{cmyk}{0.45,0.86,0,0}
\definecolor{Violet}{cmyk}{0.79,0.88,0,0}
\definecolor{Blue}{cmyk}{1,0.5,0,0}
\definecolor{ProcessBlue}{cmyk}{0.96,0,0,0}
\definecolor{GreenYellow}{cmyk}{0.6,0,1.,0}
\definecolor{Black}{cmyk}{0,0,0,1}

\definecolor{Genoa}{cmyk}{0,0,0,1}
\definecolor{Black}{cmyk}{0,0,0,1}
\definecolor{Black}{cmyk}{0,0,0,1}
\definecolor{Black}{cmyk}{0,0,0,1}

\newcommand{\blue}[1]{\textcolor{Blue}{#1}}

\newcommand{\green}[1]{\textcolor{GreenYellow}{#1}}

\def\cB{{\cal B}}
\def\cC{{\cal C}}
\def\cD{{\cal D}}

\def\cI{{\cal I}}

\def\cN{{\cal N}}
\def\cO{{\cal O}}

\def\cQ{{\cal Q}}

\def\cT{{\cal T}}
\def\cS{{\cal S}}

\def\cW{{\cal W}}

\def\half{ {1\over 2}}
\def\p{\partial}

\def\H{\mathbb{H}}
\def\bu{{\boldsymbol u}}
\def\Rbf{\mathbf{R}}
\def\D{{\Delta}}
\def\ef{\mathfrak{e}}
\def\sof{\mathfrak{so}}
\def\spf{\mathfrak{sp}}
\def\suf{\mathfrak{su}}
\def\uf{\mathfrak{u}}
\def\beq{\begin{equation}}
\def\eeq{\end{equation}}

\def\R{\mathbb{R}} 
\def\Z{\mathbb{Z}} 
\def\Sf{\mathfrak{S}}

\def\gf{\mathfrak{g}}

\newcommand\Star[3][]{%
\path[#1] (0  :#3) -- ( 36:#2) 
       -- (72 :#3) -- (108:#2)
       -- (144:#3) -- (180:#2)
       -- (216:#3) -- (252:#2)
       -- (288:#3) -- (324:#2)--cycle;}

\def\ZZ{{\mathbb Z}}

\def\NN{{\mathbb N}}
\def\CC{{\mathbb C}}
\def\QQ{{\mathbb Q}}
\def\HH{{\mathbb H}}

\def\no{\nonumber}
\def\text{\mathrm}

\newtheorem{proposition}{Proposition}

\usepackage{pifont}
\usepackage{amssymb}
\newcommand{\xmark}{\ding{55}}%

\makeatletter
\def\@fpheader{\ }
\makeatother

\title{Needles in a haystack
\\ 
{\large{An algorithmic approach to the classification of 4d $\cN=2$ SCFTs}}}
\author[1]{Justin Kaidi,}
\author[1,2,3]{Mario Martone,}
\author[2]{Leonardo Rastelli,}
\author[4]{Mitch Weaver}

\affiliation[1]{Simons Center for Geometry and Physics, Stony Brook University, Stony Brook, NY 11794-3840, USA}
\affiliation[2]{C.~N.~Yang Institute for Theoretical Physics,  Stony Brook University, Stony Brook, NY 11794-3840, USA}
\affiliation[3]{Department of Mathematics, King's College London, The Strand, London WC2R 2LS, U.K.}
\affiliation[4]{University of Cincinnati, Physics Department, PO Box 210011, Cincinnati OH 45221, USA}

\abstract{
There is a well-known map from 4d $\cN=2$ superconformal field theories (SCFTs) to 2d vertex operator algebras (VOAs). 
The 4d Schur index corresponds to the VOA vacuum character, and must be a solution with integral coefficients of a 
modular differential equation. 
This suggests a classification program for 4d $\cN=2$ SCFTs that starts with modular differential equations and proceeds by imposing all known constraints that follow from the 4d $\to$ 2d map. This program becomes fully algorithmic once one specifies the {\it order} of the modular differential equation and the  {\it rank} (complex dimension of the Coulomb branch) of the $\cN=2$ theory. 
As a proof of concept, we apply the algorithm to the study of rank-two $\cN=2$ SCFTs whose Schur indices satisfy a fourth-order untwisted modular differential equation. Scanning over a large number of putative cases, only 15 satisfy all of the constraints imposed by our algorithm, six of which correspond to known 4d SCFTs. More sophisticated constraints can be used to argue against the existence of the remaining nine cases. Altogether, this indicates that our knowledge of such rank-two SCFTs is surprisingly complete.
}

\begin{document}
\begin{minipage}{0.95\linewidth}
\begin{flushright}
YITP-SB-2022-02
\end{flushright}
\end{minipage}
\maketitle 
\section{Introduction}

Superconformal field theories (SCFTs) are an ideal laboratory for the study of quantum field theory.
The space of superconformal field theories is much more constrained than that of non-conformal, non-supersymmetric  QFTs, so much so that one might even envision a complete classification. 
While plausible classification schemes have been proposed for SCFTs with maximal supersymmetry, the case of half-maximal supersymmetry is  still wide open. In this paper we focus on four-dimensional ${\cal N}=2$ SCFTs, an extremely rich subject with deep connections to mathematics and string theory. 

Apart from rank-one SCFTs and Lagrangian gauge theories, which have been completely classified in \cite{Bhardwaj:2013qia,Argyres:2015ffa,Argyres:2015gha,Argyres:2016xmc,Argyres:2016xua},
there exists a bewildering and constantly growing list of top-down constructions  of 4d ${\cal N}=2$ SCFTs that do not (yet?) admit a conventional Lagrangian description. Two broad classes  involve compactifications of six-dimensional
SCFTs \cite{Gaiotto:2009we,Gaiotto:2009hg,Chacaltana:2010ks,Razamat:2016dpl,Bah:2017gph,Kim:2017toz,Kim:2018lfo,Razamat:2018gro,Ohmori:2018ona} 
and geometric engineering on Calabi-Yau singularities  in string theory~\cite{Katz:1996fh,Shapere:1999xr,Xie:2015rpa,Wang:2016yha,Chen:2016bzh,Chen:2017wkw,Closset:2020scj}. 

In the spirit of the bootstrap, one would like to formulate a classification program of 4d ${\cal N}=2$ SCFTs based  only on general principles, such as  unitarity and superconformal invariance. In this paper we make some modest progress in this direction. We leverage two classes of constraints:
\begin{itemize}
    \item[$a.$] \emph{VOAs and Higgs branch geometry}. It was shown  in  \cite{Beem:2013sza} that there exists a map from (unitary) four-dimensional $\cN=2$ SCFTs to a restricted subset of  (non-unitary) two-dimensional vertex operator algebras (VOAs).
    This map identifies the Schur index of the $\cN=2$ theory with the vacuum character $\chi_0$ of the VOA, up to an overall factor,
    \bea \label{vacuumSchur}
    \chi_0(q) = q^{- {c_{\mathrm{2d}}\over 24}}
    Z_{\rm Schur} (q) =  q^{- {c_{\mathrm{2d}}\over 24}}
    (1 +\sum_{n\in {\NN/ 2}} a_n q^n)~, \hspace{0.4 in} a_n \in \ZZ~.
    \eea
    Here $c_{\mathrm{2d}}$ is the 2d central charge, related to the 4d Weyl anomaly coefficient $c_{4d}$ by
    $c_{2d} = -12 c_{4d}$. The vacuum character must have integral Fourier expansion (since it counts operators in the theory), as well as unit leading coefficient (signifying the presence of a unique vacuum).
     Crucially,  the VOAs associated to 4d SCFTs are believed to be of a special type, known as ``quasi-lisse," which ensures that their vacuum characters satisfy a monic modular differential equations (MDE).\footnote{A \textit{monic} MDE is an MDE with unit leading coefficient and holomorphic coefficient functions. See Section \ref{sec:MDEconstrs} for explicit examples.} 
    The modular properties of the vacuum character allow one to control its high temperature limit, from which one can extract the other Weyl anomaly coefficient $a_{4d}$. These facts will provide the starting point for our analysis. By looking for integral solutions to monic MDEs of fixed order, we can identify Schur indices and central charges of putative $\cN=2$ SCFTs.
    
    The Schur sector of an ${\cal N}=2$ SCFT comprises the Higgs branch chiral ring. In particular, the Fourier coefficent $a_1$ counts the number of moment maps of the 4d theory, i.e. the bottom components of the flavor current supermultiplets.  In fact, according to the central conjecture of \cite{Beem:2017ooy}, the entire Higgs branch (as a complex sympletic variety) is encoded in the VOA.\footnote{However, in this work we will only discuss putative vacuum characters and MDEs, which are often insufficient to reconstruct the full VOA.}

    \item[$b.$] \emph{Coulomb branch geometry.} 
Conjecturally, every 4d $\cN=2$ theory has a branch of the moduli space with the property that the low-energy theory at a generic point is a free $\cN=2$ supersymmetric $U(1)^r$ gauge theory with no massless charged states. This is known as the Coulomb branch (CB). The quantity $r$ is called the \emph{rank} of the theory and coincides with the complex dimensionality of the CB. The CB is singular,\footnote{Purely on the basis of $\cN=2$ superconformal analysis, the CB can admit both \emph{metric} and \emph{complex structure} singularities. While the former arise when BPS states charged under the $U(1)^r$ become massless, the latter signal the fact that the CB chiral ring is not freely generated \cite{Bourget:2018ond,Argyres:2018wxu}. Throughout this manuscript we will restrict to $\cN=2$ theories with freely generated CB chiral ring and thus only consider metric singularities.} and its (complex codimension-one) singular loci encode many properties of the theory. A related fact is that a globally-defined Lagrangian description of the low-energy $U(1)^r$ theory is only possible by allowing electro-magnetically dual changes of basis. This gives rise to non-trivial monodromies (lending the CB a  \emph{Special K\"ahler} structure \cite{Seiberg:1994rs,Seiberg:1994aj,Freed:1997dp}), which constrain the properties of CB operators.
 
 The CB monodromies, which take value in the discrete electro-magnetic duality group $Sp(2r,\Z)$, can be used to restrict the scaling dimensions of CB operators to a very small set of rational values,  which depends solely on the rank of the theory \cite{Argyres:2018urp,Caorsi:2018zsq}. Furthermore, the stratification of the CB singular locus \cite{Argyres:2020wmq,Martone:2020nsy,Cecotti:2021ouq} allows us to determine  Weyl anomalies as well as flavor symmetry data in terms of the scaling dimension of CB observables. 
 
The structure of the CB geometry thus constrains some of the same SCFT data---Weyl anomalies and flavor central charges---as are captured by the VOA. The CB and VOA constraints are largely independent, making their compatibility a highly non-trivial requirement.

\end{itemize}
A fundamental open question is how these two seemingly very different sets of observables are related.
Intricate examples of mutual compatibility will be encountered throughout this work. But even more than just compatibility should be true: conjecturally, the VOA and Coulomb branch geometry separately provide complete characterizations of the 4d SCFT,\footnote{This statement concerning the Coulomb branch requires some clarification. Since the early papers on the subject, it has been known that for rank-1 theories where the Coulomb branch stratification data is trivial, the scale-invariant limit of the Coulomb branch is not enough to fully specify the SCFT \cite{Seiberg:1994aj}. However, it is still possible to enrich this geometric data with other purely Coulomb branch quantities, \emph{e.g.} the deformation pattern \cite{Argyres:2015ffa}, to provide (up to discrete gauging) a complete characterization. At higher ranks, the stratification of singular loci becomes considerably more rich and, to the authors' knowledge, no inequivalent SCFTs share the same stratification data, \emph{i.e.} assignments of low-energy theories on sub-loci of all complex codimensions. } in the sense that no two distinct SCFTs map to the same VOA or to the same CB geometry.
It should then in principle be possible to fully reconstruct the CB geometry from the VOA, and vice versa. A tantalizing hint in this direction
is the experimental observation of \cite{Cordova:2015nma}: in several cases, one can precisely relate the Schur index 
with a certain wall-crossing invariant computed from the massive  BPS spectrum on the Coulomb branch.

The current work has a more limited scope. We devise a procedure to identify candidate ${\cal N}=2$ SCFTs by generating mutually compatible pairs of VOA and Coulomb data. Our procedure becomes fully algorithmic once we specify, on the VOA side, the {\it order} $d$ of the modular differential equation obeyed by the  Schur index (more precisely, by the vacuum character (\ref{vacuumSchur})), and, on the CB side, the {\it rank} $r$ of the theory, i.e.~the complex dimension of the CB.

\begin{figure}[tbp]
\hspace{-0.3 in}\begin{tikzpicture}
     \draw[rounded corners] (0.5, 0) rectangle (2.5, 1) {};
     \node[] at (1.5,0.5) {$r,d,\chi_0$};
     \draw[ultra thick,-stealth] (2.5,0.5) to (4.2,0.5);
     \node[above] at (3.35,1) {{\footnotesize{MDE}}};
     \node[below] at (3.35,0) {{\footnotesize{Integrality (\S \ref{sec:MDEconstrs})}}};
     \draw[rounded corners] (4.2, 0) rectangle (6.8, 1) {};
      \node[] at (5.5,0.5) {$c_{\mathrm{4d}}, a_{\mathrm{4d}}, |\mathfrak{g}|$}; 
       \draw[ultra thick,-stealth] (6.8,0.5) to (8.8,0.5);
     \node[above] at (7.8,1) {{\footnotesize{Shapere-Tachikawa}}};
     \node[below] at (7.7,0) {\footnotesize{Hoffman-Maldacena (\S\ref{sec:centralcharge})}};
       \draw[rounded corners] (8.8, 0) rectangle (10.8, 1) {};
      \node[] at (9.8,0.5) {$\Delta_{u_i}, \mathfrak{g}$};
      \draw[ultra thick,-stealth] (10.8,0.5) to (12.7,0.5);
       \node[above] at (11.75,1)  {{$\substack{\mathrm{CB\,\, strata}\\ \mathrm{decomposition}}$}};
     \node[below] at (11.8,0){{\footnotesize{CB constraints (\S \ref{sec:CBconsts})}}};
       \draw[rounded corners] (12.7, 0) rectangle (15, 1) {};
        \node[] at (13.85,0.5) {$h_{\mathrm{ECB}}, k_{\mathfrak{g}},\varkappa$};
        \draw[ultra thick,-stealth] (15,0.5) to (17,0.5);
          \node[above] at (16,1){};
     \node[below] at (16,0.1)  {\footnotesize{${\substack{\mathrm{Unitarity\,\,}(\S \ref{sec:unitsubsec})  \\ \varkappa \mathrm{\,\,constraints\,\,}  (\S\ref{sec:chardim})}}$}};
     \draw[rounded corners] (17, 0) rectangle (18, 1) {};
    \Star[scale=0.1,fill=yellow!30,draw, xshift = 68.9 in, yshift = 2 in]{2}{4};
    \node[above] at (1.5,1) {IN};
     \node[above] at (17.5,1) {OUT};
    \end{tikzpicture}
    \caption{A schematic depiction of our algorithm. As described in the main text, we begin by fixing the rank $r$ and dimension $d$ of the MDE, and then inputting the first few terms of the Fourier expansion of a candidate vacuum character $\chi_0$. Various methods (given above the arrows) are then used to extract increasingly refined sets of 4d data from these inputs. At each step we impose constraints (given below the arrows) on the new data, which if failed truncate the algorithm. The output of the algorithm is the data that pass all of our constraints. }
\label{fig:outline}
\end{figure}
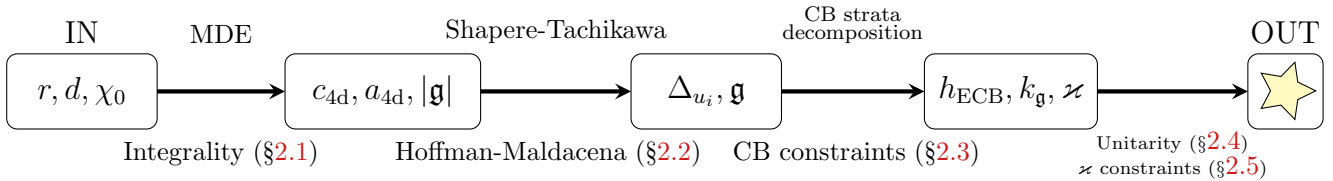

 Concretely, the algorithm takes as input $r$, $d$, and the first few terms in the Fourier expansion of a candidate vacuum character (the precise number of terms is determined by $d$, as will be reviewed in Section \ref{sec:MDEconstrs}). With this input, we begin by using the MDE to verify that these first terms can be ``completed" to a fully integral solution $\chi_0(q)$. Assuming integrality is satisfied, the MDE is then used to extract the possible central charges $(a_{\mathrm{4d}},c_{\mathrm{4d}})$ as well as the dimension of the flavor symmetry of the corresponding 4d $\cN=2$ theory. This is discussed in Section \ref{sec:centralcharge}. This data must satisfy, among other things, the Hoffman-Maldacena bounds. Assuming that it does, the Shapere-Tachikawa formula can be used to extract the Coulomb branch scaling dimensions $\Delta_{u_i}$. The flavor symmetry $\mathfrak{g}$ can also be restricted to a finite set of possiblities. The data $\Delta_{u_i}$  and $\mathfrak{g}$ must satisfy a number of stringent Coulomb branch consistency conditions, which we detail in Section \ref{sec:CBconsts}. As we will see, satisfying these conditions is roughly equivalent to constructing a consistent CB geometry. Assuming that this is possible, the stratification of the CB geometry can then be used to obtain more refined quantities such as the flavor levels $k_{\mathfrak{g}_i}$, which are subject to various unitarity constraints discussed in Section \ref{sec:unitsubsec}. Finally, a more subtle constraint is discussed in Section \ref{sec:chardim}. In the end, the algorithm outputs the sets of data which satisfy all of these constraints. A pictorial summary of this procedure is given in Figure \ref{fig:outline}. 
 
Since imposing each of these steps by hand is rather laborious, the entire process is automated on a computer.
Automation also allows for the following strategy. Instead of inputting a single vacuum character $\chi_0(q)$ and checking for a consistent 4d interpretation, we can now scan over large numbers of tentative vacuum characters (again fixing the rank $r$ and the dimension $d$ of the MDE). Since the input for the algorithm is not actually the vacuum character itself but rather just the first few terms of its Fourier expansion, this amounts to a scan over tuples of integers. As long as the integers scanned over take values in a sufficiently large range, we can be fairly confident that our algorithm will \textit{discover all 4d $\cN=2$ theories with fixed $(r,d)$}.

 As a proof of concept, we implement such a scan in what is arguably the simplest non-trivial case: namely rank-two theories whose Schur indices satisfy a fourth-order untwisted MDE. As will be reviewed below, such MDEs---as well as their solutions---are specified by three terms in the Fourier expansion of $\chi_0(q)$. We must then scan over these three parameters, and for each triplet apply our algorithm to check if the proposed vacuum character admits a 4d interpretation. Remarkably, of the order $O(10^{11})$ candidate triplets that we examine,  only 15 pass the full set of moduli space and unitarity constraints imposed in our algorithm. The data of these 15 cases is summarized in Table \ref{tab:results} and Figure \ref{fig:newHasse}. An even more careful analysis (using tools which are difficult to automate on a computer) allows us to rule out nine of the 15  theories,\footnote{More precisely, seven candidates will be rigorously ruled out, while two will be shown to be implausible.} leaving only six candidates. Comparing to the lists in e.g.~\cite{Martone:2021ixp}, all six candidates can be seen to correspond to known physical $\cN=2$ theory. 
 
 The fact that our scan does not produce any new candidate theories is perhaps surprising, as we did not anticipate our current knowledge of ${\cal N}=2$ SCFTs to be so exhaustive. However this 
 is likely an artifact of the restriction to low order and low rank. 
The extension to higher orders and higher ranks is conceptually straightforward,  but quickly becomes computationally intensive, and so we have not attempted to pursue it here. With appropriate computer implementation, we hope that these difficulties can be eventually overcome. 
We will comment on the feasibility of scaling up to these richer setups in the concluding section.

\paragraph{Organization:} The organization of this paper is as follows. We begin in Section \ref{sec:generalstrat} by introducing our general algorithm, which involves starting with a monic MDE of fixed order, searching for solutions that are integral, and imposing various four-dimensional Coulomb branch and unitarity constraints to determine whether the solutions can correspond to legitimate 4d Schur indices. A visual summary of this algorithm is given in Figure \ref{fig:outline}. In Section \ref{sec:examplesubsec} we illustrate the algorithm with a simple example---namely, we use our techniques to rediscover the rank-two $D_2(SU(5))$ theory of \cite{Cecotti:2013lda}. Then in Section \ref{sec:Results} we use the algorithm to do a computerized search for rank-two  $\cN=2$ theories with flavor symmetry having up to three simple factors, three $U(1)$ factors and with corresponding VOA satisfying a fourth-order untwisted MDE. We rediscover six known 4d $\cN=2$ theories, and  identify nine candidate novel theories. By using more refined Coulomb and Higgs branch constraints, we are able to rigorously rule out seven of the candidates, and to argue that the remaining two are also unlikely to be physical. Finally, in Section \ref{sec:conclusions} we conclude with a discussion of future directions. 

We also include a number of appendices. Appendix \ref{superconformal:subsection} gives a review of some relevant superconformal representation theory, and derives some constraints which will be used in our analysis. In Appendix \ref{app:rank2dims}, we rederive the list of allowed pairs of CB scaling dimensions for rank-two theories, with results  differing slightly from those in the literature. Appendices \ref{app:IRFree} and \ref{app:ECB} review  facts about rank-one IR-free theories and extended Coulomb branches, both of which will be relevant for our scan. We also include Appendix \ref{app:T9VOA}, which gives a tangential but interesting discussion of one of the VOAs identified in our scan.

\section{General strategy}
\label{sec:generalstrat}
In this section we describe the algorithm used to identify sets of protected data (i.e. Schur and Coulomb) that satisfy a variety of CB and unitarity constraints. An explicit example will be given in Section \ref{sec:examplesubsec}, and the results of a computer implementation will be given in Section \ref{sec:Results}. 

\subsection{Modularity and integrality constraints}
\label{sec:MDEconstrs}
 Our program begins in two dimensions and relies crucially on the fact that the Schur index of any 4d $\cN=2$ superconformal field theory can be interpreted as the vacuum character of a 2d VOA. As mentioned in the introduction, for a generic VOA there is no reason to expect that the vacuum character should have simple modular properties.  We will take as a given the central conjecture of \cite{Beem:2017ooy},  which equates the Higgs branch of the  4d $\cN=2$ theory with the {\it associated variety} of the corresponding VOA. A corollary of this conjecture is that any VOA with a 4d avatar is ``quasi-lisse'',\footnote{We remind the reader that quasi-lisse VOAs \cite{Arakawa:2016hkg} are those whose associated varieties are symplectic, or more generally symplectic singularities \cite{brieskorn1970singular, slodowy1980simple, beauville1999symplectic} with a finite number of symplectic leaves.} and as such its vacuum character must satisfy  a \textit{monic} modular differential equation (MDE) \cite{Arakawa:2016hkg}.\footnote{Explicit closed form expressions for Schur indices in terms of quasi-modular forms  were recently obtained in \cite{Pan:2021mrw, Beem:2021zvt} for a large class of SCFTs.}

Depending on whether we require modular invariance under the full modular group or just an index two subgroup \cite{Beem:2017ooy,Bae:2020xzl,Bae:2021mej}, an MDE is called respectively untwisted or twisted. The space of possible MDEs is labelled by a finite set of real parameters, the number of which depends only on the degree of the differential equation and whether it is untwisted or twisted. In this work we will focus mainly on untwisted MDEs, of which we will give some examples below.\footnote{The twisted case is required if one allows for Schur operators with half-integer chiral dimension, and as such is the most general case. The untwisted case is a specialization of the twisted one (allowing for Schur operators of only integer dimension), and thus at a given order the untwisted case involves a smaller number of real parameters than the twisted one.}

An important property of any vacuum character is that it must have unit leading coefficient, and that the remaining Fourier coefficients must be integral. Not every choice of MDE, i.e.~of the real parameters which determine it, is compatible with such integrality. A first step in our classification program is to identify the MDEs which do give rise to a sensible vacuum character. In the context of  2d CFT, this idea goes back to the classic work of Mathur, Mukhi, and Sen~\cite{Mathur:1988rx, Mathur:1988na, Mathur:1988gt}, 
see e.g.~\cite{Gaberdiel:2008pr,Hampapura:2015cea, Gaberdiel:2016zke, Hampapura:2016mmz, Mukhi:2017ugw, Chandra:2018pjq, Mukhi:2019xjy, Mukhi:2020gnj, Bae:2020xzl, Das:2020wsi, Mason:2021xfs, Kaidi:2020ecu,Kaidi:2021ent,Das:2021uvd,Bae:2021mej,Bae:2021jkc} for recent developments. As scanning over the set of real numbers parameterizing the MDE would clearly be impossible, our first step is to map the problem to a scan over integers. This can be done as follows:
\begin{itemize}
    \item[1.] We fix the order and the type of the MDE, i.e.~untwisted or twisted. This choice determines the number $n$ of real parameters that must be specified to define the MDE.
    
    \item[2.] Instead of specifying the $n$ real parameters directly, we specify an equal number of integral Fourier coefficients of a putative vacuum character $\chi_0$, and then fix the parameters of the MDE by demanding that the given vacuum character is a solution up to the appropriate order in $q$. As will be elaborated on below, certain parameters of $\chi_0$ have immediate physical interpretation and thus are subject to additional constraints. For example, the leading exponent of $\chi_0$ is related to the central charge $c_{\mathrm{4d}}$ of the 4d $\cN=2$ SCFT, while the $O(q)$ Fourier coefficient gives the number of moment map operators, i.e.~the dimension of the flavor symmetry. The higher-order Fourier coefficients do not have such straightforward physical interpretations, but they must be consistent with the fact that they count operators of the 4d theory, and in particular they must organize  into representations of the flavor symmetry. This is discussed in detail in Appendix \ref{superconformal:subsection}.
    
    \item[3.] Since each coefficient of $\chi_0$ fixes one real parameter of the MDE,  step 2  fixes the MDE completely. We now take the MDE constructed above as an input and ``complete'' $\chi_0$ by assuming that it is a solution to all orders in $q$. If all Fourier coefficients (at least up to some reasonable cutoff which will depend on the order of the MDE) are integral, we will take $\chi_0$ to be a candidate vacuum character and proceed. If the Fourier coefficients are \textit{not} integral, then $\chi_0$ is discarded. 
    
\end{itemize}

Let us now briefly discuss the lowest-order monic untwisted MDEs, which will be the focus of the rest of this work.
Denoting the order of the MDE by $d$, the most general MDE is given by
\bea
\label{eq:basicMDEs}
&d=2:& \hspace{0.5 in} \left[ D^{(2)} + \mu_1 E_4 \right] \chi = 0~,
\no\\
 &d=3:&\hspace{0.5 in}  \left[ D^{(3)} + \mu_1 E_4 D^{(1)} + \mu_2 E_6\right] \chi = 0~,
\no\\
&d=4:&\hspace{0.5 in} \left[ D^{(4)} + \mu_1 E_4 D^{(2)} + \mu_2 E_6 D^{(1)} + \mu_3 E_4^2 \right] \chi = 0~,
\eea
where $E_i$ are the holomorphic Eisenstein series, and the $D^{(k)}$ are the $k$-th order modular covariant derivatives acting on weight-zero modular forms, 
\bea
D^{(k)} : = \prod_{s=1}^k D_{2s-2}~, \hspace{0.8 in} D_w := {1\over 2 \pi i }{d\over d \tau} - {w\over 12} E_2(\tau)~.
\eea
Clearly the equations (\ref{eq:basicMDEs}) depend respectively on one, two, and three real parameters, denoted by~$\m_i$. 

Every solution of an untwisted MDE admits a Fourier expansion in integer powers of $q:=e^{2 \pi i \tau}$. In particular, the vacuum character will admit an expansion 
\bea
\chi_0(q)  = q^{- {c_{\mathrm{2d}}\over 24}}(1+\sum_{n >0} a_n q^n) = q^{- {c_{\mathrm{2d}}\over 24}}(1 + a_1 q + a_2 q^2 + a_3 q^3 + \dots)~,
\eea
with $c_{\mathrm{2d}}$ the 2d central charge. As we have already mentioned, we will not take the full vacuum character as an input, but rather only the first few terms in its Fourier expansion. We then assume that these first few terms provide a solution to one of the MDEs in (\ref{eq:basicMDEs}). Since the $d=2,3,4$ MDEs are specified by respectively $1,2,3$ free parameters, only this number of terms in the expansion of $\chi_0(q)$ are needed to fix the MDE uniquely. In other words, the input data will be taken to be
\bea
\label{eq:initdata}
&d=2:& \hspace{0.5 in} \{c_{\mathrm{2d}}\}~,
\no\\
 &d=3:&\hspace{0.5 in} \{c_{\mathrm{2d}}, a_1\}~,
\no\\
&d=4:&\hspace{0.5 in} \{c_{\mathrm{2d}}, a_1, a_2\}~.
\eea
Solving the MDE order-by-order in $q$, we may then fix the parameters $\m_i$ in terms of this input data as
\bea
&d=2:& \hspace{0.5 in} \mu_1 = -{1\over 576} c_{\mathrm{2d}}(c_{\mathrm{2d}}+4)~,
\no\\\no\\\no
 &d=3:&\hspace{0.5 in} \mu_1 = {c_{\mathrm{2d}}(21c_{\mathrm{2d}}^2 + 240\,c_{\mathrm{2d}} + 704) - a_1(3c_{\mathrm{2d}}^2 - 48\, c_{\mathrm{2d}} + 320) \over 576(a_1 - 31 c_{\mathrm{2d}})}~,
 \no\\
 &\vphantom{.}&\hspace{0.5 in} \mu_2 = - {c_{\mathrm{2d}}^2(5 c_{\mathrm{2d}}^2 + 66\, c_{\mathrm{2d}} + 144)+ a_1\,c_{\mathrm{2d}}(c_{\mathrm{2d}}^2 - 30 \,c_{\mathrm{2d}} + 144) \over 6912(a_1 - 31c_{\mathrm{2d}})}~,
\eea
and similarly for $d=4$. 

Having fixed the MDE completely in terms of the data (\ref{eq:initdata}), we now solve the MDE order-by-order to obtain the vacuum character to arbitrarily high order in $q$. Assuming that it remains integral at all orders (in practice, up to some reasonably high cut-off) we now ask whether there can exist a 4d $\cN=2$ SCFT with that vacuum character as its Schur index.

\subsection{Central charges} 
\label{sec:centralcharge}
Assume that we have found an MDE compatible with integrality. This gives us a candidate Schur index for a candidate 4$d$ $\cN=2$ SCFT. In addition to the vacuum character, an order-$d$ MDE will generically have $(d-1)$ non-vacuum solutions, each with an associated chiral dimension $h_i$. These dimensions are obtained as follows. Assume that the solutions to the MDE take the form 
\bea
\chi_i(q) = q^{s_i} \sum_{n} a_n^{(i)} q^n~,
\eea
for $i = 0,\dots, d-1$. The leading exponent of the vacuum character $\chi_0(q)$ is $s_0 = -{c_{2d} \over 24}$, while the leading exponents of the remaining characters are $s_i := h_i - {c_{2d} \over 24}$. These leading exponents are given by the roots of the ``indicial equation,"  i.e. the order $O(q^s)$ portion of the MDE,
\bea
\label{eq:indicialeq}
&d=2:& \hspace{0.5 in}  s^2 - {s \over 6} + \mu_1 = 0~,
\no\\
 &d=3:&\hspace{0.5 in}  s^3 - \half s^2  + \left( \mu_1 + {1\over 18}\right) s + \mu_2 = 0~,
\no\\
&d=4:&\hspace{0.5 in}  s^4 - s^3 + \left( \mu_1 + {11\over 36}\right) s^2 -\left({1\over 36} + {\mu_1 \over 6} - \mu_2\right)s + \mu_3= 0~.
\eea
Hence the chiral dimensions can be identified by computing the indicial roots. 

With the chiral dimensions in hand, it is now possible to restrict the central charges of the putative 4d $\cN=2$ theory to a discrete set of possibilities. Indeed, the following standard formulae can be used \cite{Beem:2013sza,Cecotti:2015lab,Beem:2017ooy},\footnote{To be precise, in the twisted case the $h_*$ appearing in (\ref{eq:ca4d}) is one of the chiral dimensions of the \textit{$S$-tranformed} VOA---see e.g. \cite{Beem:2017ooy}. Here we are focussing on the untwisted case.}
\bea
\label{eq:ca4d}
c_{{\mathrm{4d}}} = -{ c_{{\mathrm{2d}}} \over 12}~, \hspace{0.5 in} a_{\mathrm{4d}} = {1\over 48}(24 h_* - 5 c_{\mathrm{2d}})~.
\eea
When $c_{{\mathrm{4d}}} > a_{{\mathrm{4d}}}$, the $h_*$ appearing above is required to be the \textit{minimal} chiral dimension $h_* = \mathrm{min} \{ h_i\}$ \cite{DiPietro:2014bca}. When $c_{{\mathrm{4d}}} < a_{{\mathrm{4d}}}$ this need no longer be the case, and it is in fact unclear if $h_* \in \{h_i\}$ at all. But it is still seems reasonable to expect that $h_*$ takes values in $\{h_i\}$ mod 1. The argument here relies on the study of the high-temperature behavior of the index \cite{ArabiArdehali:2015ybk,DiPietro:2014bca}. For Lagrangian $\cN=2$ theories, it can be shown that there always exists a contribution to the high-temperature behavior of the Schur index with exponent proportional to $c_{\rm 4d} - a_{\rm 4d}$. Since the Schur index (or really the vacuum character) forms a vector-valued modular function together with the other solutions to the MDE, its $S$ transform is a linear combination of these other solutions, and thus every exponent of the high-temperature index is of the form $h_i -{c_{\mathrm{2d}} \over 24}+ m$ for some $h_i \in \{h_i\}$ and $m \in \NN$. Together these imply that $h_* = h_i + m$, and hence for Lagrangian theories we indeed have $h_* \in \{h_i\}$ mod 1. All known examples suggest that this result holds for non-Lagrangian theories as well, and we will assume it to be true here. 

We are then left with a discrete but potentially infinite set of possibilities for $( a_{\mathrm{4d}},c_{\mathrm{4d}})$. This set is made finite by imposing the Hofman-Maldacena bounds \cite{Hofman:2008ar},
\bea
\half \leq {a_{\mathrm{4d}} \over c_{{\mathrm{4d}}}} \leq {5\over 4}~.
\eea
Indeed, in the case that $c_{{\mathrm{4d}}} < a_{{\mathrm{4d}}}$, this will restrict the quantity $m \in \NN$ to a finite number of possible values (since $a_{\mathrm{4d}}$ increases linearly with $m$ while $c_{\mathrm{4d}}$ is independent). 

\subsection{Coulomb geometry constraints}
\label{sec:CBconsts}

We now have a finite list of possible $( a_{\mathrm{4d}},c_{\mathrm{4d}})$ for the candidate 4d $\cN=2$ theory. We next consider constraints imposed by consistency of the Coulomb branch (CB). Let's start by reviewing some basic facts about CBs of $\cN=2$ SCFTs. If the CB is spanned by $r$ complex coordinates $\bu=\{u_i\}$, $i=1,...,r$, then the $\cN=2$ theory is said to be a \emph{rank}-$r$ SCFT. Henceforth we will assume that the CB chiral ring of the 4d $\cN=2$ theory is freely-generated, in which case it is possible to choose the coordinates of the CB to have globally defined scaling dimensions $ \Delta_{u_i}$ for $i=1,...,r$. There exist constraints on the possible values of these scaling dimension. For example, at rank one it turns out that there are only seven allowed values \cite{Argyres:2018urp,Caorsi:2018zsq},
\bea
\label{eq:rank1list}
\Delta \in \left \{6, 4, 3, 2, {3\over 2}, {4\over 3}, {6 \over 5} \right\}~.
\eea
At rank two, there are again a finite number of possibilities for $\Delta_{u_1}$ and $\Delta_{u_2}$. In fact, now there are constraints not only on the individual scaling dimensions, but also on the allowed pairs. In total, there are 79 allowed pairs, the full list of which is provided in Table \ref{tab:CBSDr2} of Appendix \ref{app:rank2dims}  (note that the results here follow the revised analysis given in \cite{Argyres:2022yet}, differing slightly from those in \cite{Caorsi:2018zsq}). 
This analysis can be extended, with enough effort and dedication, to any desired rank, though the number of allowed \emph{r}-tuples increases considerably.

\label{sec:CBsubsec}
We now fix the rank of the putative theory to some particular $r$. The CB scaling dimensions are known to be related to the central charges via the Shapere-Tachikawa formula \cite{Shapere:2008zf},\footnote{There are known exceptions to this formula in the case of theories obtained by discrete gauging of   ``more fundamental" theories.  We will assume that exceptions can only arise in this way. } 
\bea
\label{eq:ShapTach}
\sum_{i=1}^{r}\Delta_{u_i} = 2 (2 a_{\mathrm{4d}} - c_{\mathrm{4d}})+{r\over 2}~.
\eea

Returning to our algorithm, we may now use the finite list of allowed $( a_{\mathrm{4d}},c_{\mathrm{4d}})$ obtained from the MDE to compute a finite list of candidate sums of CB scaling dimensions. Recall that the CB dimensions themselves take values in a finite list, c.f. Appendix \ref{app:rank2dims}, which means that their {sums} do as well. For example, in the case of rank-two the sums $\Delta_{u_1}+ \Delta_{u_2}$ are constrained to take one of only 68 distinct values. 
If the value computed via the Shapere-Tachikawa formula is not on the list of allowed rank-$r$ sums, then that particular pair $( a_{\mathrm{4d}},c_{\mathrm{4d}})$ can be discarded. If every one of the finite number of possible pairs $( a_{\mathrm{4d}},c_{\mathrm{4d}})$ is discarded, then there is no 4d theory corresponding to the input data. If on the other hand at least one pair $( a_{\mathrm{4d}},c_{\mathrm{4d}})$ gives rise to a sum on the list,   we proceed. 

Our next step is to read off all possible  $\D_{u_i}$ adding up to the desired sum. For low enough rank there is usually a unique choice, though in some cases there are multiple possibilities. 
As examples, at rank two one might have
\bea
\Delta_{u_1} + \Delta_{u_2} &=&  {78 \over 11} \hspace{0.5 in} \Rightarrow \hspace{0.5 in}(\Delta_{u_1}, \Delta_{u_2}) = \left({12 \over 11}, 6 \right) ~,
\no\\
\Delta_{u_1} + \Delta_{u_2} &=& 18 \hspace{0.5 in} \Rightarrow \hspace{0.5 in}(\Delta_{u_1}, \Delta_{u_2}) \in\left\{  (6,12) , (8, 10) \right\} ~,
\eea
as the reader may confirm by consulting Appendix \ref{app:rank2dims}. We thus have identified a relatively small (or empty) list of potential $\{\Delta_{u_i}\}$ for each possible $(a_{\mathrm{4d}},c_{\mathrm{4d}})$.

Next we will need more detailed knowledge of the geometric structure of the Coulomb branch. The Coulomb branch can have singular loci $\{\cI_a\}$ falling into two broad classes \cite{Argyres:2018zay}: 
\begin{itemize}
\item Unknotted: Locus defined by $u_i=0$~,
\item Knotted: Locus defined by $P(\bu) = 0$, with $P$ an irreducible homogeneous polyonomial in $\bu$~.
\end{itemize}
We may associate to each singular locus $\cI_a$ a value  $\Delta_a^{\mathrm{sing}}$ which is the degree of homogeneity of the polynomial specifying the stratum, e.g. for unknotted strata $\D^{\rm sing}\equiv \D_{u_i}$. Each singular locus can host a rank-one theory, either conformal or IR-free. The full set of conformal rank-one theories is known \cite{Argyres:2015ffa,Argyres:2015gha,Argyres:2016xua,Argyres:2016xmc}, and since we will make heavy use of them we include the relevant data in Table \ref{Theories}. On the other hand, the full set of IR-free rank-one theories is infinite, but each such theory takes the form of $U(1)$ or $SU(2)$ gauge theory coupled to sufficient numbers of hypermultiplets in arbitrary representations. In practice, when we implement this algorithm on a computer it is necessary to restrict to a finite subset of these IR-free theories. The particular subset which we choose is discussed in Appendix \ref{app:IRFree}. We believe this to be a well-motivated representative sample of the full space of IR-free theories.   

\begin{table}[tp]
\hspace*{-0.8cm}
\def\arraystretch{1}%
\begin{tabular}{c|c|c|c|c|c|c|c|cc}
Name & \multicolumn{1}{c|}{$12\, c_{\mathrm{4d}}$ } & $\D_u$ &$h_{\mathrm{ECB}}$&$\mathbf{R}$ & $T(\mathbf{R})$& $b$&$\mathfrak{g}$&$ k_{\mathfrak{g}}$  \\
\hhline{=========}
$ \cT^{(1)}_{E_8,1}$ &62&6&0&${\mathbf1}$&0& 10&$\ef_8$&12 
\\
$\cT^{(1)}_{E_7,1}$ &38&4&0&${\mathbf1}$&0&9&$\ef_7$&8\\
 $\cT^{(1)}_{E_6,1} / [\cT^{(1)}_{E_6,1}]_{\ZZ_2}$ &26&3&0&${\mathbf1}$&0&8&$\ef_6/ \mathfrak{f}_4$&6\\
$\cT^{(1)}_{D_4,1}/[\cT^{(1)}_{D_4,1}]_{\ZZ_2}/[\cT^{(1)}_{D_4,1}]_{\ZZ_3}$ &14&2&0&${\mathbf1}$&0&6&$\sof(8)/ \mathfrak{so}(7)/\mathfrak{g}_2$&4\\
$\cT^{(1)}_{A_2,1}/[\cT^{(1)}_{A_2,1}]_{\ZZ_2}/[\cT^{(1)}_{A_2,1}]_{\ZZ_3}$ &8&$\frac32$&0&${\mathbf1}$&0&4&$\suf(3)/\suf(3)/\suf(2)$&3\\
$\cT^{(1)}_{A_1,1}/[\cT^{(1)}_{A_1,1}]_{\ZZ_2}$ &6&$\frac43$&0&${\mathbf1}$&0&3&$\suf(2)/\suf(2)$&$\frac83$\\
 $\cT^{(1)}_{\varnothing,1}$ &$\frac{22}5$&$\frac65$&0&${\mathbf1}$&0&2&$\varnothing$&$\star$\\
\cdashline{1-9}
$\cS^{(1)}_{E_6,2}$&49&6&5&${\mathbf{10}}$&1&7&$\spf(10)$&7&\\
$\cS^{(1)}_{D_4,2}$&29&4&3&$({\mathbf{6,1}})$&1&6&$\spf(6){\times} \suf(2)$&(5,8)&\\
$\cS^{(1)}_{A_2,2}/[\cS^{(1)}_{A_2,2}]_{\ZZ_2}$&19&3&2&${\mathbf4}_0$&1&5&$\spf(4){\times} \uf(1)/ \mathfrak{sp}(4)$&$(4,\star)$\\
$\blue{\cS^{(1)}_{\varnothing,2}}$&9&2&1&${\mathbf2}$&1& 3&$\suf(2)$&3\\
\cdashline{1-9}
$\cS^{(1)}_{D_4,3}$&42&6&4&${\mathbf4}\oplus\bar{\mathbf 4}$&2&6&$\suf(4)$&14\\
$\cS^{(1)}_{A_1,3}$&24&4&3&${\mathbf2}_+\oplus{\mathbf 2}_-$&2&5&$\suf(2){\times}\uf(1)$&$(10,\star)$\\
$\green{\cS^{(1)}_{\varnothing,3}}$&15&3&1&${\mathbf1}_+\oplus{\mathbf 1}_-$&0&4&$\uf(1)$&$\star$ \\
\cdashline{1-9}
$\cS^{(1)}_{A_2,4}$&38&6&3&${\mathbf3}\oplus\bar{\mathbf 3}$&2&$\frac{11}2$&$\suf(3)$&14\\
$\green{\cS^{(1)}_{\varnothing,4}}$&21&4&1&${\mathbf1}_+\oplus{\mathbf 1}_-$&0&$\frac92$&$\uf(1)$&$\star$
\end{tabular}
\caption{\small The list of rank-one SCFTs which can describe the low-energy physics on codimension-one singular loci of the Coulomb branch. Additional information can be found in \cite{Argyres:2016yzz}. Theories in green are $\cN=3$, while the  theory in blue is $\cN=4$.
\label{Theories}}
\end{table}%

The singular loci are further constrained by the following three results \cite{Argyres:2018urp,Martone:2020nsy,Martone:2021ixp}: 

\begin{proposition}
\label{thm:nonzeroknot}
If a theory of rank $r>1$ is irreducible, i.e.~not the product of lower rank theories, then there must  be at least one knotted stratum. 
\end{proposition}

\begin{proposition}
\label{thm:rank1thm}
An unknotted stratum ${u_i}=0$ is allowed only if the $r-1$ scaling dimensions $(\Delta_{u_1},...,\Delta_{u_{i-1}},\Delta_{u_{i+1}},...,\Delta_{u_r})$ give an allowed $(r-1)$-tuple of rank-$(r-1)$ Coulomb branch dimensions. 
\end{proposition}

\begin{proposition}[UV-IR Flavor constraint]
\label{thm:UVIR}
For a theory with flavor symmetry $\mathfrak{g} = \mathfrak{u}(1)^n \bigoplus_i \mathfrak{g}_i$ with $\mathfrak{g}_i$ simple, each $\mathfrak{g}_i$ must be realized on at least one codimension-one stratum. 
\end{proposition}

These results prove highly constraining. To see this, let us begin by considering the flavor symmetry of our putative $\cN=2$ theory. Recall that the  Fourier coefficient $a_1$ of $\chi_0$ counts the number of moment map operators, and thus fixes the dimension of the flavor algebra of our putative theory. This quantity will either be one of the inputs used to define the MDE (for $d=4$) or will be an output obtained by solving the MDE at order $q$ (for $d=2,3$); c.f. (\ref{eq:initdata}).  

Clearly $a_1$ alone does not determine the flavor symmetry. Indeed, there can be many Lie algebras with a given dimension $a_1$. For example, if $a_1=248$, then we can have the following decompositions
\bea
\{248\},\, \{24,224\}, \,\{80,168\},\, \{3,21,224\},\,\dots
\eea
 which would be interpretable as the following Lie algebras,
 \bea
 \mathfrak{e}_8, \,\,\mathfrak{su}(5) \oplus \mathfrak{su}(15), \,\,\mathfrak{su}(9) \oplus \mathfrak{su}(13), \,\,\mathfrak{su}(2) \oplus \mathfrak{sp}(6) \oplus \mathfrak{su}(15),\,\, \dots
 \eea
 If we allow for arbitrary numbers of $\mathfrak{u}(1)$ factors, then clearly we should consider all integer partitions of $a_1$. In practice we will set an upper bound on the number of $\mathfrak{u}(1)$ factors.

Given a choice of flavor symmetry, we may now attempt to reconstruct parts of the CB geometry. To begin, Proposition \ref{thm:UVIR} tells us that each simple flavor factor must be realized on a codimension-one locus of the CB. Hence if we take the flavor symmetry to be e.g. $\mathfrak{e}_8$, then we must identify a rank-one theory which supports that flavor symmetry. Consulting Table~\ref{Theories}, we see that the only possibility is $\cT_{E_8,1}^{(1)}$. On the other hand, if we take the flavor symmetry to be $\mathfrak{su}(2) \oplus \mathfrak{sp}(6) \oplus \mathfrak{su}(15)$, then $\mathfrak{su}(15)$ must be realized by a $[I_{15}, \mathfrak{su}(15)]$ stratum, whereas $\mathfrak{su}(2)$ can be realized in three possible ways via $\cT_{A_1,1}^{(1)}, \cS_{0,2}^{(1)},$ and $[I_2, \mathfrak{su}(2)]$, and $\mathfrak{sp}(6)$ in two possible ways via $[I_6, \mathfrak{su}(6)]_{\ZZ_2}$ and $[I_2^*, \mathfrak{sp}(6)]$. In total then, we see that the $\{248\}$ decomposition is realizable  on the Coulomb branch in only a single way, whereas the $\{3,21,224\}$ decomposition is realizable in six ways. The total number of Coulomb branch geometries that can capture a given $a_1$ is given by summing over the different integer partitions of $a_1$, weighted by the number of different ways of realizing a given decomposition on the CB strata.

Note that simple flavor factors cannot enhance their ranks on any strata, except perhaps for the appearance of additional $\mathfrak{u}(1)$ factors \cite{Martone:2020nsy}. This means, for example, that if we had a theory with $\mathfrak{e}_7$ flavor symmetry, we would require a $\cT_{E_7,1}^{(1)}$ stratum to satisfy Proposition \ref{thm:UVIR}, as opposed to a $\cT_{E_8,1}^{(1)}$ stratum, even though the latter technically contains $\mathfrak{e}_7 \subset \mathfrak{e}_8$ flavor symmetry. On the other hand, rank-preserving enhancements are technically allowed. For simplicity we will assume that such enhancements do not take place.

With these data about the strata, together with the lists of possible $\{\Delta_{u_i}\}$ identified before, we are now ready to proceed to the next step. 
Let us denote the set of all Coulomb branch strata as $\cI = \{ \cI_a\}$, and the set of strata carrying the UV simple flavor factor $\mathfrak{g}_i$ by $\cI_{\mathfrak{g}_i} \subset \cI$. We may now make use of the following formulas relating the data of the theory at the origin to the data of the theories living on each stratum (in the absence of flavor symmetry enhancements) \cite{Martone:2020nsy,Martone:2021ixp}:
\bea
\label{eq:mainstrateqs}
h_{\mathrm{ECB}}  &=& 12 c_{\mathrm{4d}}- 2r - \sum_{\cI_a \in \cI} b_a \,\Delta_a^{\mathrm{sing}}~,
\no\\
k_{\mathfrak{g}_i} &=&  \sum_{\cI_a \in \cI_{\mathfrak{g}_i}} {\Delta_a^{\mathrm{sing}} \over \Delta_a} [k_a - T^i(\mathbf{R}_a)]+ T^i(\mathbf{R})~,\,\,\,\,\,
\eea 
where $h_{\mathrm{ECB}} $ is the dimension of the extended Coulomb branch and $\Delta_a^{\mathrm{sing}}$ was defined above Proposition \ref{thm:nonzeroknot}.\footnote{A theory with a non-trivial extended Coulomb branch (ECB) is one such that a generic point of the Coulomb branch contains free hypermultiplets. The quantity $h_{\mathrm{ECB}}$ is the quaternionic dimension of the ECB, meaning that the Coulomb branch of a theory with ECB of dimension $h_{\mathrm{ECB}}$ is effectively enlarged to an $r+2h_{\rm ECB}$ complex-dimensional space, which looks locally like a product CB$\times \mathbb{H}^{h_{\mathrm{ECB}}}$ with $\mathbb{H}$ representing a free hypermultiplet. One must be careful in distinguishing the case where the CB looks like CB$\times \mathbb{H}^{h_{\mathrm{ECB}}}$ at \textit{all} points (a \textit{free} ECB), or only away from singular loci (a \textit{coupled} ECB)---see Appendix \ref{app:ECB} for details.} The quantity $T^i(\mathbf{R})$ is somewhat more subtle, and will be described in Appendix \ref{app:ECB}.  Finally, the quantities $b_a,\, \Delta_a,$ and $T^i(\mathbf{R}_a)$ depend only on the rank-one theories living on $\cI_a$, and are collected in Table \ref{Theories}.

For each choice of decomposition of $a_1$, and each choice of strata realizing this decomposition, we can now compute $h_{\mathrm{ECB}}$ and $k_{\mathfrak{g}_i}$. Note that $h_{\mathrm{ECB}}$ must be a non-negative integer. If we compute $h_{\mathrm{ECB}}$ and find $h_{\mathrm{ECB}}\notin \NN$, we declare that case ruled out.  Otherwise, we take the values of $k_{\mathfrak{g}_i}$ and proceed to the next step.

Before moving on, let us summarize what we have done so far. We began by inputting the first few Fourier coefficients of the vacuum character and using them to determine a finite list of possible $( a_{\mathrm{4d}},c_{\mathrm{4d}})$ for the tentative 4d theory. For each element of this list, we then used the Shapere-Tachikawa formula to obtain $\sum_{i=1}^r \Delta_{u_i}$, from which we obtained a finite number 
of possible $r$-tuples $(\Delta_{u_1}, \dots, \Delta_{u_r})$. At the same time, the quantity $a_1$ was used to obtain a finite number of possible flavor symmetries $\mathfrak{g} = \mathfrak{u}(1)^n\bigoplus_i \mathfrak{g}_i$, each of which could be realized on the CB strata in a finite number of ways. 
We may then consider all possible combinations of  $r$-tuples $(\Delta_{u_1}, \dots, \Delta_{u_r})$ and CB stratifications and compute $h_{\mathrm{ECB}}$ and $k_{\mathfrak{g}_i}$ for each. Our goal is now to impose enough constraints to whittle this large number of candidate cases down to a small number of physical cases. Some first constraints are provided by Propositions \ref{thm:nonzeroknot}, \ref{thm:rank1thm}, and \ref{thm:UVIR}, as well as by requiring that $h_{\mathrm{ECB}} \in \NN$. The rest of this section will describe additional constraints.

\subsection{Unitarity constraints}
\label{sec:unitsubsec}
Four-dimensional unitarity can be used to obtain stringent constraints on the data obtained thus far. As we now review, these constraints include lower bounds on $k_{\mathfrak{g}} $, as well as upper and lower bounds on $c_{{\mathrm{2d}}}$ \cite{Beem:2017ooy,Beem:2018duj}. Beginning with the former, unitarity demands that 
\bea
\label{eq:kunitbd}
k_{\mathfrak{g}} \geq \widetilde k_{\mathfrak{g}}~,
\eea
where $\widetilde k_{\mathfrak{g}}$ is a fixed value associated to each $\mathfrak{g}$. The values of $\widetilde k_{\mathfrak{g}}$ are collected in Table \ref{tab:kunit}. If we have $\mathfrak{g} = \mathfrak{u}(1)^n \bigoplus_i \mathfrak{g}_i$, these bounds must be satisfied for each simple factor $\mathfrak{g}_i$. 

\begin{table}[tp]
\begin{center}
\begin{tabular}{cc|c|c}
\multicolumn{2}{c|}{$\mathfrak{g}$} &$\widetilde k_{\mathfrak{g}}$& $h^\vee$
\\\hline\hline
$\mathfrak{su}(N)$& $N \geq 3$,& $ N$& $N$
\\
$\mathfrak{so}(N)$&$N = 4,\dots,8$ & $4$& 
\\
$\mathfrak{so}(N)$& $N \geq 8$ &  $N-4$&$N-2$
\\
$\mathfrak{sp}(2N)$ & $N \geq 3$ &  $N+2$&$N+1$
\\
\multicolumn{2}{c|}{$\mathfrak{g}_2$ } &${10 \over 3}$&4
\\
\multicolumn{2}{c|}{$\mathfrak{f}_4$} & $ 5$&9
\\
\multicolumn{2}{c|}{$\mathfrak{e}_6$} & $6$&12
\\
\multicolumn{2}{c|}{$\mathfrak{e}_7$} & $8$&18
\\
\multicolumn{2}{c|}{$\mathfrak{e}_8$} &$12$&30
\end{tabular}
\end{center}
\caption{Values of $\widetilde k_{\mathfrak{g}}$ and $h^\vee$ appearing in the unitarity constraints. Algebras not appearing in this table do not have any bounds on $k_\mathfrak{g}$.}
\label{tab:kunit}
\end{table}%

As for the bounds on $c_{{\mathrm{2d}}}$, there is a general upper bound
\bea
\label{eq:cunitbd}
c_{{\mathrm{2d}}} \leq -{11 \over 5} \left(1 + \sqrt{1+ {180 \over 121} \sum_{i}{ {k_{\mathfrak{g}_i} }\, {\mathrm{dim}}\mathfrak{g}_i \over 3 k_{\mathfrak{g}_i}- h^\vee_i}} \right) ~. 
\eea
Furthermore, when there is a sub-critical simple factor in the flavor group, then one also has the lower bound\footnote{Recall that a critical factor is one such that $k_\mathfrak{g} = 2 h^\vee$, while a sub-critical factor is one such that $k_\mathfrak{g} < 2 h^\vee$.} 
\bea
\label{eq:Sugdef}
c_{{\mathrm{2d}}} \geq c_{\mathrm{Sug}} ~, \hspace{0.5in }c_{\mathrm{Sug}}  =\sum_i { k_{\mathfrak{g}_i} \,\mathrm{dim} \mathfrak{g}_i \over k_{\mathfrak{g}_i} -2 h^\vee_i}~.
\eea 
For convenience the dual Coxeter numbers $h^\vee_i$ of $\mathfrak{g}_i$ are tabulated in Table \ref{tab:kunit}.

In addition to the bounds on $k_{\mathfrak{g}_i}$ and $c_{{\mathrm{2d}}}$, unitarity also requires the following more subtle constraints \cite{Beem:2018duj}:
 \begin{enumerate}
\item If the flavor symmetry has critical factors, there can be no sub-critical factors.
\item If the flavor symmetry has no critical factors, there can be at most one sub-critical factor. 
\end{enumerate}
Imposing all of these constraints will reduce the number of candidate cases drastically, as will be seen explicitly when we turn to our computerized scan.

\subsection{Characteristic dimension constraints}
\label{sec:chardim}
There is one final constraint which we impose. To introduce it, we begin by writing 
\bea
(\Delta_{u_1}, \dots, \Delta_{u_r}) = \lambda (d_1, \dots, d_r)
\eea
where $\lambda \in \QQ$ and $d_i$ are the unique integers with $\mathrm{gcd}(d_1, \dots, d_r) =1$ satisfying the above.
We then introduce a quantity $\varkappa$ known as the ``characteristic dimension,'' defined in  \cite{Cecotti:2021ouq} to be:
\beq
\varkappa:=\frac1{\{\l^{-1}\}}~,
\eeq
where for any $x\in\R$, $\{x\}$ is the unique real number equal to $x$ mod 1 such that $0<\{x\}\leq1$. In terms of this quantity we have the following result,
\begin{proposition}
\label{prop:kappa}
The characteristic dimension can only take one of eight rational values $\varkappa\in\{1,6/5,4/3,3/2,2,3,4,6\}$, and if $\varkappa\neq1,2$ then there can be no IR-free theories living on any Coulomb branch strata. 
\end{proposition} 
\noindent
This proposition can be argued for as follows.  First, when $\varkappa\neq1,2$ the CB geometry is such that at a generic point the $U(1)_r$ symmetry is only broken to a $\Z_n$ subgroup, with
\bea
\label{eq:ndef}
n = \left\{ \begin{matrix}
6 &\,\,\, \mathrm{if}\,\,\, \varkappa = {6\over 5},6~,
\\
4 &\,\,\, \mathrm{if}\,\,\, \varkappa = {4\over 3},4~,
\\
3 &\,\,\, \mathrm{if}\,\,\, \varkappa = {3\over 2},3~.
\end{matrix} \right.
\eea
This puts severe restriction on the allowed BPS spectra, which in particular constrains the full CB geometry to be \emph{isotrivial} and \emph{diagonal}. Isotriviality means that the total space of the CB is locally a product $A \times \CC^2$ with $A$ a Jacobian variety, while diagonality means that $A$ factorizes into a product of $r$ genus-one varieties $A = E_\tau \times\dots \times E_\tau$ at the same coupling. 
Thus for a given value of the rank $r$ and $\varkappa \neq 1,2$, the total structure of the Coulomb branch can be one of only a handful possibilities---see Section \ref{Sec:N2theories} for more detail.

It is a non-trivial result that isotriviality implies that all of the monodromies around singular loci have to be of finite order \cite{Cecotti:2021ouq}. Since the hallmark of IR-free theories is monodromy of infinite order (i.e. of parabolic-type), isotriviality also implies that no IR-free theory can live on any CB strata. This gives the desired statement. 

In practice, this constraint is implemented in our algorithm as follows. If the steps outlined in the previous subsections produce a candidate theory with a set of $\{\D_{u_i}\}$ for which $\varkappa\neq1,2$, and the only compatible Coulomb branch stratification includes IR-free theories, we may discard that case.

\section{An Example: Rediscovering $D_2(SU(5))$}
\label{sec:examplesubsec}
We now illustrate the algorithm outlined above by means of a simple rank-two example. As we shall see, even in this simple example the analysis is lengthy, and one is well-advised to implement the algorithm via a computer. 

When restricting to rank two, the CB analysis outlined in Section \ref{sec:CBconsts} simplifies significantly. For example,  Proposition \ref{thm:rank1thm} reduces to the statement that an unknotted stratum on $u_1=0$ can appear only if $\Delta_{u_2}$ takes values in (\ref{eq:rank1list}), and vice versa. A further simplifying feature is that at rank-two, knotted singularities are specified by a single type of irreducible homogeneous polynomial, taking the generic form
\beq
P(u_1,u_2)=u_1^p+\a \,u_2^q~,
\eeq 
where $\alpha$ is a free parameter and $p,q$ are chosen such that the left-hand side is homogeneous under a $\CC^*$-action, \emph{i.e.}  $p/q := \Delta_{u_2} / \Delta_{u_1}$ such that $(p,q)=1$.  Knotted singularities have $\Delta^{\mathrm{sing}}=p \Delta_{u_1}$.

Let us consider the following input data,
\bea
\label{eq:exampledata}
(c_{\mathrm{2d}}, a_1, a_2)= \left(-24 , \,24, \, 300\right) \hspace{0.2 in}\Rightarrow \hspace{0.2in} \chi_0(q) = q + 24 q^2 + 300 q^3 + O(q^4)~ ,
\eea 
and assume that the vacuum character solves an order-4 untwisted MDE.
We now ask if there exists any rank-two unitary $\cN=2$ SCFT compatible with this data. First, we solve the corresponding MDE order-by-order and check if the corresponding vacuum character has integer Fourier expansion and unit leading coefficient. Indeed, one finds the closed form solution 
\bea
\chi_{0}(q) = \left(\eta(q^2) \over \eta(q) \right)^{24} =  q\,(1 + 24 q + 300 q^2 + 2624 q^3 + 18126 q^4 + 105504 q^5+ \dots)\,
\eea
which is integral to all orders in $q$.\footnote{This can be seen, for example, by relating it to the McKay-Thompson series for Co$_0$ \cite{Duncan:2014eha}.}
We next note that the indicial equation (\ref{eq:indicialeq}) can be used to obtain the chiral dimensions $\{h_i\} = \left\{0,\,-\half,\, -1,\, -{3\over 2}\right\}$. These may be used to compute the 4d central charges via (\ref{eq:ca4d}). We start with the choice $h_* = -{3\over2}$, giving
\bea
\label{eq:acexample}
(a_{\mathrm{4d}}, c_{\mathrm{4d}}) = \left({7\over 4}, \,2\right)~.
\eea
Since $c_{\mathrm{4d}}> a_{\mathrm{4d}}$, this is only legitimate if $h_*$ is the minimal chiral dimension, which is indeed the case. These values also satisfy the Hoffman-Maldacena bounds. If we choose other values for $h_*$ we will get other candidates for $(a_{\mathrm{4d}}, c_{\mathrm{4d}})$, and all of them must be checked. For the moment let us simply focus on (\ref{eq:acexample}). 

We next use the Shapere-Tachikawa formula (\ref{eq:ShapTach}) to obtain the sum of the CB dimensions, which gives $\Delta_{u_1} + \Delta_{u_2} = 4$.  We then decompose the sum into allowed pairs of dimensions, giving
\bea
\left(\Delta_{u_1}, \Delta_{u_2} \right) \in \left\{ \left({4\over 3} ,{8\over 3} \right),  \left({3\over 2} ,{5\over 2} \right),  \left({8\over 5} ,{12\over 5} \right),  \left(2,2\right)\right\} ~,
\eea
\emph{c.f.} Appendix \ref{app:rank2dims}. 
We also note that $a_1=24$ admits the following decompositions into dimensions of simple Lie algebras, 
\bea
&\{ 24\}: &\hspace{0.5 in}\mathfrak{su}(5)
\no\\
&\{ 3, 21\}: & \hspace{0.5 in}\mathfrak{su}(2) \oplus \mathfrak{so}(7) \,\,\, \mathrm{or} \,\,\, \mathfrak{su}(2) \oplus \mathfrak{sp}(6)
\no\\
& \{10, 14\}: & \hspace{0.5 in}\mathfrak{so}(5) \oplus \mathfrak{g}_2
 \no\\
& \{8,8,8\}:&  \hspace{0.5 in}\mathfrak{su}(3) \oplus \mathfrak{su}(3) \oplus \mathfrak{su}(3)
\eea
among others. If we allow for $\mathfrak{u}(1)$ factors, then we must allow for arbitrary integer partitions of 24, of which there are 1575.

With these decompositions in hand, we now proceed by constructing possible CB stratifications. Let us begin by assuming flavor symmetry $\mathfrak{su}(5)$ and scaling dimensions $\left(\Delta_{u_1}, \Delta_{u_2} \right) =  \left({4\over 3} ,{8\over 3} \right)$. From Proposition \ref{thm:UVIR}, we know that the full flavor symmetry must be realized on at least one codimension-one stratum, which according to Table \ref{Theories} can only be a $[I_5, \mathfrak{su}(5)]$ stratum. We may still allow for unflavored strata $\cT_{\varnothing, 1}^{(1)}$ and $[I_1,  \varnothing]$.\footnote{The careful reader might notice that $[I_1,  \varnothing]$ has a $U(1)$ flavor symmetry, and thus might object to use of the word ``unflavored.'' However, here we are only interested in flavor factors which act on the CB by ``splitting'' the singular locus. Because the $U(1)$ mass deformation does not split the $I_1$ singularity (see \emph{e.g.} \cite{Argyres:2015ffa}) the $[I_1,  \varnothing]$ stratum is effectively unflavored. This is to be contrasted with the cases of  $\cS_{\varnothing,3}^{(1)}$ and $\cS_{\varnothing,4}^{(1)}$, which also have $U(1)$ flavor symmetry, but for which the $U(1)$ splits the singular locus.} From Proposition \ref{thm:rank1thm}, we know that unknotted strata can only be of the form $u_2=0$, since $\Delta_{u_2} = {8 \over 3}$ is not on the list of allowed rank-one Coulomb branch dimensions. 
Thus both knotted and unknotted strata must have  $\Delta^{\mathrm{sing}}_i = {8\over 3}$. 

Using (\ref{eq:mainstrateqs}) to compute $k_{\mathfrak{su}(5)}$, we find that $k_{\mathfrak{su}(5)}= {16\over 3}$ regardless of whether the flavored stratum is unknotted or knotted. This value of the level satisfies the unitarity bound (\ref{eq:kunitbd}). It also satisfies the bounds in (\ref{eq:cunitbd}) and (\ref{eq:Sugdef}), which read $-27 \lesssim c_{\mathrm{2d}} \lesssim -11.6$. However, note that the characteristic dimension $\varkappa$ in this case is $\varkappa = {4\over 3}$. Thus by Proposition \ref{prop:kappa}, there can be no IR-free theories living on any Coulomb branch strata. Since $[I_5, \mathfrak{su}(5)]$ is IR-free, it is not allowed, and so we conclude that there is no consistent Coulomb branch geometry with $\mathfrak{su}(5)$ and $\left(\Delta_{u_1}, \Delta_{u_2} \right) =  \left({4\over 3} ,{8\over 3} \right)$.

Next consider the choice of flavor symmetry $\mathfrak{su}(5)$ and $\left(\Delta_{u_1}, \Delta_{u_2} \right) =  \left({3\over 2} ,{5\over 2} \right)$. In this case the unknotted strata can again only be of the form $u_2=0$. We see that unknotted strata have  $\Delta^{\mathrm{sing}}_i = {5 \over 2} $, while knotted strata have $\Delta^{\mathrm{sing}}_i = {15\over 2} $. Turning again to the levels, we find that $k_{\mathfrak{su}(5)} = {5}$ (resp. ${15}$) if the flavored stratum is unknotted (resp. knotted). Both of these satisfy the unitarity bound (\ref{eq:kunitbd}), but only one satisfies the bounds (\ref{eq:cunitbd}) and (\ref{eq:Sugdef}). Indeed, for the case $k_{\mathfrak{su}(5)} = {5}$ the bounds read $-24 \leq c_{\mathrm{2d}} \lesssim -11.7$ and are satisfied (in fact, we see that the lower bound is saturated, signaling that the theory is Sugawara). So for consistency we must have the flavored stratum be unknotted. 

To complete the construction of the Coulomb branch geometry, we must now identify the relevant unflavored strata. First we note that the characteristic dimension is $\varkappa=1$, so by Proposition \ref{prop:kappa} IR-free strata are allowed. Since the unknotted $u_2=0$ stratum is already ``occupied'' by the $[I_5,\suf(5)]$ theory, the unflavored strata must be knotted. The most general scenario is to assume $n_1$ knotted $[I_1,  \varnothing]$  strata, and $n_2$ knotted $\cT_{\varnothing,1}^{(1)}$ strata.
 Then from (\ref{eq:mainstrateqs}) we compute the dimension of the extended Coulomb branch to be
\bea
h_{\mathrm{ECB}}  = {15\over 2} - {15\over 2}(n_1+ 2 n_2)~.
\eea
This is required to be a non-negative integer. Furthermore, from Proposition \ref{thm:nonzeroknot} we see that we must have at least one knotted stratum, and hence $n_1 + 2 n_2 \geq 1$. There is a single non-trivial solution to these equations, namely $(n_1,n_2) = (1,0)$ and $h_{\mathrm{ECB}}=0$. The fact that there exists a solution at all is an extremely non-trivial consistency check. Finally, the two unitarity bounds mentioned after (\ref{eq:Sugdef}) are trivially satisfied since the flavor symmetry is simple.

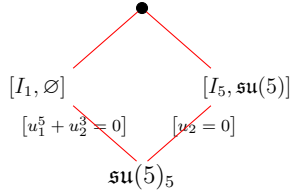
\begin{figure}[tbp]
\begin{center}
\begin{tikzpicture}[decoration={markings,
mark=at position .5 with {\arrow{>}}}]
\begin{scope}[scale=1.5]
\node[bbc,scale=.5] (p0a) at (0,0) {};
\node[scale=.5] (p0b) at (0,-1.4) {};
\node[scale=.8] (t0b) at (0,-1.5){$\mathfrak{su}(5)_5$};
\node[scale=.7] (p1) at (-.8,-.7) {$[I_1,\varnothing]$\quad\ \ };
\node[scale=.7] (p2) at (.8,-.7) {\quad\ $[I_5,\mathfrak{su}(5)]$};
\node[scale=.7] (p3) at (0,-.6){};
\node[scale=.7] (p4) at (0,-.8){};
\node[scale=.8] (t2b) at (-.6,-1.05) {{\scriptsize$\big[u_1^5+u_2^3=0\big]$}};
\node[scale=.8] (t3b) at (.55,-1.05) {{\scriptsize$\big[u_2=0\big]$}};
\draw[red] (p0a) -- (p1);
\draw[red] (p0a) -- (p2);
\draw[red] (p1) -- (p0b);
\draw[red] (p2) -- (p0b);
\end{scope}
\end{tikzpicture}
\caption{Hasse diagram for the {CB} of the theory with $\mathfrak{su}(5)_5$ flavor symmetry.}
\label{fig:Hasse}
\end{center}
\end{figure}

In summary, based on the input data $(c_{\mathrm{2d}}, a_1, a_2)= \left(-24 , \,24, \, 300\right)$, we have used two- and four-dimensional consistency conditions to gain evidence for the existence of a rank-two $\cN=2$ theory with $( a_{\mathrm{4d}}, c_{\mathrm{4d}})= ({7 \over 4},2)$, flavor symmetry $\mathfrak{su}(5)_{5}$, Coulomb branch dimensions   $(\Delta_{u_1},\Delta_{u_2}) =  ({3\over 2},{5\over 2})$, and extended Coulomb branch dimension $h_{\mathrm{ECB}} =0$. Furthermore, the Coulomb branch of the tentative theory is required to have the particular form given in Figure \ref{fig:Hasse}. Such a theory would satisfy all consistency conditions of which we are aware.

In fact, precisely such a theory is known---it is the $D_2(SU(5))$ theory of \cite{Cecotti:2013lda}. What we have seen is that, rather remarkably, the data of this theory is fixed almost entirely in terms of three parameters (\ref{eq:exampledata}). Of course, we have not yet shown that this is the \textit{only} consistent $\cN=2$ theory that can follow from this 2d data. In other words, though it is conjecturally believed that the map from $\cN=2$ theories to VOAs is one-to-one, this is something which must be checked. In order to check it, we must show that the remaining combinations of $(a_{\mathrm{4d}}, c_{\mathrm{4d}})$, of the flavor symmetry ($\mathfrak{su}(2) \oplus \mathfrak{so}(7)$ etc.), and of the Coulomb branch scaling dimensions do not give rise to any other seemingly legitimate theories. By repeating the steps above for all of the remaining case (with the help of a computer), one can indeed verify that this is the case.  

\section{Rank-two scan}
\label{sec:Results}
In the previous sections we introduced an algorithm for identifying mutually consistent sets of VOA and CB data which could correspond to legitimate 4d $\cN=2$ theories. In particular, the algorithm allowed us to determine the central charges and Coulomb branch data of the $\cN=2$ theories, including scaling dimensions and possible stratifications. In this section we will apply these techniques to search for rank-two theories with Schur indices satisfying an order-four untwisted MDE. 
In particular, by scanning over a wide range of values for $(c_{\mathrm{2d}},a_1, a_2)$ and applying our algorithm in each case, we aim to obtain a chart of the landscape of such 4d theories.

One obvious drawback of this strategy is that it is quite computationally intensive. Indeed, our goal is to scan over a large enough range of values $c_{\mathrm{2d}} = - {p \over q}$, $a_1$, and $a_2$, with $p,q,a_1,a_2$  positive integers, such that we are confident that all 4d $\cN=2$ theories would fall into that range. While there is no guarantee that any finite scan will be exhaustive, we find it reasonable to restrict to $p,q \leq 200$,  $a_1\leq 248$, and $a_2 \leq 1000\,a_1$---the latter in particular seems reasonable since the coefficient $a_2$ must be decomposable into representations of a flavor symmetry of dimension $a_1$, which for $a_1 \leq 248$ will not be exorbitantly larger than $a_1$ itself, unless rather baroque representations make an appearance. Restricting to $\mathrm{gcd}(p,q)=1$, the total number of cases which we will scan over is then of order $O(10^{11})$.
Furthermore, for each case one must consider all ways of partitioning $a_1$ into dimensions of simple Lie algebras. If we truly allow for all possible decompositions, the rough number of cases which must be checked is of order $O(10^{25})$. Scanning over such a large number of cases is clearly impractical, even with a computer. 

This number may be reduced drastically by restricting the allowed partitions of $a_1$.  One simplification is to restrict to flavor symmetries with small numbers of $U(1)$ factors---for example, only up to two $U(1)$ factors. Another is to restrict to flavor symmetries with only a certain number of simple factors---for example, up to 3 simple factors. Since these two conditions are satisfied by all known rank-two theories \cite{Martone:2021ixp}, we find it reasonable to impose them both, though it is certainly possible for there to exist unknown theories violating them. These conditions cut the number of cases down to something of order $O(10^{12})$---a large number to be sure, but one which is now accessible with advanced enough computing. 

\subsection{Results}

We now report the results of the scan described above. Running a number of Mathematica tasks in parallel, we carried out a scan over $O(10^{11})$ triplets $(c_{\mathrm{2d}},a_1, a_2)$. In each case we applied our algorithm to search for data consistent with a 4d $\cN=2$ interpretation. Remarkably, amongst all of these cases we identify only 15 candidate theories, of which six have appeared in the literature \cite{Martone:2021ixp}. We give the data of all of these candidate theories in Table \ref{tab:results}, as well as their CB Hasse diagrams in Figure \ref{fig:newHasse}.

Before providing information about the putative new theories, let us briefly summarize the known theories that our algorithm rediscovers. These theories are usually denoted by $D_1^{20}(E_8)$, $(A_1,A_4)$, $\mathrm{USp}(4) + 6 F$, $D_2(SU(5))$, $(A_1,D_5)$, and $\cT_{\varnothing,1}^{(1)} \times \cT_{\varnothing,1}^{(1)}$.  Their data, including  Coulomb and Higgs branch geometries, can be found in e.g. Table 1 of \cite{Martone:2021ixp}. In Section \ref{sec:examplesubsec}, we showed explicitly how our algorithm reconstructs the $D_2(SU(5))$ theory. The other cases proceed similarly.  

Clearly, we have only reproduced  a small fraction of the known rank-two theories catalogued in \cite{Martone:2021ixp}. This is of course to be expected, since we searched only for theories whose Schur indices satisfy untwisted order-four MDEs. A preliminary analysis shows that the overwhelming majority of known rank-two theories satisfy \textit{twisted} fourth-order MDEs.\footnote{It remains an open question whether there is a universal upper bound on the order of the MDE for a fixed rank SCFT.} For this reason, we anticipate that extending the current scan to the order-four twisted case would give even more interesting results.

We have found that in all cases, for a given set of legitimate $(c_{\mathrm{2d}}, a_1, a_2)$, there is always a \textit{unique} possibility for the data of the corresponding 4d $\cN=2$ theory. This gives non-trivial evidence for the injective nature of the map from 4d $\cN=2$ theories to chiral algebras.

\begin{table}[!t]
\begin{center}
\hspace*{-1.3cm}\begin{tabular}{cc|c|c|c|c|c|c|c}
&$(c_{\mathrm{2d}}, a_1,a_2)$ & $c_{\mathrm{4d}}$ & $a_{\mathrm{4d}}$ & $(\Delta_{u_1}, \Delta_{u_2})$ &$h_{\mathrm{ECB}}$& $\varkappa$ & $\mathfrak{g}_k$ & Status
\\\hline\hline
1.& $(-124,\, 248,\,31124)$ & ${31 \over 3} $&$ {101 \over 12} $& $(4,10)$ & 0 & 2& $(\mathfrak{e}_8)_{20}$ & $D_1^{20}(E_8)$
\\\hline
2.& $(-{68 \over 7}, \, 0, \, 1)$ & ${17 \over 21}$ & ${67\over 84}$ & $({8\over 7}, {10 \over 7})$ & 0 & 2 & $\varnothing$ & $(A_1,A_4)$
\\\hline
3.& $(-44, \,66, \, 2200)$ & $11 \over 3$ & $37 \over 12$ & $(2,4)$& 0 &  2& $\mathfrak{so}(12)_8$ & $\mathrm{USp}(4) + 6 F$
\\\hline
4.& $(-{24}, \, 24,\, 300)$ & $2$ & $7 \over 4$ & $({3\over 2},{5\over2})$& 0 &1 & $\mathfrak{su}(5)_5$ & $D_2(SU(5))$
\\\hline
5.& $(-{12},\, 3, \, 9)$ & $1$ & $19 \over 20$ & $({6\over 5},{8 \over 5})$& 0 &2&  $\mathfrak{su}(2)_{16\over 5}$ & $(A_1,D_5)$
\\\hline 
6.& $(-{44 \over 5}, \, 0, \, 2)$ & $11 \over 15$ & $43 \over 60$ & $({6\over 5},{6 \over 5})$& 0 & $6\over 5$& $\varnothing$ & $\cT_{\varnothing,1}^{(1)}\times \cT_{\varnothing,1}^{(1)}$
\\\hline\hline
7.& $(-64,\, 64, \, 2080)$ & $16 \over 3$ & $14 \over 3$ & $(4,5)$& 0 &1& $\mathfrak{su}(8)_{10} \times \mathfrak{u}(1)$ & \textbf{?}
\\\hline
8.& $(-{44 \over 5}, \, 2, \, 2)$ & $11 \over 15$ & $43 \over 60$ & $({6\over 5},{6 \over 5})$& 0 &$6\over 5$& $\mathfrak{u}(1)^2$ & \textbf{?}
\\\hline
9.& $(-{44 \over 3},\, 0,\,3)$ & $11 \over 9$ & $49 \over 36$ & $({4\over 3},{8 \over 3})$& 0 &$4\over 3$& $\varnothing$ & \xmark
\\\hline
10.& $(-{36}, \, 18,\,189)$ & $3$ & $11 \over 4$ & $(3,3)$& 5 & 3 & $\mathfrak{su}(2)_4 \times \mathfrak{su}(4)_8$ & \xmark
\\\hline
11.& $(-{28}, \, 8,\,38)$ & $7\over 3$ & $29 \over 12$ & $(3,3)$& 0 & 3& $\mathfrak{su}(3)_6$ & \xmark
\\\hline
12.& $(-{24}, \, 8,\,44)$ & $2$ & $15 \over 8$ & $({3\over 2},3)$& 2 & $3\over 2$& $\mathfrak{su}(2)_4\times\mathfrak{su}(2)_4 \times \mathfrak{u}(1)^2$ & \xmark
\\\hline
13.& $(-{20},\, 4,\,17)$ & $5\over 3$ & $19 \over 12$ & $({3\over 2},{5\over 2})$& 1 & 1 &  $\mathfrak{su}(2)_{7/2} \times \mathfrak{u}(1)$ & \xmark
\\\hline
14.& $(-{10}, \, 2,\,7)$ & $5\over 6$ & $11 \over 12$ & $({3\over 2},{3\over 2})$& 0 & $3\over 2$& $\mathfrak{u}(1)^2$ & \xmark
\\\hline
15.& $(-{84 \over 5}, \, 2,\,7)$ & $27\over 20$ & $7 \over 5$ & $({6\over 5},{12\over 5})$& 2 & $6\over 5$& $\mathfrak{u}(1)^2$ & \xmark
\end{tabular}
\end{center}
\caption{The 15 tentative theories identified in our scan. The first six are already known, c.f. Table 1 of  \cite{Martone:2021ixp}. The remaining nine theories are discussed in the rest of this work. Those marked with an \xmark\, will be shown not to exist, while those marked with a \textbf{?} have not yet been ruled out but seem unlikely to exist.}
\label{tab:results}
\end{table}%

\subsection{New rank-two theories?}\label{Sec:N2theories}
 
We now discuss the nine candidate 4d $\cN=2$ SCFTs which our scan identifies but which have not appeared in the literature. We will denote these tentative theories by $\cT_i$, with the index $i=7,\dots,15$ corresponding to the label in Table \ref{tab:results}. It is useful to split them  into two categories: those with characteristic dimension $\varkappa \neq 1,2$ and those with $\varkappa=1,2$.

\subsubsection{Cases with $\varkappa \neq 1,2$}
\label{sec:Isotrivialcases}

We start with the six candidate theories that have characteristic dimension $\varkappa \neq 1,2$.
As discussed in in Section \ref{sec:chardim}, theories of this kind satisfy stronger CB constraints, which allow us to make more sophisticated statements about the proposed CB stratifications. We warn the reader that the arguments in this section are somewhat technical, and can be skipped on a first reading.

\begin{figure}[!tbp]
\begin{center}
\begin{tikzpicture}[decoration={markings,
mark=at position .5 with {\arrow{>}}}]
\begin{scope}[scale=1.5]
\node[bbc,scale=.5] (p0a) at (0,0) {};
\node[scale=.5] (p0b) at (0,-1.4) {};
\node[scale=.8] (t0b) at (0,-1.5){$\cT_{7}$};
\node[scale=.7] (p1) at (-.8,-.7) {$[I_8, \mathfrak{su}(8)]$\ \ };
\node[scale=.7] (p2) at (.8,-.7) {\ $[I_1, \varnothing]$};
\node[scale=.7] (p3) at (0,-.6){};
\node[scale=.7] (p4) at (0,-.8){};
\node[scale=.8] (t2b) at (-.6,-1.05) {{\scriptsize$\Delta^{\mathrm{sing}} = {5}$}};
\node[scale=.8] (t3b) at (.55,-1.05)  {{\scriptsize$\Delta^{\mathrm{sing}} = {20}$}};
\draw[red] (p0a) -- (p1);
\draw[red] (p0a) -- (p2);
\draw[red] (p1) -- (p0b);
\draw[red] (p2) -- (p0b);
\end{scope}
\begin{scope}[scale=1.5,xshift=2.7cm]
\node[bbc,scale=.5] (p0a) at (0,0) {};
\node[scale=.5] (p0b) at (0,-1.4) {};
\node[scale=.8] (t0b) at (0,-1.5){$\cT_8$};
\node[scale=.7] (p1) at (-.8,-.7) {$\cT_{\varnothing,1}^{(1)}$\ \ };
\node[scale=.7] (p2) at (.8,-.7) {\ $\cT_{\varnothing,1}^{(1)}$};
\node[scale=.7] (p3) at (0,-.6){};
\node[scale=.7] (p4) at (0,-.8){};
\node[scale=.8] (t2b) at (-.6,-1.05) {{\scriptsize$\Delta^{\mathrm{sing}} = {6 \over 5}$}};
\node[scale=.8] (t3b) at (.55,-1.05)  {{\scriptsize$\Delta^{\mathrm{sing}} = {6 \over 5}$}};
\draw[red] (p0a) -- (p1);
\draw[red] (p0a) -- (p2);
\draw[red] (p1) -- (p0b);
\draw[red] (p2) -- (p0b);
\end{scope}
\begin{scope}[scale=1.5,xshift=5.4cm]
\node[bbc,scale=.5] (p0a) at (0,0) {};
\node[scale=.5] (p0b) at (0,-1.4) {};
\node[scale=.8] (t0b) at (0,-1.5){$\cT_9$};
\node[scale=.7] (p1) at (-.8,-.7) {$\cT_{\varnothing,1}^{(1)}$\ \ };
\node[scale=.7] (p2) at (.8,-.7) {\ $\cT_{\varnothing,1}^{(1)}$};
\node[scale=.7] (p3) at (0,-.6){};
\node[scale=.7] (p4) at (0,-.8){};
\node[scale=.8] (t2b) at (-.6,-1.05) {{\scriptsize$\Delta^{\mathrm{sing}} = {8 \over 3}$}};
\node[scale=.8] (t3b) at (.55,-1.05)  {{\scriptsize$\Delta^{\mathrm{sing}} = {8 \over 3}$}};
\draw[red] (p0a) -- (p1);
\draw[red] (p0a) -- (p2);
\draw[red] (p1) -- (p0b);
\draw[red] (p2) -- (p0b);
\end{scope}
\newline
\begin{scope}[scale=1.5,yshift=-2.1cm]
\node[bbc,scale=.5] (p0a) at (0,0) {};
\node[scale=.5] (p0b) at (0,-1.4) {};
\node[scale=.8] (t0b) at (0,-1.5){$\cT_{10}$};
\node[scale=.7] (p1) at (-.8,-.7) {$\cS_{\varnothing,2}^{(1)}$\ \ };
\node[scale=.7] (p2) at (.8,-.7) {\ $\cS_{D_4,3}^{(1)}$};
\node[scale=.7] (p3) at (0,-.6){};
\node[scale=.7] (p4) at (0,-.8){};
\node[scale=.8] (t2b) at (-.6,-1.05) {{\scriptsize$\Delta^{\mathrm{sing}} = {3}$}};
\node[scale=.8] (t3b) at (.55,-1.05)  {{\scriptsize$\Delta^{\mathrm{sing}} = {3}$}};
\draw[red] (p0a) -- (p1);
\draw[red] (p0a) -- (p2);
\draw[red] (p1) -- (p0b);
\draw[red] (p2) -- (p0b);
\end{scope}
\begin{scope}[scale=1.5,yshift=-2.1cm,xshift=2.7cm]
\node[bbc,scale=.5] (p0a) at (0,0) {};
\node[scale=.5] (p0b) at (0,-1.4) {};
\node[scale=.8] (t0b) at (0,-1.5){$\cT_{11}$};
\node[scale=.7] (p1) at (-.8,-.7) {$\cT_{\varnothing,1}^{(1)}$\ \ };
\node[scale=.7] (p2) at (.8,-.7) {\ $\cT_{A_2,1}^{(1)}$};
\node[scale=.7] (p3) at (0,-.6){};
\node[scale=.7] (p4) at (0,-.8){};
\node[scale=.7]  at (.15,-.7) {$\cT_{\varnothing,1}^{(1)}$\quad\ \ };
\node[scale=.8] (t2b) at (-.5,-1.05) {{\scriptsize$\Delta^{\mathrm{sing}} = {3}$}};
\node[scale=.8] (t3b) at (.55,-1.05)  {{\scriptsize$\Delta^{\mathrm{sing}} = {3}$}};
\draw[red] (p0a) -- (p1);
\draw[red] (p0a) -- (p2);
\draw[red] (p1) -- (p0b);
\draw[red] (p2) -- (p0b);
\draw[red] (p4) -- (p0b);
\draw[red] (p0a) -- (p3);
\end{scope}
\begin{scope}[scale=1.5,yshift=-2.1cm,xshift=5.4cm]
\node[bbc,scale=.5] (p0a) at (0,0) {};
\node[scale=.5] (p0b) at (0,-1.4) {};
\node[scale=.8] (t0b) at (0,-1.5){$\cT_{12}$};
\node[scale=.7] (p1) at (-.8,-.7) {$\cS_{\varnothing,2}^{(1)}$ };
\node[scale=.7] (p2) at (.8,-.7) { $\cS_{\varnothing,2}^{(1)}$};
\node[scale=.7] (p3) at (0,-.6){};
\node[scale=.7] (p4) at (0,-.8){};
\node[scale=.8] (t2b) at (-.6,-1.05) {{\scriptsize$\Delta^{\mathrm{sing}} = {3}$}};
\node[scale=.8] (t3b) at (.55,-1.05)  {{\scriptsize$\Delta^{\mathrm{sing}} = {3}$}};
\draw[red] (p0a) -- (p1);
\draw[red] (p0a) -- (p2);
\draw[red] (p1) -- (p0b);
\draw[red] (p2) -- (p0b);
\end{scope}
\newline
\begin{scope}[scale=1.5,yshift=-4.1cm]
\node[bbc,scale=.5] (p0a) at (0,0) {};
\node[scale=.5] (p0b) at (0,-1.4) {};
\node[scale=.8] (t0b) at (0,-1.5){$\cT_{13}$};
\node[scale=.7] (p1) at (-.8,-.7) {$\cS_{\varnothing,2}^{(1)}$ };
\node[scale=.7] (p2) at (.8,-.7) { $[I_1, \varnothing]$};
\node[scale=.7] (p3) at (0,-.6){};
\node[scale=.7] (p4) at (0,-.8){};
\node[scale=.8] (t2b) at (-.6,-1.05) {{\scriptsize$\Delta^{\mathrm{sing}} = {5\over 2}$}};
\node[scale=.8] (t3b) at (.55,-1.05)  {{\scriptsize$\Delta^{\mathrm{sing}} = {15\over 2}$}};
\draw[red] (p0a) -- (p1);
\draw[red] (p0a) -- (p2);
\draw[red] (p1) -- (p0b);
\draw[red] (p2) -- (p0b);
\end{scope}
\begin{scope}[scale=1.5,yshift=-4.1cm,xshift=2.7cm]
\node[bbc,scale=.5] (p0a) at (0,0) {};
\node[scale=.5] (p0b) at (0,-1.4) {};
\node[scale=.8] (t0b) at (0,-1.5){$\cT_{14}$};
\node[scale=.7] (p1) at (-.8,-.7) {$\cT_{\varnothing,1}^{(1)}$ };
\node[scale=.7] (p2) at (.8,-.7) { $\cT_{\varnothing,1}^{(1)}$};
\node[scale=.7] (p3) at (0,-.6){};
\node[scale=.7] (p4) at (0,-.8){};
\node[scale=.8] (t2b) at (-.6,-1.05) {{\scriptsize$\Delta^{\mathrm{sing}} = {3\over 2}$}};
\node[scale=.8] (t3b) at (.55,-1.05)  {{\scriptsize$\Delta^{\mathrm{sing}} = {3\over 2}$}};
\draw[red] (p0a) -- (p1);
\draw[red] (p0a) -- (p2);
\draw[red] (p1) -- (p0b);
\draw[red] (p2) -- (p0b);
\end{scope}
\begin{scope}[scale=1.5,yshift=-4.1cm,xshift=5.4cm]
\node[bbc,scale=.5] (p0a) at (0,0) {};
\node[scale=.5] (p0b) at (0,-1.4) {};
\node[scale=.8] (t0b) at (0,-1.5){$\cT_{15}$};

\node[scale=.7] (p3) at (0,-.6){};
\node[scale=.7] (p4) at (0,-.8){};
\node[scale=.7]  at (.15,-.7) {$\cS_{\varnothing,4}^{(1)}$\quad\ \ };
\node[scale=.8] (t2b) at (0,-1.05) {{\scriptsize$\Delta^{\mathrm{sing}} = {12 \over 5}$}};

\draw[red] (p4) -- (p0b);
\draw[red] (p0a) -- (p3);
\end{scope}
\end{tikzpicture}
\caption{Hasse diagrams for the {CB} of the candidate new theories $\cT_i$ for $i=7,\dots,15$, as output by the algorithm.}
\label{fig:newHasse}
\end{center}
\end{figure}
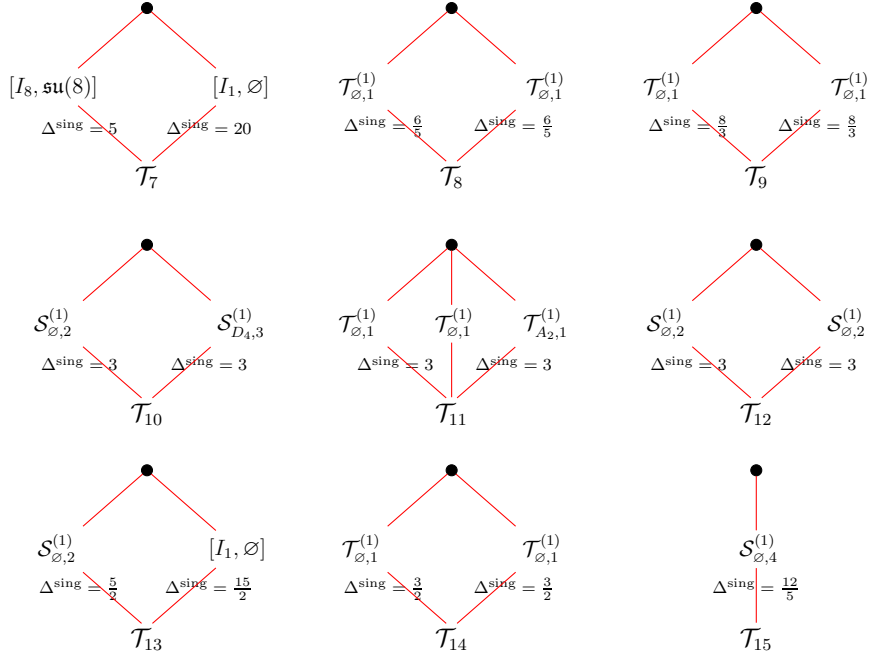

To begin, theories with $\varkappa \neq 1,2$ cannot support any IR-free strata on their CBs, and this is indeed reflected in the Hasse diagrams shown in Figure \ref{fig:newHasse}. This property stemmed from the more fundamental fact that the CBs of such theories must be \textit{isotrivial} and \textit{diagonal}. The condition of isotriviality means that the total space of the CB is locally a product $A \times \CC^2$, while the diagonality means that $A$ factors into a product of two genus-one varieties $A = E_\tau \times E_\tau$. Furthermore, the value of $\tau$ is fixed by the value of $\varkappa$: in particular, we have (locally)
\bea
\varkappa&\in& \left\{4,{4\over 3}\right\}: \hspace{0.82 in} \tau = i~,
\no\\
\varkappa&\in&\left\{ 3, {3\over 2},6, {6\over 5}\right\}: \hspace{0.5 in} \tau = \rho~, 
\eea
with $\rho := e^{\pi i /3}$. The global structure of the Coulomb geometry $X$ is then given by choosing a discrete group $G$ and a group homomorphism 
\bea
\label{eq:sigmadef}
\sigma:\,G \rightarrow \mathrm{Aut}(A)
\eea
and defining 
\bea
X = A \times \CC^2 / [(a,x) \sim (\sigma(g) a , \, g x)~,~~g \in G]~.
\eea
As mentioned before, we assume that the CB chiral ring is freely-generated (or alternatively that the CB has no complex singularities \cite{Argyres:2017tmj,Argyres:2018wxu}), and thus $G$ is restricted to be a complex reflection group by the Chevalley-Shephard-Todd theorem \cite{Shephard:1954,Chevalley:1955}. We will also for the moment assume that it is indecomposable,  i.e. that the rank-two Coulomb branch is not simply a product of two rank-one Coulomb branches. We will lift this assumption later. 

It is useful to introduce notation $G'$ for the image of the map $\sigma$,
\bea
G' := \sigma(G) \subset \mathrm{Aut}(A)~.
\eea
The unbroken R-symmetry generated by $e^{2 \pi i/n}$ (with $n$ as given in (\ref{eq:ndef})) must be contained in $G'$, so we have the chain of inclusions 
\bea
\langle e^{2 \pi i/n} \rangle \subset G' \subset \mathrm{Aut}(A)~.
\eea
It is also clear that $G'$ must be crystallographic, since it should leave the lattice of $A$ invariant.
Noting that the maximal automorphism groups of $E_\tau \times E_\tau$ at the points $\tau= i, \rho$ are given by \cite{Fujiki:1988}
\bea
A &=& E_i \times E_i: \hspace{0.53 in} \mathrm{Aut}(A) \in \left\{G_8, \, G(4,1,2) \right\}~,
\no\\
A &=& E_\rho \times E_\rho: \hspace{0.5 in} \mathrm{Aut}(A) \in \left\{G_5, \, G(6,1,2) ,\, G(6,2,2)\right\}~,
\eea
it is then possible to determine all possible groups $G'$. Indeed, one finds
\bea
n&=&4: \hspace{0.65in} G' \in \left\{G_8,\, G(2,1,2), \, G(4,1,2),\, G(4,2,2) \right\}~,
\no\\
n&=&3,6: \hspace{0.5in} G' \in \left\{G_4,\, G_5,\, G(3,1,2), \, G(6,1,2),\, G(6,2,2) \right\}~.
\eea
Note that the full group $G$ is potentially a non-trivial extension of $G'$ by a group $H$, with $H$ such that the extension preserves the rank of $G'$. It is actually possible to list all valid $G$ and $H$ for a given $G'$, as will be done in upcoming work \cite{Cecotti:2021yet}. Here we will only quote the portion of those results relevant for the current analysis.

We begin with the case of $n=4$, which in particular means that $\varkappa \in \{4, {4\over 3}\}$. The only such case in our list of tentative theories is $\cT_9$, which has $(\Delta_{u_1}, \Delta_{u_2})=({4\over 3}, {8\over 3})$ and two $\cT_{\varnothing, 1}^{(1)}$ Coulomb branch strata. From the above discussion, we conclude that the group $G$ must be a (potentially trivial) extension of $G'=G_8,\, G(2,1,2),\, G(4,1,2),$ or $G(4,2,2)$. It is an important fact that the pair of Coulomb branch dimensions $(\Delta_{u_1}, \Delta_{u_2})$ is proportional to the pair of degrees $(d_1, d_2)$ of invariant polynomials of $G$ (see the discussion around (\ref{eq:SWdiff}) for more on this). Since we would like $(\Delta_{u_1}, \Delta_{u_2})=({4\over 3}, {8\over 3})$, it is clear that we must have $d_2 = 2d_1$. Not every complex reflection group satisfies this property---indeed, the only ones which do are 
\bea
\label{eq:d22d1list}
G \in \left\{G(m,1,2), \, G_5, \, G_{10},\,G_{15}, \, G_{18}\right\}~.
\eea
If $G'=G(2,1,2)$ or $G(4,1,2)$, then it is consistent to have $G=G'$ since these appear in the list (\ref{eq:d22d1list}). In fact, these cases do not admit any extensions to alternative entries in (\ref{eq:d22d1list}), and thus having $G=G'$ is the only possibility for them. On the other hand, $G_8$ and $G(4,2,2)$ do not appear in the list, nor can they be extended to elements of the list, and thus they must be ruled out. 

We are thus left with two choices  for $G$. We would now like to see if either of these choices can reproduce the desired Coulomb branch stratification. Recall that we want two $\cT_{\varnothing, 1}^{(1)}$ strata, which are of Kodaira-type $II$. In other words, the monodromy around these singular loci must be of order 6. For a quotient of $\CC^2$ by $G$, the singular loci are in one-to-one correspondence with conjugacy classes of maximal cyclic subgroups of $G$ generated by reflections. If we consider the fixed locus of a generator $g$ of one of the maximal cyclic subgroups, then the monodromy around that locus is given by $\sigma(g)$. Noting that neither $G(2,1,2)$ nor $G(4,1,2)$ have any elements of order 6 and recalling that $\sigma$ is a group homomorphism, we conclude that there cannot be any monodromy of order 6, and hence that the proposed structure of the Coulomb branch is inconsistent. 

Before moving on, it is worth noting that though a 4d interpretation for this case can be ruled out, this does not mean that the corresponding VOA is ill-defined. Indeed, in the current case the VOA can be identified with a simple-current extension of a $\ZZ_3$ gauging of the $(4,9)$ $\cW_3$ minimal model, as shown in Appendix \ref{app:T9VOA}. What we have shown here is simply that this VOA does not lie in the image of the 4d $\to$ 2d map.

Next consider the case of $n=3,6$, which means that $\varkappa \in \left\{ 3, {3\over 2},6, {6\over 5}\right\}$. Looking at Table \ref{tab:results}, we see that there are two subcases here: those with $\Delta_{u_2} = 2 \Delta_{u_1}$ (namely $\cT_{12}$ and $\cT_{15}$) and those with $\Delta_{u_2} = \Delta_{u_1}$ (namely $\cT_8, \cT_{10},\cT_{11},$ and $\cT_{14}$). 
Let us begin with the former. In this case we again require $G$ to be an element of the list (\ref{eq:d22d1list}). If $G'=G_5,\, G(3,1,2),$ or $G(6,1,2)$, then we can have $G=G'$, and indeed this is the only choice since there does not exist any non-trivial extension to other elements in the list. On the other hand, for $G_4$ and $G(6,2,2)$ the extension must be non-trivial. For $G_4$ there is in fact no valid extension, so this case must be ruled out, whereas for $G(6,2,2)$ there does exist an extension by the quaternion group $H=\cQ_8$ to $G_{15}$. There are then a total of four options for $G$, namely $G=G_5,\, G(3,1,2),G(6,1,2),$ or $G_{15}$. 

Next note that the candidate theory $\cT_{12}$ has two $\cS_{\varnothing, 2}^{(1)}$ strata, which are of Kodaira-type $I_0^*$. The monodromy around each must be of order 2. For $G_5$ the only  cyclic subgroups are of order 3 (see e.g. Table D.1 of \cite{lehrer2009unitary}) and hence this cannot give the desired structure of the Coulomb branch. Likewise for $G(3,1,2)$ or $G(6,1,2)$, each has only a single cyclic subgroup of order 2, so these cannot work either. Finally for $G_{15}$, there do in fact exist two  cyclic subgroups of order 2, but there is also a subgroup of order 3. In this case we would obtain three singular loci, which is incompatible with the proposed Coulomb branch stratification. 

As for the candidate theory $\cT_{15}$, this has a single $\cS_{\varnothing,4}^{(1)}$ stratum of Kodaira type $III^*$, and hence must be of order 4. By the comments in the previous paragraph, there is again no way to realize the desired Coulomb branch. We may thus rule this case out as well.

We now return to the case of $\Delta_{u_2} = \Delta_{u_1}$. In this case we must consider complex reflection groups such that the degrees are equal, $d_1 = d_2$. The only complex reflection groups for which this holds are
\bea
\label{eq:identicalDelt}
G(m,2,2), \,\, G_7, \,\, G_{11}, \,\, G_{19}~.
\eea
For $G'=G_4, \,G_5,\, G(3,1,2)$ or $G(6,1,2)$, we conclude that we must have a non-trivial extension to a group in this list. For $G'=G_4,\, G_5$ no such extension is possible, but for $G' = G(3,1,2),\, G(6,1,2)$ we can do a $\ZZ_2$ extension to obtain $G=G(6,2,2),\, G(12,2,2)$ respectively. On the other hand, for $G'=G(6,2,2)$ we can take $G=G'$, and indeed there exists no extension giving an alternative element of (\ref{eq:identicalDelt}). Hence we have a total of two possibilities for $G$, namely $G=G(6,2,2), G(12,2,2)$ (and for the case of $G=G(6,2,2)$ we have seen that there are two choices for $G'$). 

We now ask whether any of these  cases can give rise to the desired Coulomb branch stratifications. The Kodaira type of the strata of the various theories are as follows:
\bea
&\cT_{8}:& \hspace{0.3 in} (II,II) \hspace{0.85 in} \cT_{10}:\hspace{0.3 in} (I_0^*, II^*)
\no\\
&\cT_{11}:&\hspace{0.3 in} (II,II,IV) \hspace{0.57 in}\cT_{14}:\hspace{0.3 in} (II,II)
\eea
The strata of type $I_0^*$ have monodromy of order 2, while those of type $II/II^*$ have monodromy of order 6, and those of type $IV$ have monodromy of order 3. Note that both $G(3,1,2)$ and $G(6,2,2)$ have two conjugacy classes of maximal cyclic subgroups generated by reflections, of order 2 and 3 respectively, while $G(12,2,2)$ has subgroups of order 2 and 6 instead. It is thus clear that the Coulomb branch stratifications proposed for the $\cT_8, \cT_{11}$, and $\cT_{14}$ theories can never be realized in this way. These theories may thus be ruled out as indecomposable rank-two theories (we will discuss the possibility of them being decomposable below). 

On the other hand, it is seemingly possible for $\cT_{10}$ to be realized by $G=G(12,2,2)$. To rule this out, we will need additional information about the form of the Seiberg-Witten differential. First, note that there are two possibilities for $\sigma$ in (\ref{eq:sigmadef}), namely $(i)$ the natural map $G(12,2,2) \rightarrow G(6,1,2)$ and $(ii)$ the conjugate to the natural map \cite{Cecotti:2021yet}.  Denoting the coordinates on $A \times \CC^2$ as $(a,x)$, the Seiberg-Witten differential may then be written as 
\bea
\label{eq:SWdiff}
&(i):& \qquad \lambda =  x^1 da_1 + x^2 da_2~,
\no\\
&(ii):& \qquad \lambda = {\p p_1 \over \p x^1} da_1 +  {\p p_2 \over \p x^1}  da_2~,
\eea
where $da_i$ are a basis of holomorphic differentials on $A$ and the coefficient functions were chosen such that $\lambda$ is well-defined on the orbifold space.
 This in particular means that the $p_i(x)$ are built from invariant polynomials of $G(12,2,2)$. Let us say that these  $p_i(x)$ are degree-$k$ polynomials in $x$, and since we are restricting to complex reflection groups whose invariant polynomial have the same dimensions $d_1=d_2=d$, we can assume without loss of generality that $k=n\cdot d$ for some $n\in \NN$. Since the integrals of $\lambda$ along the cycles of $A$ give rise to masses, we expect that $[\frac{\partial p_1(x)}{\partial x}] = [p_1(x)]-[x]= 1$ and hence that $[x]={1 \over k-1}$. The Coulomb branch scaling dimensions are obtained by the resulting dimension of the invariant polynomials in $x$:
\bea
&(i):& \qquad (\Delta_{u_1}, \Delta_{u_2}) = (d,d)~,
\no\\
&(ii):& \qquad (\Delta_{u_1}, \Delta_{u_2}) = {1\over n\, d-1}(d,d)~.
\eea
In our case $d=12$, since the degree of the invariant polynomials of $G(12,2,2)$ are $(12,12)$, and it is immediate to see that no $n$ exists for which $(\Delta_{u_1}, \Delta_{u_2}) =(3,3)$. Hence we arrive at a contradiction and  $\cT_{10}$ must be ruled out as well.

We have thus seemingly ruled out \textit{all} of our candidate theories with $\varkappa \neq 1,2$. However, it should be noted that the previous analysis assumed that the Coulomb branch is \textit{indecomposable}. We may now consider lifting this assumption. For this to be possible, the four-dimensional central charges of the rank-two SCFT should be decomposable into the sum of two rank-one central charges. By referring to Table \ref{Theories}, it is easy to check that this is only possible for the theory $\cT_8$. Indeed, in this case $c_{\mathrm{4d}}=-{44\over 5}$ is simply twice the central charge of the $\cT_{\varnothing,1}^{(1)}$ SCFT. This suggests that the Coulomb branch is simply the product of two copies of the Coulomb branch of $\cT_{\varnothing,1}^{(1)}$. We will discuss this case in more detail now.

\paragraph{$\cT_8$:}

The candidate theory $\cT_{8}$ is specified by 2d data $(c_{\mathrm{2d}}, a_1, a_2) = (-{44 \over 5}, 2, 2)$. As shown in Table \ref{tab:results}, one finds that this 2d data is consistent with a 4d theory with central charges $( a_{\mathrm{4d}},  c_{\mathrm{4d}})= ({43 \over 60}, {11 \over 15})$, Coulomb branch scaling dimensions $(\Delta_{u_1}, \Delta_{u_2}) = ({6\over 5},{6 \over 5})$, and $U(1)^2$ flavor symmetry. This looks nearly identical to the data of the $\cT_6= \cT_{\varnothing,1}^{(1)}\times \cT_{\varnothing,1}^{(1)}$ theory, except for the presence of two spin-one currents. The Schur index of the theory is given by 
\bea \label{ZT8}
Z_{\cT_8} &=&1+2 q+2 q^2+4 q^3+5 q^4+8 q^5+11 q^6+16 q^7+21 q^8+30 q^9+39 q^{10}+ \dots~,
\no\\
\eea
to be contrasted with that of $\cT_6$,
\bea
Z_{\cT_6} &=& \mathrm{PE}\left[ {2 (q^2 -q^4)\over (1-q)(1-q^{5})}\right]
\no\\
&=& 1+2 q^2+2 q^3+3 q^4+4 q^5+7 q^6+8 q^7+13 q^8+16 q^9+23 q^{10} + \dots~.
\eea
In the current case the minimal MDE satisfied by the vacuum character is of order four, with the tentative chiral dimensions of the corresponding VOA being 
\bea
\{ h_i\}=\left\{0,\,-{1\over 3},\,-{2\over 5},\,{4 \over 15} \right\}~.
\eea

The general lore that no two SCFTs should have the same CB geometry seems to suggest that this candidate theory is invalid, since it has the same geometry as $\cT_6$, which is known to exist. 
What's more, we could not come up with a natural ansatz for the strong generators of the VOA associated to the putative index~(\ref{ZT8}). Our hunch is that this candidate theory is inconsistent, but unfortunately we  we will not be able to say anything more conclusive.

\subsubsection{Cases with $\varkappa = 1,2$}
We next consider the theories with $\varkappa = 1,2$. Amongst the candidate new theories, only two satisfy this property, namely $\cT_7$ and $\cT_{13}$. Because in these cases the CB is no longer required to be isotrivial and diagonal, the powerful tools used in the previous subsection are no longer applicable.

\paragraph{$\cT_{13}$\,:}
The candidate theory $\cT_{13}$ is defined by 2d data  $(c_{\mathrm{2d}}, a_1, a_2) = (-20,4,17)$. As shown in Table \ref{tab:results}, one finds that this 2d data is consistent with a 4d theory with central charges $( a_{\mathrm{4d}},  c_{\mathrm{4d}})= ({19 \over 12}, {5 \over 3})$, Coulomb branch scaling dimensions $(\Delta_{u_1}, \Delta_{u_2}) = ({3\over 2},{5 \over 2})$, and flavor symmetry $\mathfrak{su}(2)_{7/2}\times \mathfrak{u}(1)$. The Schur index of the tentative theory is obtained by solving the MDE, giving
\bea
\label{eq:T13index}
Z_{\cT_{13}}
&=& 1+4 q+17 q^2+56 q^3+163 q^4+428 q^5+1063 q^6
\no\\
&\vphantom{.}&\hspace{1 in}+2472 q^7+5515 q^8+11792 q^9+24404 q^{10}+ \dots~
\eea
Let us check whether this putative Schur index is compatible with the sum rule (\ref{a2sumrule}). First note that there are no Kac-Moody nulls at level two, so $a_2^{\rm KM} = 14$. 
It is simple to check that the Sugawara condition is obeyed, so $n^{\rm T} = 0$. Additional  $h=2$ states must be affine Kac-Moody (AKM) primaries. An AKM primary in the spin-$j$ representation of $\mathfrak{su}(2)_{-7/4}$ and with charge $q$ under $\mathfrak{u}(1)$ has holomorphic dimension
\be
h (j, q) = \frac{j(j+1)}{-\frac{7}{4}+2} - q^2 = 4 j (j+1) - q^2 \,,
\ee
and thus in order to have $h=2$ we need both a non-trivial $\mathfrak{su}(2)$ representation and a non-zero $\mathfrak{u}(1)$ charge (for example, $j=1/2$ and $q = \pm 1$). CPT invariance implies that both signs of the $\mathfrak{u}(1)$ charge must be present. We then conclude that $\#  \hat {\cal B}_2$ must be even, and thus from 
 (\ref{a2sumrule}) that $a_2$ itself is even, in contradiction with the value  $17$ in (\ref{eq:T13index}). We conclude that our candidate character does not have a consistent 4d interpretation.

It is intriguing that an actual $\cN=2$ SCFT with the same flavor symmetry (including the level) and  central charges  does exist, but at rank three. This is  the $(A_1,D_8)$ theory with CB scaling dimensions $(\frac32,\frac54,\frac74)$ and Schur index \cite{Buican:2015ina}
\bea\label{eq:A1D8index}
Z_{(A_1,D_8)}
&=& 1+4 q+18 q^2+56 q^3+167 q^4+436 q^5+1086 q^6
\no\\
&\vphantom{.}&\hspace{1 in}+2520 q^7+5631 q^8+12024 q^9+24906 q^{10}+ \dots~
\eea
Comparing with \eqref{eq:T13index}, the two indices look remarkably similar. Indeed the $O(q^2)$ term is off by only one (allowing the coefficient to circumvent the issue that led to the previous inconsistency) and the $O(q^3)$ term matches. 
It is unclear whether this coincidence has any meaning.

\paragraph{$\cT_7$\,:} The candidate theory $\cT_{7}$ is defined by 2d data  $(c_{\mathrm{2d}}, a_1,a_2) = (-64, 64, 2080)$. As shown in Table \ref{tab:results}, one finds that this 2d data is consistent with a 4d theory with central charges $( a_{\mathrm{4d}},  c_{\mathrm{4d}})= ({14 \over 3}, {16 \over 3})$, Coulomb branch scaling dimensions $(\Delta_{u_1}, \Delta_{u_2}) = (4,5)$, and flavor symmetry $\mathfrak{su}(8)_{10} \times \mathfrak{u}(1)$.
The Schur index of the tentative theory is obtained by solving the corresponding MDE, giving
\bea
Z_{\cT_{7}}
&=& \mathrm{PE}\left[64(q+q^3) \over 1-q^4\right] = 1+64 q+2080 q^2+45824 q^3+770576 q^4\dots
\eea
We have found it  difficult to obtain a compelling guess for the Higgs branch of this theory. Indeed, the simplest guess would be that the first stratum is located on the minimal nilpotent orbit of $\mathfrak{a}_7$, and that the theory eventually Higgses to free hypermultiplets. This simplest guess turns out to be inconsistent.

To see this, begin by noting the following. Say that upon moving onto the Higgs branch, we spontaneously break a $\mathfrak{g}$ flavor symmetry, thereby taking a putative theory $\cT$ to a new SCFT $\cT_{\rm Higgs}$. Denote the sublocus (i.e. symplectic leaf) of the HB where this specific Higgsing takes place $\Sf$. Using anomaly matching, the 
central charges of the two theories can be related by \cite{Giacomelli:2020jel,CCLMW2021}
\beq
\label{eq:Higgsbranchcrel}
12 c_{\cT}=12 c_{\cT_{\rm Higgs}}+2\left(\frac32 k_{\mathfrak{g}}-1\right)+{\rm dim}_\H\Sf-1~.
\eeq
In the current case, we have $12 c_{\cT_{7}} = 64$, $k_{\mathfrak{su}(8)}={10}$, and with the assumption that the theory Higgses to free hypers,  we also conclude that $d_{\mathrm{HB}}^{\cT_{7}} =24(c_{\mathrm{4d}}- a_{\mathrm{4d}})= 16$. Since ${\rm dim}_\H\Sf = d_{\mathrm{HB}}^{\cT_{7}} -d_{\mathrm{HB}}^{\mathrm{Higgs}}$, where $d_{\mathrm{HB}}^{\mathrm{Higgs}}$ is the dimension of the HB of $\cT_{\rm Higgs}$, we then conclude that the theory on $\Sf$ must have 
\bea
\label{eq:neededforT6}
12 c_{\cT_{\mathrm{Higgs}}} - d_{\mathrm{HB}}^{\mathrm{Higgs}} = {21}~.
\eea
Furthermore, by definition $h_{\mathrm{ECB}}=0$ implies that Higgsing $\cT_7$ should necessarily lead to theories of rank lower than two. Under the assumption that the only rank-zero theories are free hypermultiplets and discrete gauging thereof---which are inconsistent with the analysis above---the theory on the lowest stratum should have rank one. We are thus led to search for a rank-one theory satisfying (\ref{eq:neededforT6}). The only such theory is  $\cS_{A_1,3}^{(1)}$, with flavor level $k_{IR}$ given in Table \ref{Theories}. Noting that the flavor level $k_{\mathfrak{su}(8)}$ of the UV theory must be related to that on the minimal nilpotent orbit by \cite{Giacomelli:2020jel,Beem:2019tfp,Beem:2019snk}
\bea
k_{\mathfrak{su}(8)}= {2 + k_{IR} I_{\mathfrak{f}_{\mathrm{IR}}\hookrightarrow \mathfrak{a}_7} \over I_{\mathfrak{a}_5 \hookrightarrow \mathfrak{a}_7}
}~,
\eea 
where $I_{\mathfrak{h}\hookrightarrow \mathfrak{g}}$ is the embedding index of $\mathfrak{h}$ in $\mathfrak{g}$, we see that in no case does this combination reproduce the correct value of $k_{\mathfrak{su}(8)}={10}$. 

Of course this alone does not rule this out as a consistent theory---it simply rules out the simplest guess for the structure of the Higgs branch. We may for example consider more complicated guesses in which the lowest stratum is not located on the minimal nilpotent orbit of $\mathfrak{a}_7$, or cases in which the theory does not Higgs completely down to free hypermultiplets. 
It would be interesting to use more sophisticated Higgs branch technology to reach a definitive conclusion about  the status of this putative theory. The candidate data also look somewhat implausible from the point of view of the sum rule (\ref{a2sumrule}). Substituting $a_2=2080$, $a_2^{\rm KM}=2144$ and  $n^{\rm T}=1$ we find
 \be
 \#  \hat {\cal B}_2  - 2\, \#   \cD_{1(0,0)}  + 2\, \#  \cD_{\frac{1}{2}(0,\frac{1}{2})} = -65\, .
\ee
There are of course many ways to assign flavor representations to the additional generators so that this sum rule is satisfied, but there appears to be no economical or natural choice.

\section{Outlook}
\label{sec:conclusions}

In this paper we have introduced an algorithm that  takes as input the first few terms of a putative Schur index 
and gives as output a (potentially empty) set of mutually consistent Schur and Coulomb data.
The  algorithm uses constraints encoding the general principles of unitarity and superconformal invariance. These constraints turn out to be extremely powerful,
vastly winnowing down the number of consistent possibilities.
 As a proof of concept, we performed a  search for rank-two $\cN=2$ superconformal field theories with Schur indices satisfying an untwisted modular differential equation of order four. 
  Our analysis reproduces a variety of known examples, but perhaps surprisingly it does not conclusively produce any new candidate theories. This may be evidence that our understanding of rank-two theories is more complete than expected,
  or  just an artifact of our restriction to low order and low rank.

Only six of the nearly seventy known rank-two theories \cite{Martone:2021ixp} have a vacuum character satisfying an untwisted order-four MDE; most of them have characters satisfying a \textit{twisted} order-four MDE. So even the simplest extension of our scan to twisted fourth-order is expected to have a large payoff.  This extension is conceptually straightforward, but rather challenging from a computational viewpoint.  A solution of a fourth-order twisted MDE is specified by eight parameters, and thus the number of cases that serve as input would be too large for a scan as thorough as the one given here. One might hope to implement a more limited scan (e.g.~searching for cases with no or small-dimensional flavor symmetry)
or parallelize on a cluster. More speculatively,  since we now have a method for generating large data sets of consistent versus inconsistent inputs (in the form of tuples of integers), one might envision the use of 
machine learning techniques.

The other obvious  extension is to perform searches for 4d SCFTs of higher rank. 
Unfortunately, it is not in general understood how the rank is related to the order of the MDE. Both are in some sense measures of the complexity of the SCFT, and experimentally one notices that the order grows at least linearly with the rank.
An order-four untwisted MDE scan is thus likely to produce sparser and sparser candidate theories at higher rank, and it will be necessary to consider higher-order MDEs. Apart from the increased computational cost, there are also some conceptual obstacles that must be overcome starting at rank three. For example, the scaling dimension $\D^{\rm sing}$ for a {knotted} stratum is no longer set solely by the scaling dimension of the unknotted ones. This difficulty arises from the known fact that the set of irreducible homogenous polynomial in three complex variables is infinite. To explore theories of rank three or higher, one would thus need to find an efficient way to get around this limitation.

Altogether, we have seen that the space of consistent 4d $\cN=2$ theories is subject to a stringent set of constraints, which, when imposed in their entirety, admit a surprisingly small set of solutions. The approach of this paper is similar to that used in the conformal bootstrap to identify candidate theories from general principles, though in the current case the complete set of principles remains unknown. Furthermore, the principles which are known are likely still not phrased in their most economical form. Once the full set of compatibility conditions between CB and VOA data is uncovered, one could imagine carrying out an even more stringent scan over candidate theories. If some of the new constraints are imposable already at the level of the inputs to the algorithm, scans could be done over smaller numbers of input parameters, thereby allowing for significant computational improvement. We hope that this work provides a first step in a new, computational approach to the classification of 4d $\cN=2$ theories.

\section*{Acknowledgements}

 JK would like to thank Kavli IPMU for their hospitality---a significant portion of this work was completed there. MM is supported in part by the NSF grant PHY-1915093, by the Simons Foundation grant 815892 and STFC grant ST/T000759/1. LR is supported in part by the NSF grant PHY-1915093. MW is supported in part by DOE grant SC0011784.

\begin{appendix}

\section{Superconformal representation theory constraints}

\label{superconformal:subsection}

In order for a vacuum character to be identifiable with a 4d Schur index, one must be able to interpret its Fourier coefficients as counting Schur operators.  Superconformal representation theory imposes certain constraints.
 In Table \ref{tab:SchurOp} we list the 4d superconformal multiplets that contain Schur operators (one for each multiplet).  It will be useful to recall the interpretation of a few of them (see~e.g.~\cite{Dolan:2002zh, Beem:2013sza} for more details):
\begin{itemize}
\item $\hat{\cB}_{R}$: Higgs branch multiplets. Their bottom components are Higgs chiral ring operators, with  conformal dimension $\Delta = 2 R$. The corresponding VOA operators have chiral dimension $h = R$. The cases $R=\frac{1}{2}$ and $R=1$ are special: $\hat{\cB}_{\frac{1}{2}}$ are free hypermultiplets, whereas $\hat{\cB}_{1}$ are moment-map multiplets, which contain the conserved flavor currents of the theory. The VOA counterparts of the latter are affine Kac-Moody currents.
\item  
$\cD_{0(0,j_2)}$ and $\overline{\cD}_{0(j_1,0)}$ are
free field multiplets. For  $j_1 = j_2 = 0$ they correspond to the standard free vector multiplet,  while for higher $j_1$, $j_2$ they are (exotic) higher-spin free fields.
\item $\hat{\cC}_{0 (j_1 ,j_2)}$  contain conserved currents of spin $2 + j_1 + j_2 $. In particular, $\hat{\cC}_{0 (0 ,0)}$ is the standard stress tensor multiplet.
\item $\cD_{\frac{1}{2}(0, 0)}$,  $\overline{\cD}_{\frac{1}{2}(0,0)}$ are ``additional'' supercurrent multiplets, on top of the supercurrent contained in the stress tensor multiplet. Their presence signals supersymmetry enhancement.
In the VOA, they correspond to fermionic $h=\frac{3}{2}$ operators (i.e.~2d supercurrents).\footnote{
4d ${\cal N}=3$ SCFTs map to VOAs containing an ${\cal N}=2$ super-Virasoro algebra, while 4d ${\cal N}=4$ SCFTs map to VOAs containing a   (small) ${\cal N}=4$ super-Virasoro algebra.}
\end{itemize}
We wish to assume that
the 4d SCFT is fully interacting,
i.e.~that it doesn't contain any free field subsector.
It follows that:
\begin{enumerate}
    \item[(i)] No free field supermultiplets are allowed. This forbids $\hat{\cB}_{\frac{1}{2}}$, $\cD_{0(0,j_2)}$, and $\overline{\cD}_{0(j_1,0)}$ multiplets. 
    \item[(ii)] No supermultiplets containing conserved currents of spin greater than two are allowed. Indeed,    such higher spin currents would imply the existence of a free subsector \cite{Bhardwaj:2013qia}.
    This forbids $\hat{\cC}_{0 (j_1 ,j_2)}$  with 
    $j_1 + j_2 >0$.
\end{enumerate}
It is also natural to assume:
\begin{enumerate}
\item[(iii)] The theory contains a {\it single} spin-two conserved current, which is identified with the stress tensor operator. This demands the presence of precisely one $\hat{\cC}_{0 (0 ,0)}$ multiplet.
\end{enumerate}
This last assumption is meant to capture locality (the existence of a  local stress tensor) and indecomposibility (a direct product of SCFTs would have multiplet stress tensors).\footnote{Strictly speaking, we are not aware of a rigorous argument showing the converse, i.e.~that a theory with multiple spin two conserved currents is necessarily decomposable. Nevertheless it is a standard assumption. \label{indecomposibility:footnote}}

\begin{table}[!t]
\begin{center}
\renewcommand{\arraystretch}{1.2}
\hspace*{-0.1cm}\begin{tabular}{c|c|c}
Multiplet& $h$&$\bf{r}$
\\\hline\hline
$\hat{\cB}_R$&$R$&$0$\\
\hline
$\cD_{R(0,j_2)}$&$R+j_2+1$&$j_2+\frac12$\\
\hline
$\overline{\cD}_{R(j_1,0)}$&$R+j_1+1$&$-j_1-\frac12$\\
\hline
$\hat{\cC}_{R(j_1,j_2)}$&$R+j_1+j_2+2$&$j_2-j_1$\\
\end{tabular}
\end{center}
\caption{List of 4d ${\cal N}=2$ superconformal multiplets that contain Schur operators (one for each multiplet). Here $R$ and $(j_1, j_2)$ denote the Cartan of the $SU(2)_R$ symmetry and the Lorentz quantum numbers of the superconformal primary of the multiplet, while
$h$ is the chiral dimension and $\bf{r}$ the $U(1)_{\bf r}$ charge of the Schur operator. 
In our conventions, $R$, $j_1$, and $j_2$ take non-negative half-integer values.  Even and odd values of $2 \bf{r}$ correspond to bosonic and fermionic Schur operators, respectively.
\label{tab:SchurOp}}
\end{table}

The Schur index only gives us information about the chiral dimension $h$, and we see from Table \ref{tab:SchurOp}  that for fixed $h$ there is a certain ambiguity in identifying the corresponding 4d supermultiplet, which gets worse for increasing $h$. 
Nevertheless, leveraging the physical assumptions above, some  useful statements can be made for low $h$. A general Schur index has a series expansion of the form\footnote{In most of the paper we focus on untwisted modular equations, which yield characters with an integer power expansion, but here we discuss  the general case.}
\be
Z_{\rm Schur}( q) = \sum_{k \in \frac{\mathbb{N}}{2} } a_k q^k=
1 + a_{\frac{1}{2}} q^{\frac{1}{2}} + a_1 q +  a_{\frac{3}{2}} q^{\frac{3}{2}}+ a_2 q^2 + \dots \,, \quad a_k \in \mathbb{Z} \, .
\ee
Let us interpret the first few coefficient:
\begin{itemize}
\item
We see at once that $a_{\frac{1}{2}} = 0$, because the only supermultiplet that could yield $h=\frac{1}{2}$ states is the hypermultiplet $\hat{\cB}_{\frac{1}{2}}$, violating assumption (i). 
\item $a_1$ counts the number of $\hat{\cB}_{1}$ multiplets, which are the only non-free multiplets with $h=1$. This basic fact is used throughout the paper.

\item $a_{\frac{3}{2}}$ gets a positive contribution from the number of $\hat{\cB}_{\frac{3}{2}}$ multiplets  and  a negative contribution from the number of additional supercurrent multiplets  (if any). One often has an \emph{a priori} opinion about the number of supersymmetries of the theory, and thus can read off the number of Higgs  operators  with $R = \frac{3}{2}$. (They are necessarily generators of the chiral ring since they cannot be composites of operators with lower $R$). 
\item $a_2$ gets a variety of contributions. There are first of all the affine Kac-Moody algebra states $\{ J^a_{-2} |0 \rangle,  J_{-1}^a J_{-1}^b | 0 \rangle \}$,
where $a, b$ are adjoint indices of the flavor symmetry algebra. We denote their number 
 by $a_2^{\rm KM}$.\footnote{To determine $a_2^{\rm KM}$, one must of course take care of subtracting possible null states, which are determined if one knows the levels of the (simple factors of the) affine current algebra. In the absence of null states, $a_2^{\rm KM} = a_1 + a_1 (a_1+1)/2$.} If the central charge satisfies the Sugawara condition (\ref{eq:Sugdef}), the 4d stress tensor multiplet $\hat{\cC}_{0 (0 ,0)}$ is already accounted for by the 2d Sugawara construction \cite{Beem:2013sza, Beem:2018duj}. If the Sugawara condition is not obeyed, there must be an independent generator with $h=2 $ to account for the 4d stress tensor.  Finally, there can be additional generators of type 
$\hat {\cal B}_2$ (bosonic), $\cD_{1(0,0)}$,  $\overline{\cD}_{1(0,0)}$ (fermionic), or
$\cD_{\frac{1}{2}(0,\frac{1}{2})}$,  $\overline{\cD}_{\frac{1}{2}(\frac{1}{2},0)}$ (bosonic). It is important to note that
CPT invariance of the 4d theory implies that  $\cD_{R(0,j)}$ and $\overline{\cD}_{R(j,0)}$ multiplets must come in pairs. 
All in all then, we have\footnote{Here  $\#  \hat {\cal B}_2$ denote the number of {\it additional}  $\hat {\cal B}_2$ multiplets, on top of the ones that correspond to the composites  $J_{-1}^a J_{-1}^b | 0 \rangle$. These additional  $\hat {\cal B}_2$ operators are generators of the Higgs chiral ring.}
\be \label{a2sumrule}
a_2 =  a_2^{\rm KM} + n^{\rm T} + \#  \hat {\cal B}_2  - 2\, \#   \cD_{1(0,0)}  + 2\, \#  \cD_{\frac{1}{2}(0,\frac{1}{2})} \, ,
\ee
where  $n^{\rm T}\equiv 0$ if the Sugawara condition is obeyed and $n^{\rm T}\equiv 1$ if it is not obeyed.
\end{itemize}
The sum rule (\ref{a2sumrule}) can sometimes be used to rigorously rule out putative Schur indices. In the context of this paper, our algorithm yields the requisite information to compute $a_2^{\rm KM}$ and $n^{\rm T}$. If we happen to also have an opinion about the Higgs branch of the theory, in particular on the number of $\hat {\cal B}_2$ generators, we get a sharp constraint on the {\it parity} (even or odd) of $a_2$. A more heuristic use of the sum rule is as a plausibility check. The additional level-two generators must form representations of the flavor algebra. There are usually ways to satisfy the sum rule, but if they  require baroque choices of flavor decomposition, we may conclude that the character is unlikely to have a 4d interpretation.

\section{Allowed rank-two Coulomb branch dimensions}
\label{app:rank2dims}

In this appendix we rederive the allowed pairs of Coulomb branch dimensions for rank-two $\cN=2$ SCFTs, which will be necessary data for our rank-two scan. We begin by noting that, based solely on the fact that CB monodromies must take values in the discrete group $Sp(2r,\ZZ)$, the allowed scaling dimensions of Coulomb branch operators are constrained to take values in \cite{Argyres:2018urp,Caorsi:2018zsq}:
\beq
\D\in\left\{\frac n m\,\Big|\,\,n,m \in \mathbb{N},\,0<m\leq n,\,{\rm gcd}(n,m)=1,\,\varphi(n)\leq 2r\right\}~,
\eeq
where $\varphi(n)$ is Euler's totient function. This set depends only on the rank $r$ of the SCFT. The maximum allowed value scales as $\D_{\rm max}\sim r \ln \ln r$, while the total numbers of allowed scaling dimensions $\boldsymbol{N}$ and integer scaling dimensions $\boldsymbol{N}_{\rm int}$ scale as \cite{Caorsi:2018zsq}
\begin{align}
\boldsymbol{N}\sim 2 r^2 \frac{\zeta(2)\zeta(3)}{\zeta(6)}~, \hspace{0.5 in}\boldsymbol{N}_{\rm int}\sim 2 r \frac{\zeta(2)\zeta(3)}{\zeta(6)}~,
\end{align}
with $\zeta(s)$ the Riemann zeta function. Naively, the number of allowed $r$-tuples would be
\beq\label{eq:counting}
{\rm Number\, of \, \emph{r-}tuples}=\left(\begin{array}{c}\boldsymbol{N}+r-1\\\boldsymbol{N}-1\end{array}\right)~. 
\eeq 
However, we will now see that this number can be cut down significantly \cite{Argyres:2022yet}.

\begin{table}[!t]
\begin{center}
\renewcommand{\arraystretch}{1.2}
\hspace*{-0.1cm}\begin{tabular}{c|c}
\multicolumn{2}{c}{\textsc{Genuinely rank-two pairs}}\\
\hline\hline
$P_i(z)$& $\{\D_1,\D_2\}$
\\\hline\hline
\multirow{2}{*}{$\Phi_5(z)$}&$\{\frac{4}{3},\frac53\}\ \{\frac54,\frac32\}\ \{\frac32,\frac52\}\ \{\frac54,3\}\ \{\frac53,3\}\ \{\frac52,3\}\ \{3,5\}\ \{\frac54,4\}$\\
&$\{\frac53,4\}\ \{\frac52,4\}\ \{4,5\}\ \{\frac54,8\}\ \{\frac53,8\}\ \{\frac52,8\}\ \{5,8\}$\\
\hline
\multirow{3}{*}{$\Phi_8(z)$}&$\{\frac65,\frac85\}\ \{\frac43,\frac83\}\ \{\frac87,\frac{10}7\}\ \{\frac87,\frac{12}7\}\ \{\frac85,\frac{12}5\}\ \{\frac83,\frac{10}3\}\ \{\frac87,4\}\ \{\frac85,4\}$\\
&$\{\frac83,4\}\ \{4,8\}\ \{\frac87,6\}\ \{\frac85,6\}\ \{\frac83,6\}\ \{6,8\}\ \{\frac87,12\}\ \{\frac85,12\}$\\
&$\{\frac83,12\}\ \{8,12\}$\\
\hline
\multirow{2}{*}{$\Phi_{10}(z)$}&$\{\frac{10}{9},\frac43\}\ \{\frac43,\frac{10}3\}\ \{\frac{10}9,4\}\ \{\frac{10}7,4\}\ \{\frac{10}3,4\}\ \{4,10\}\ \{\frac{10}9,8\}\ \{\frac{10}7,8\}$\\
&$\{\frac{10}3,8\}\ \{8,10\}$\\
\hline$\Phi_{12}(z)$&$\{\frac65,\frac{12}5\}\ \{\frac{12}{11},6\}\ \{\frac{12}7,6\}\ \{\frac{12}5,6\}\ \{6,12\}\ \{\frac{12}{11},8\}\ \{\frac{12}7,8\}\ \{\frac{12}5,8\}$\\
\hline\hline
\multicolumn{2}{c}{\textsc{Non-Genuine Rank-Two Pairs}}\\
\hline\hline
\multicolumn{2}{c}{$\{\D_1,\D_2\}$}\\
\hline\hline
\multicolumn{2}{c}{$\{\frac65,\frac65\}\ \{\frac65,\frac43\}\ \{\frac65,\frac32\}\ \{\frac65,2\}\ \{\frac65,3\}\ \{\frac65,4\}\ \{\frac65,6\}\ \{\frac43,\frac43\}\ \{\frac43,\frac32\}\ \{\frac43,2\}$}\\
\multicolumn{2}{c}{$\{\frac43,3\}\ \{\frac43,4\}\ \{\frac43,6\}\ \{\frac32,\frac32\}\ \{\frac32,2\}\ \{\frac32,3\}\ \{\frac32,4\}\ \{\frac32,6\}\ \{2,2\}\ \{2,3\}$}\\
\multicolumn{2}{c}{$\{2,4\}\ \{2,6\}\ \{3,3\}\ \{3,4\}\ \{3,6\}\ \{4,4\}\ \{4,6\}\ \{6,6\}$}\\
\hline\hline
\end{tabular}
\end{center}
\caption{Full list of allowed pairs of scaling dimensions at rank-two. The \emph{genuinely rank-two} pairs are those which include at least one entry which is not allowed at rank-one. In the first column we report the form of the characteristic polynomial where $\Phi_d(z)$ denotes a degree-$d$ cyclotomic polynomial. 
}
\label{tab:CBSDr2}
\end{table}%

First observe that the product of a rank-$n$ SCFT and a rank-$(r-n)$ SCFT (for $n<r$) gives a consistent rank-$r$ $\cN=2$ theory. This implies that the union of any allowed $(r-n)$-tuple with an allowed $n$-tuple gives an allowed $r$-tuple. Thus, without loss of generality, we can focus on determining the \emph{genuine} $r$-tuples, i.e. those which contain at least one Coulomb branch scaling dimension which is not allowed at rank $r-1$. Now consider the following locus of the Coulomb branch,
\beq
\cI_i:=\{u_1=0\,,\dots,\,u_{i-1}=0,\,u_i\neq0,\, u_{i+1}=0,\dots,\,u_r=0\}
\eeq
and assume that this locus is regular, i.e. $\cI_i$ is not part of the singular locus. In particular, by Proposition \ref{thm:rank1thm}, if we choose $i$ such that $\D_i$ is not an allowed scaling dimension at rank-$(r-1)$, then $\cI_i$ is necessarily non-singular. 

Since $\cI_i$ is non-singular, the fiber $X^i_u$ is a rank-$r$ Abelian variety for all $u\in \cI_i$. That is,
\beq
X^i_u = \CC^r/\Lambda^i_u~,\qquad \forall u\in \cI_i
\eeq
with $\Lambda^i_u$ a rank-$2r$ lattice. The following $U(1)_r$ transformation
\beq\label{autIi}
\xi^i:\ u_k\to \exp\left(2 \pi i \D_k/\D_i\right) u_k~,\hspace{0.5 in} \xi^i \in U(1)_r
\eeq
acts trivially on $\cI_i$ and thus generates an automorphism of $X^i_u$, i.e. an element $\boldsymbol{\xi}^i\in GL(r,\CC)$ such that $\boldsymbol{\xi}^i(\La^i_u)=\La^i_u$. Using the fact that the holomorphic symplectic form on $X\to \cC$ is non singular \cite{Donagi:1995cf}, given \eqref{autIi} it is possible to derive an explicit representation of $\boldsymbol{\xi}^i$:
\begin{eqnarray}\label{ana}
\boldsymbol{\xi}^i&&:=\rho_a(\xi^i)\in GL(r,\CC)\\\nonumber
&&:={\rm diag}\left(e^{2\pi i(\D_1-1)/\D_i},...,e^{2\pi i(\D_{i-1}-1)/\D_i},e^{-2\pi i/\D_i},e^{2\pi i(\D_{i+1}-1)/\D_i},...,e^{2\pi i(\D_r-1)/\D_i}\right)
\end{eqnarray}
which explicitly depends on the $r$-tuple of Coulomb branch scaling dimensions $\{\D_1,...,\D_r\}$. We now observe that there is a general relation between $\boldsymbol{\xi}^i$, which is also called the \emph{analytic representation} of $\xi^i$, and the induced action $\xi^i_H\in Sp(2r,\Z)$ on the homology of $X^i_u$, i.e. the \emph{homological representation} of $\xi^i$. In particular, the following isomorphism of complex representations holds:
\beq\label{anahomo}
\xi^i_H\otimes \CC \cong \boldsymbol{\xi}^i\oplus (\boldsymbol{\xi}^i)^*
\eeq
where $^*$ denotes complex conjugation. Given \eqref{anahomo} and \eqref{ana}, we obtain the restriction that the following quantities
\beq
\exp\left(2\pi i\frac{\D_j-1}{\D_i}\right),\quad i=1,2,...,r
\eeq
must be roots of a degree-$2r$ polynomial $P_i(z)$ (the characteristic polynomial of $\xi_H^i$) which is a product of degree-$d$ cyclotomic polynomials $\Phi_d(z)$:
\beq
P_i(z)=\prod_d\Phi_d(z)^{n(d)}~,\hspace{0.5 in} \sum_d\varphi(d)n(d)=2r~,
\eeq
where again $\varphi(d)$ is Euler's totient function. Imposing that $\D_i$ is not a rank-$(r-1)$ dimension further constraints the form of $P_i(z)$, giving
\beq
P^{\rm new}_i(z)=\Phi_d(z)~,\hspace{0.5in} \varphi(d)=2r~.
\eeq
Carrying out the analysis for $r=2$, we then obtain 51 genuinely rank-2 pairs of CB scaling dimensions. Together with 28 pairs having both entries being allowed at rank-one, this gives a total of 79 allowed pairs of scaling dimensions at rank-two. These are tabulated in Table \ref{tab:CBSDr2}. As promised, this provides a great improvement on the naive counting in \eqref{eq:counting}, which would have predicted a total of 276 allowed pairs.

Let us conclude by noting that for cases in which one of the  Coulomb branch scaling dimensions is $\Delta_{u_i}=2$, the theory has an exactly marginal operator parameterizing a conformal manifold. In \cite{Beem:2014zpa,Perlmutter:2020buo} it was conjectured that any $\cN=2$ $n$-dimensional conformal manifold arises by gauging $n$ simple factors in the global symmetry of isolated $\cN=2$ SCFTs. This means that if $\Delta_{u_i}=2$ the corresponding theory is either Lagrangian, or an $SU(2)$ conformal gauging of a lower rank theory. Since all Lagrangian theories have already been classified, we will for the most part exclude such cases from our scan.

\section{Rank-one IR-free theories}
\label{app:IRFree}

In this appendix we discuss the set of rank-one IR-free theories, which will serve as necessary data for our rank-two scan. The full set of rank-one IR-free is obtained by considering $U(1)$ or $SU(2)$ gauge theories coupled to sufficient numbers of hypermultiplets (or discrete gaugings thereof). Let us begin with the case of $U(1)$ gauge theory. The most general matter content consists of $n_i$ hypermultiplets of charge $q_i$ for $i=1, \dots, N$. The beta-function for this theory is given by $\beta\propto\sum_i q_i^2 n_i$. The relevant Kodaira fiber is then of type $I_{\sum_i q_i^2 n_i}$, and the corresponding flavor symmetry is  $\bigoplus_i \mathfrak{su}(n_i)$. By nature of being a $U(1)$ theory we have $\Delta_u = 1$, and it is easy to see that  $12c_{\mathrm{4d}} = 2 + \sum_i n_i$ and $h_{\mathrm{ECB}}=0$. This is enough data to compute $b= \sum_i n_i$ and $k_i = 2$. A summary of the CFT data for $U(1)$ IR-free theories is given in Table \ref{tab:U1IRFree}.

\begin{table}[t!]
\begin{center}
\renewcommand{\arraystretch}{1.2}
\hspace*{-0.1cm}\begin{tabular}{|c|c|c|c|c|c|c|c|}
\multicolumn{8}{c}{\large{$\left[I_{\sum\limits_i q_i^2n_i }, \bigoplus\limits_i \suf(n_i)\right]$}}\\
\hline
\hline
$12 c_{\mathrm{4d}}$& $\Delta_u$ &$h_{\mathrm{ECB}}$&$\mathbf{R}$& $T(\mathbf{R})$&$b$&$\mathfrak{g}$&$k_{\mathfrak{g}}$\\
\hline
$2 + \sum_i n_i$& 1&  0&$(\mathbf{1}, \dots, \mathbf{1})$& 0 &$ \sum_i n_i $& $\bigoplus_i \suf(n_i)$&  $(2, \dots, 2)$\\
\hline
\end{tabular}
\end{center}
\caption{CFT data of $\cN=2$ IR-free theories with $U(1)$ gauge group and $n_i$ hypermultiplets of charge $q_i$.}
\label{tab:U1IRFree}
\end{table}%

One may also consider discrete gaugings of this class of theories. In particular, when one of the $n_i$ is even a $\ZZ_2$ gauging can be done. This results in a theory with the same data as above, but with the corresponding $\mathfrak{su}(2n_i)$ flavor factor replaced by $\mathfrak{sp}(2n_i)$.

Next consider IR-free $SU(2)$ gauge theories. The most general matter content consists of $n_i$ hypermultiplets in real representations $\mathbf{R_i}$ for $i=1, \dots, N_R$, and $n_i$ hypermultiplets in pseudoreal representations $\mathbf{R_i}$ for $i = N_R+1, \dots, N_R+N_{PR}$. The beta-function is given by $\beta \propto \sum_i n_i T(\mathbf{R_i}) - 4$. The relevant Kodaira fiber is of type $I^*_{ \sum_i n_i T(\mathbf{R_i}) - 4}$, and the corresponding flavor symmetry is $\bigoplus_{i=1}^{N_R} \mathfrak{sp}(2n_i) \bigoplus_{i=N_R +1}^{N_{R}+N_{PR}} \mathfrak{so}(2n_ i)$. Since it is a $SU(2)$ gauge theory, we must have $\Delta_u = 2$. Note that each $ \mathfrak{sp}(2n_i)$ flavor factor contributes a factor of $n_i$ to the dimension of the extended Coulomb branch, i.e. $h_{\mathrm{ECB}} = \sum_{i=1}^{N_R} n_i$, and the central charge receives contributions from all matter fields $12 c_{\mathrm{4d}} = 6 + \sum_{i} n_i\, \mathrm{dim}\, \mathbf{R_i}$. From this information we may then compute the quantity $b$ as well as the flavor levels. A summary of the CFT data for $SU(2)$ theories is reported in Table \ref{tab:SU2IRFree}.

Note that in the $SU(2)$ case there is an additional subtlety which is that changing the normalization of the generators can lead to different monodromies---see e.g. Section 4.2 of \cite{Argyres:2015ffa}. However, this change in normalization does not affect any of the data in Table \ref{tab:SU2IRFree}, and hence will be irrelevant for the purposes of this paper. One could also consider discrete gaugings of this class of theories, but we will not do so here.

\begin{table}[!t]
\begin{center}
\begin{adjustbox}{center,max width=.7\textwidth}
\renewcommand{\arraystretch}{1.2}
\begin{tabular}{|c|c|c|c|c|c|c|c|}
\multicolumn{8}{c}{\large{$\left[I^*_{ \sum_i n_i T(\mathbf{R_i}) - 4},\, \bigoplus\limits_{i=1}^{N_R} \mathfrak{sp}(2n_i) \bigoplus\limits_{i=N_R +1}^{N_{R}+N_{PR}} \mathfrak{so}(2n_ i)\right]$}}\\
\hline
\hline
$12 c_{\mathrm{4d}}$& $\Delta_u$ &$h_{\mathrm{ECB}}$&$\mathbf{R}$& $T(\mathbf{R})$&$b$&$\mathfrak{g}$&$k_{\mathfrak{g}}$\\
\hline
$6\text{+}\hspace{-7pt} \sum\limits_{i=1}^{\scaleto{N_R+N_{PR}}{4pt}}\hspace{-5pt} n_i\, d_\mathbf{R_i}$&
2&  
$\sum\limits_{i=1}^{N_R} n_i$&
$(\mathbf{2n_1},...,\mathbf{2n}_{N_R}\underbrace{\mathbf{1}, ...,\mathbf{1}}_{N_{PR}\ {\rm times}}\hspace{-6pt})$& 
$N_R$ &
$ 2\text{+} \hspace{-3pt}\sum\limits_{i=1}^{N_R} \hspace{-2pt}\frac{n_i (d_{\mathbf{R}_i}-1)}{2}\text{+} \hspace{-7pt} \sum\limits_{\scaleto{i=N_R+1}{4pt}}^{\scaleto{N_R+N_{PR}}{4pt}}\hspace{-2pt} \frac{n_i \,d_{\mathbf{R}_i}}{2}  $& 
$\bigoplus\limits_{i=1}^{N_R} \mathfrak{sp}(2n_i) \hspace{-2pt}\bigoplus\limits_{\scaleto{i=N_R +1}{4pt}}^{\scaleto{N_{R}+N_{PR}}{4pt}} \mathfrak{so}(2n_ i)$& 
$(\underbrace{3, \dots, 3}_{N_R\ {\rm times}} , \underbrace{4, \dots, 4}_{N_{PR}\ {\rm times}})$\\
\hline
\end{tabular}
\end{adjustbox}
\caption{CFT data of $\cN=2$ IR-free theories with a $SU(2)$ gauge group and with $N_R$ real and $N_{PR}$ pseudoreal representations. $d_{\Rbf_i}$ indicates the dimension of the representation $\Rbf_i$.}
\label{tab:SU2IRFree}
\end{center}
\end{table}%

For the purposes of our computerized scan, it will be necessary to restrict to a finite subset of rank-one IR-free theories. As will be discussed in the main text, our scan will restrict to theories with at most three simple factors in their flavor symmetry, and hence we will impose the same on the allowed IR-free theories. In the case of $U(1)$ gauge theories and their discrete gaugings, this gives rise to infinite families of theories labelled by charges $(q_1,q_2,q_3)$ and degeneracies $(n_1,n_2,n_3)$. Since the charges do not appear anywhere in the data in Table \ref{tab:U1IRFree}, and since we do not discuss the monodromy of IR-free theories anywhere in this paper, we can neglect them for the purposes of the scan. As for the degeneracies, we simply restrict them to a finite range---in practice $0 \leq n_i \leq 10$.

Turning next to the case of $SU(2)$ gauge theories, these are labelled by representations $(\mathbf{R}_1, \mathbf{R}_2,\mathbf{R}_3)$ and degeneracies $(n_1,n_2,n_3)$. As before, we restrict the degeneracies to take values in a finite range $0 \leq n_i \leq 10$. However, in the current case the dependence on the representations $(\mathbf{R}_1, \mathbf{R}_2,\mathbf{R}_3)$ does not drop out, and hence we must place restrictions on these as well. We find it well-motivated to restrict $\mathbf{R}_i$ to be either the fundamental or adjoint representation. This gives the finite set of theories which is fed into the algorithm.

\section{General structure of the extended Coulomb branch}
\label{app:ECB}

In this appendix we collect a number of observations on the generic structure of the extended Coulomb branch (ECB) of $\cN=2$ SCFTs and the way it transforms under the flavor symmetry $\gf$. 
These observations are necessary to properly run our scan and, to the authors' knowledge, have not yet appeared in the literature. After introducing the general idea, we will provide two concrete Lagrangian examples where these features arise. For simplicity we focus on rank-two theories, though the entire discussion straightforwardly generalizes to higher rank (and non-Lagrangian) SCFTs.

\subsection*{General discussion}

We begin with some general remarks. At rank-one there is only one way in which the ECB arises, namely as a series of free hypermultiplets with perhaps global quotienting structure, thus allowing for $\gf\neq \spf(2n)$. At rank-two there are instead two conceptually distinct ways:
\begin{itemize}

\item The ECB arises from free hypermultiplets tensored over every point of the CB (up to global twists). This means that the CB has locally a CB$\times \HH^n$ structure over \emph{all} points. This is the way in which the ECB of only one of the two examples below ($\cN=4$ $\suf(3)$ theory) behaves as well as all ECBs of rank-one theories.

\item The ECB might arise from rank-one theories supported on singular strata carrying a non-trivial ECB themselves. In this case, the local CB$\times \HH^n$ structure breaks down on the complex codimension-one locus where the rank-one theory is supported, and the $n$ free hypers in question become strongly-coupled there. This is a distinctively higher rank feature and in fact applies to both rank-two cases discussed below.

\end{itemize}

We will call the former possibility a \emph{free ECB} and the latter a \emph{coupled ECB}. There is a major difference among these two possibilities:
\begin{itemize}

\item[(1)] A \emph{free ECB} carries a flavor symmetry which \emph{does not act} on BPS states which are both massless and charged under the low-energy $\mathfrak{u}(1)^2$. As we will soon argue, a free ECB necessarily implies a coupled ECB, and that the ECB does not transform irreducibly under $\gf$.

\item[(2)] A \emph{coupled ECB} carries a flavor symmetry which instead \emph{acts} on charged states which become massless where the rank-one theory is supported.

\end{itemize}

The UV-IR simple flavor condition \cite{Martone:2020nsy} forces any simple flavor factor to act on massless charged states somewhere on the CB, and thus free ECBs seem to contradict this condition. However there are situations in which a free ECB can be realized: namely, when one of the rank-one theories on the codimension-one strata carries the same flavor symmetry which acts on the free ECB (either $\spf(2n)$ or $\suf(n)/\mathfrak{u}(n)$) and the flavor symmetry of the rank-two SCFT is the diagonal combination of the two. Thus a free ECB \emph{necessarily requires} a coupled one with the same flavor symmetry factor. This implies that the complex scalars of the free hypermultiplets of the ECB transform in a reducible representation of the flavor symmetry:
\beq\label{ECBrep}
\Rbf= \underbrace{\oplus_a \Rbf_a}_\text{Coupled\ ECB} \overbrace{\oplus \Rbf_\H}^\text{Free\ ECB}~,
\eeq
where the sum is over all the irreducible singular complex codimension-one strata which realize the flavor symmetry $\gf$. Here $\Rbf_\H$ is the contribution to the ECB which  is decoupled on the whole CB$\setminus\{0\}$.

How is this discussion relevant for the algorithm outlined in the main text? Recall that the way in which we compute the flavor central charge for a given simple flavor symmetry factor $k_{\gf_i}$ was to use \eqref{eq:mainstrateqs}, which we reproduce here for convenience:
\beq\label{applev}
k_{\mathfrak{g}_i} =  \sum_{\cI_a \in \cI_{\mathfrak{g}_i}} {\Delta_a^{\mathrm{sing}} \over \Delta_a} [k_a - T^i(\mathbf{R}_a)]+ T^i(\mathbf{R})~.
\eeq
This sum is over all the irreducible singular complex co-dimension one strata which realize the flavor symmetry $\gf_i$ and we have introduced an indexed Dynkin index $T^i(\cdot)$, which indicates the Dynkin index of the corresponding irreducible representaion with respect to the flavor factor $\gf_i$, i.e.~acting on a representation $\Rbf=(\Rbf_1,\Rbf_2)$ of $\gf_1\oplus \gf_2$, we have
\begin{align}
T^1(\Rbf)&:={\rm dim} \Rbf_2\, T(\Rbf_1)~,\\
T^2(\Rbf)&:={\rm dim} \Rbf_1\, T(\Rbf_2)~.
\end{align}
The Dynkin index also satisfies $T^i(\Rbf_1\oplus\Rbf_2)=T^i(\Rbf_1)+T^i(\Rbf_2)$ and thus, using \eqref{ECBrep} and assuming that each $\gf_i$ is realized on a single irreducible stratum, we find: 
\beq
T^i(\Rbf)=\left\{
\begin{array}{cl}
    T^i(\Rbf_a)&\hspace{0.2 in} \text{Coupled\ ECB}~,\\
    2\,T^i(\Rbf_a) &\hspace{0.2 in}\text{Free\ ECB}~.
\end{array}
\right.
\eeq
The upshot is then that a free versus coupled ECB affects the computation of the quantities $k_{\mathfrak{g}_i}$ appearing in our algorithm. In cases in which one cannot rigorously exclude a free ECB based on flavor symmetries, our algorithm checks both the free and coupled cases. In the scan carried out in Section \ref{sec:Results}, the only case that we identify which has a free ECB is $\cT_{15}$.

The discussion above generalizes straightforwardly to the ECB of rank-$r$ SCFTs, but with caveat that there are $r-1$ different types of free ECB depending on the co-dimensionality, strictly larger than one, of the locus where the free hyper is decoupled.

\subsection*{Examples}

Let us now see how these features arise in two rank-two Lagrangian examples, namely $\cN=4$ $\suf(3)$ SYM and $\cN=2$ $\spf(4)$ gauge theory with four hypers in the ${\bf 4}$ and one in the ${\bf 5}$. Both of these theories have a non-trivial ECB, but the former has a \emph{free ECB} while the latter has a \emph{coupled} one. 

\paragraph{$\boldsymbol{\cN=4\ \suf(3)}:$} From the $\cN=2$ point of view, this theory is an $\suf(3)$ vector multiplet with a single hypermultiplet transforming in the adjoint representation. The globally well-defined Coulomb branch coordinates are $u_1=\left\langle {\rm Tr}\left(\Phi^2\right)\right\rangle$ and $u_2=\left\langle {\rm Tr}\left(\Phi^3\right)\right\rangle$, where $\Phi$ is the complex scalar of the $\cN=2$ vector multiplet. It follows that $\D_{u_1}=2$ and $\D_{u_2}=3$.

A generic point on the Coulomb branch is parametrized, up to Weyl transformations, by:
\beq
\langle\Phi\rangle=\left(
\begin{array}{ccc}
a_1&0&0\\
0&a_2&0\\
0&0&-a_1-a_2
\end{array}
\right)~.
\eeq
As expected, this breaks $\suf(3)\to \mathfrak{u}(1)^2$. Notice that on the generic point of the CB not all components of the adjoint hypermultiplet are massive. In fact, the massless components are those which are uncharged under both $\mathfrak{u}(1)$ factors. It is then a straightforward exercise (involving decomposing the adjoint representation) to show that at a generic point of the CB there are two free hypermultiplets giving, locally, a CB$\times \H^2$ structure and $h_{\rm ECB}=2$. This matches our expectation that any rank-$r$ $\cN\geq 3$ theory has an $r$-(quaternionic-)dimensional ECB.

Now let's extend the analysis to the singular strata. Extra massless states appear on special loci where either there is an enhancement of the unbroken gauge group, or some component of the hypermultiplet become massless. The only relevant enhancement in this case is when $a_1=a_2$ (and their Weyl equivalents) which corresponds to an enhancement $\mathfrak{u}(1)^2\to\suf(2)\oplus \mathfrak{u}(1)$. This locus can be written in terms of the CB coordinates as
\beq
u_1^3+u_2^2=0\qquad \Rightarrow \qquad \D^{\rm sing}=6~,
\eeq
and the ${\bf 8}$ decomposes as:
\beq
{\bf 8}\to {\bf3}_0\oplus {\bf 2}_{\pm2}\oplus {\bf1}_0~,
\eeq
where the subscript denotes the $\mathfrak{u}(1)$ charge. To understand which component charged under the $\suf(2)$ is massless, we need to look at its charge under the commutant of $\suf(2)$ inside $\suf(3)$. This is simply $\mathfrak{u}(1)$. All in all we find that 
\begin{itemize}
    \item [$a$)] ${\bf 1}_0$ denotes a massless component of the adjoint hyper which is a singlet under both the $\suf(2)$ and the $\mathfrak{u}(1)$ and which is therefore massless not just over the generic point of the CB but over the entire CB$\setminus\{0\}$ - the aforementioned hallmark of a free ECB.
    
    \item[$b$)] The ${\bf 3}_0$ instead gives rise to $\cN=4$ $\suf(2)$ SYM, denoted ${\cS^{(1)}_{\varnothing,2}}$ in what follows, which is then identified as the rank-one theory describing the low-energy effective theory along the singularity. Notice that ${\cS^{(1)}_{\varnothing,2}}$ itself has a one dimensional ECB, which gives rise to the second free hyper which is present at a generic point on the CB.
\end{itemize} 

Let's now discuss the flavor structure of the theory. Since the theory has a single eight-dimensional real representation, the flavor symmetry is $\suf(2)_8$. We would like to reproduce this result from the analysis of the CB stratification. Under the flavor $\suf(2)$, the two-(quaternionic-)dimensional ECB transforms as $\Rbf={\bf 2}\oplus{\bf 2}$, one ${\bf 2}$ coming from the free hyper in $(a)$ and one from $(b)$. In particular, the ECB does not transform irreducibly. The UV-IR simple flavor condition is satisfied as the $\suf(2)$ is realized on the singular stratum---i.e. the $\cN=4$ $\suf(2)$ theory also has an $\suf(2)$ flavor symmetry. To compute the level we can use \eqref{applev} which simplifies since there is a single singular stratum:
\beq\label{levelSU3}
k_{\suf(2)} = {\Delta^{\mathrm{sing}} \over \Delta_{{\cS^{(1)}_{\varnothing,2}}}} \left[k_{{\cS^{(1)}_{\varnothing,2}}} - T(\mathbf{R}_{{\cS^{(1)}_{\varnothing,2}}})\right]+ T(\mathbf{R})
\eeq
where $T(\cdot)$ indicates the Dynkin index of the corresponding representation, normalized in such a way that $T({\bf N})=1$ for ${\bf N}$ being the fundamental representation of $\suf(N)$. 
 To compute $T(\mathbf{R}_{{\cS^{(1)}_{\varnothing,2}}})$ observe that the ECB of the $\cN=4$ $\suf(2)$ theory transforms as ${\bf 2}$ under the $\suf(2)$ flavor symmetry. Similarly,
 \beq
T(\Rbf)=T({\bf 2}\oplus{\bf 2})=T({\bf 2})+T({\bf 2})=2~.
\eeq
The remaining quantities can be directly read off from Table \ref{Theories}. Plugging this in to \eqref{levelSU3}, we obtain the desired $k_{\suf(2)}=8$.

\paragraph{$\boldsymbol{\spf(4)}$ with $4({\bf4})\oplus{\bf 5}$} Let's now discuss the second example. We will be somewhat brief since most of the analysis follows closely the one above. In this case the globally defined Coulomb branch coordinates are $u_1=\left\langle {\rm Tr}\left(\Phi^2\right)\right\rangle$ and $u_2=\left\langle {\rm Tr}\left(\Phi^4\right)\right\rangle$, hence $\D_{u_1}=2$ and $\D_{u_2}=4$.

In the $\spf(4)$ case there are two inequivalent directions, up to Weyl tranformation, where an $\suf(2)$ is left unbroken (corresponding to the long and short simple roots) and which therefore give rise to singularities. The loci can be written in both cases in terms of the global coordinates as $u_2+\l u_1^2=0$, $\l\in\CC^*$, giving $\D^{\rm sing}=4$. 

The commutant, whose charges determine which component is massless, is different in the two cases. Let us denote the case with commutant $\suf(2)$ by $(i)$ and the case with commutant $\mathfrak{u}(1)$ by $(ii)$.  We may now analyze how the hypermultiplets decompose:
\begin{align}
    (i)&: \qquad\left\{
    \begin{array}{l}
        {\bf 4}\to({\bf2},{\bf1})\oplus({\bf1},{\bf2})~,\\
        {\bf 5}\to({\bf2},{\bf2})\oplus({\bf1},{\bf1})~,\\
    \end{array}
    \right.\\
    (ii)&: \qquad\left\{
    \begin{array}{l}
        {\bf 4}\to{\bf2}_{\pm1}~,\\
        {\bf 5}\to{\bf3}_{\pm0}\oplus{\bf1}_{\pm2}~.\\
    \end{array}
    \right.
\end{align}
These imply that along $(i)$, each hyper in the ${\bf4}$ contributes a massless hyper in the fundamental of the $\suf(2)$, while the ${\bf 5}$ contributes only a decoupled free hyper.  In contrast, along $(ii)$ all the components of the hyper in the ${\bf 4}$ are massive but the ${\bf 5}$ contributes a massless hyper in the adjoint. We conclude that along $(i)$ the low-energy effective theory looks like $\cT^{(1)}_{D_4,1}\times\H$, \emph{i.e.} an $\suf(2)$ $\cN=2$ theory with $N_f=4$ times a free hyper, while along $(ii)$ the low-energy theory is simply ${\cS^{(1)}_{\varnothing,2}}$. 

Let us now discuss the flavor symmetry, which in this case is $\sof(8)_4\oplus \suf(2)_5$. Since only the $\suf(2)_5$ acts on the ECB, we will not discuss the $\sof(8)_4$ factor. Notice three important things:
\begin{itemize}
    \item[1.] The $\suf(2)$ only acts on charged massless states along $(ii)$---again ${\cS^{(1)}_{\varnothing,2}}$ has a $\suf(2)$ flavor symmetry. Along $(i)$, the $\suf(2)$ instead acts on the free hyper which is massless but decoupled.

    \item[2.] Relatedly, in this case there is no component of the ECB which is decoupled on the entire CB$\setminus\{0\}$. The free hyper on $(i)$ couples on $(ii)$ to form a low-energy ${\cS^{(1)}_{\varnothing,2}}$.
    
    \item[3.] The ECB in this case transform irreducibly under the $\suf(2)$, i.e. $\Rbf={\bf2}$.
\end{itemize}
To conclude our discussion, let's compute the flavor level. Since the $\suf(2)$ is realized only one of the two strata where the rank-one theory describing the massless states is ${\cS^{(1)}_{\varnothing,2}}$, the relevant equation is analogous to \eqref{levelSU3}, but now
\beq
\begin{array}{c}
\suf(3)\\
\hline
\D^{\rm sing}=6\\
T(\Rbf)=2
\end{array}\qquad\to\qquad
\begin{array}{c}
\spf(4)\\
\hline
\D^{\rm sing}=4\\
 T(\Rbf)=1
\end{array}
\eeq
which, with the new values, gives $k_{\suf(2)}=5$.

\section{The VOA for $\cT_9$}
\label{app:T9VOA}
In this appendix we give additional information about the VOA of the candidate theory $\cT_9$. Despite the fact that $\cT_9$ was shown to not exist as a legitimate 4d $\cN=2$ theory in Section \ref{sec:Isotrivialcases}, here we will see that the corresponding VOA is a perfectly well-defined minimal model. To begin, note that in this case the inidicial roots give the following chiral dimensions 
\bea
\label{eq:ourhs}
\{h_i\} = \left \{0, - {1\over 3}, -{ 2\over 3} , -{4\over 9}\right\}~.
\eea
We note further that the central charge $ c_{\mathrm{2d}} = -{44\over 3}$ is the same as that of the $(4,9)$ $\cW_3$ minimal model. The latter theory has a total of 28 modules, with the list of distinct chiral dimensions being (see e.g. \cite{Bilal:1991eu,Iles:2014gra,Iles:2013jha})
\bea
\label{eq:W3hs}
\{h_i\} = \left \{0, - {1\over 3}, -{ 2\over 3} , -{4\over 9}, 2, {1\over 3}, {2\over 3}, {1\over 27},{4\over 27}, {22\over 27}, {25\over 27},-{8\over 27},-{11\over 27}, -{14 \over 27}, -{17 \over 27} \right\}~.
\eea 
 We will denote the corresponding operators by $O_{h_i}$. For some $h_i$ there are in fact two operators with conjugate $W_0$ eigenvalues, in which case we denote them as $O_{h_i}$ and $\bar O_{h_i}$. 

We see that (\ref{eq:ourhs}) is a subset of (\ref{eq:W3hs}), and that $\cW_3(4,9)$ has primaries of integer dimension.
Indeed there are two spin-two primaries, $O_2$ and $\bar O_2$.  Inserting the modular data of $\cW_3(4,9)$ (which can be found in \cite{Beltaos:2010ka}) into the Verlinde formula, we may compute the fusion of these spin-two primaries to be
\bea
O_2 \times O_2 = \bar O_2~,\hspace{0.5in} \bar O_2 \times \bar O_2 = O_2~, \hspace{0.5 in} O_2 \times \bar O_2 = 1~.
\eea
This means that the corresponding Verlinde lines $L_2$ and $L_{\bar 2}$ satisfy $L_{ 2}^3 = L_2 L_{\bar 2} =1$, and thus generate a $\ZZ_3$ symmetry. 

In order to go from $\cW_3(4,9)$ to the VOA corresponding to $\cT_9$, we begin by gauging this $\ZZ_3$ symmetry. This projects out all but the operators of chiral dimension 
\bea
\{h_i\} |_{\ZZ_3 \,\, \mathrm{even}}= \left \{0, - {1\over 3}, -{ 2\over 3} , -{4\over 9}, 2, {1\over 3}, {2\over 3} \right\}~,
\eea
and leads to a theory with a block-diagonal modular invariant. This modular invariant can be made diagonal by extending the chiral algebra by operators of integer spin---in this case, $O_2$ and $\bar O_2$.\footnote{Note that this entire discussion has a direct analog in the more familiar realm of Virasoro minimal models. Indeed, it is well-known that the 3-states Potts model can be obtained by starting with the tetracritical Ising model and gauging a $\ZZ_2$ symmetry. This gives rise to a block-diagonal modular invariant, which can be made diagonal by extending by a single spin-three current.} That such an extension is legitimate can be checked by explicitly closing the OPEs of $T$, $W$, $O_2$, and $\bar O_2$, where $W$ is the spin-three current of the $\cW_3$ algebra. Since we will need these OPEs in a moment, we give the relevant portions here:
\bea
O_2(z) O_2(w) &\sim& {2 \bar O_2 \over (z-w)^2} +  { \p \bar O_2 \over z-w} +\dots
\no\\
 \bar O_2(z) \bar O_2(w)&\sim& {2   O_2 \over (z-w)^2} +  {\p  O_2 \over z-w}+\dots
\no\\
T(z) O_2(w) &\sim& {2 O_2 \over (z-w)^2} + {\p O_2  \over z-w} + \dots
\no\\
T(z) \bar O_2(w) &\sim& {2 \bar O_2 \over (z-w)^2} + {\p \bar O_2  \over z-w} + \dots
\no\\
O_2(z) \bar O_2(w) &\sim& -{77 \over 16}{1 \over (z-w)^4} + {21 \over 16}{ T \over (z-w)^2} + {{21\over 32} \,\p T + {3 i \over 16} \sqrt{77 \over 2}\,W \over z-w} + \dots
\no\\
W(z) O_2(w) &\sim& {2 i\over 3}\sqrt{22\over 7}{ O_2 \over (z-w)^3} + i \sqrt{11\over 14}{ \p O_2 \over (z-w)^2} + {4 i \sqrt{2\over 77}\, \p^2 O_2 -6 i \sqrt{2 \over 77}\, :T  O_2:  \over z-w} +\dots
\no\\
W(z) \bar O_2(w) &\sim& - {2 i\over 3}\sqrt{22\over 7}{ \bar O_2 \over (z-w)^3} - i \sqrt{11\over 14} { \p \bar O_2 \over (z-w)^2} + {-4 i \sqrt{2\over 77}\,  \p^2 \bar O_2  +6 i \sqrt{2 \over 77}\,  :T \bar O_2: \over z-w} +\dots
\no\\
\eea
where $\dots$ denote terms regular in $(z-w)$, and we have refrained from writing the OPEs involving only $T$ and $W$, since these take the standard $\cW_3$ form.

Noting that the fusion rules of $O_2$ and $\bar O_2$ on the surviving primaries take the form 
\bea
\label{eq:OPEs}
&\vphantom{.}&O_2 \times O_{-2/3} = \bar O_{1/3}~, \hspace{0.8 in}\bar O_2 \times O_{-2/3} =  O_{1/3}~,
\no\\
 &\vphantom{.}&O_2 \times O_{-4/9} = \bar O_{-4/9}~, \hspace{0.7 in}\bar O_2 \times  O_{-4/9} = \bar O_{-4/9} ~,
 \no\\
&\vphantom{.}& O_2 \times O_{2/3}= O_{-1/3}~,\hspace{0.8 in}\bar O_2 \times  O_{2/3} = \bar O_{-1/3} ~,
 \no\\
&\vphantom{.}& O_2 \times O_{-1/3} = \bar O_{-1/3}~,\hspace{0.71 in}\bar O_2 \times O_{-1/3} = O_{2/3}~,
 \no\\
&\vphantom{.}&  O_2 \times \bar O_{-1/3} = \bar O_{2/3}~,\hspace{0.8 in}\bar O_2 \times\bar O_{-1/3} = O_{-1/3}~,
\no\\
&\vphantom{.}&O_2 \times O_{1/3} = O_{-2/3}~,\hspace{0.8 in}\bar O_2 \times O_{1/3} =\bar O_{1/3}~,
\no\\
&\vphantom{.}&O_2 \times \bar O_{1/3} = O_{1/3}~,\hspace{0.896 in}\bar O_2 \times \bar O_{1/3} = O_{-2/3}~,
\eea
we see that such an extension identifies $O_{-2/3} \leftrightarrow \bar O_{1/3} \leftrightarrow O_{1/3}$ as well as $O_{2/3} \leftrightarrow \bar O_{-1/3} \leftrightarrow O_{-1/3}$. At the level of characters, we then expect the result to be a theory with \textit{four} unrefined characters, namely\footnote{One could ask about twisted sectors, but these also have character $\chi_{-4/9}(\tau)$. Similar to the case of 3-state Potts, this manifests itself in a non-unit coefficient for $\chi_{-4/9}(\tau)$ in the modular invariant, 
\bea
Z_{\mathrm{inv} }= |\chi_0 |^2 + |\chi_{-1/3}|^2 + |\chi_{-2/3}|^2 + 3|\chi_{-4/9}|^2~.
\eea} 
\bea
\chi_0(\tau) &=& \tilde \chi_0(\tau) +2 \widetilde \chi_2(\tau)~,
\no\\
\chi_{-1/3}(\tau) &=& 2 \widetilde \chi_{-1/3}(\tau) +\widetilde \chi_{ 2/3}(\tau)~,
\no\\
\chi_{-2/3}(\tau) &=& \widetilde \chi_{-2/3}(\tau) + 2 \widetilde \chi_{1/3}(\tau)~,
\no\\
\chi_{-4/9}(\tau) &=& \widetilde \chi_{-4/9}(\tau) ~,
\eea
where $\widetilde \chi_{i}(\tau)$ are the characters of $\cW_3(4,9)$. These results match precisely with the four solutions to the original MDE.

To summarize, we have seen that upon gauging the $\ZZ_3$ of the  $\cW_3(4,9)$ theory and extending by a pair of spin-two currents, we obtain the VOA for the (failed) $\cT_9$ theory. In Section \ref{sec:Isotrivialcases}, we used a variety of Coulomb branch constraints to show that $\cT_9$ could not correspond to a legitimate $\cN=2$ theory. We close this appendix by showing how, even just from the viewpoint of the VOA data, the putative 4d theory would be rather exotic.

Assuming that $\cT_9$ did exist, we could us ask which 4d Schur multiplets the 2d operators $O_2$ and $\bar O_2$ could arise from. Both have chiral dimension $h =2$, and cannot have a $U(1)_r$ charge, since this would be incompatible with the OPEs, c.f. (\ref{eq:OPEs}). Thus from Table \ref{tab:SchurOp} we conclude that the only possible multiplets are $\hat \cB_2$  or $\hat \cC_{0(0, 0)}$. The $\hat \cB_2$ multiplet is a Higgs chiral ring multiplet with corresponding Schur operator being a superprimary with $R=2$, while $\hat \cC_{0(0, 0)}$ is the conserved stress tensor multiplet with Schur operator being the R-current with $R=1$. 

We now combine this information with some further constraints  that follow from 4d unitarity. It
 can be shown~\cite{Beem:2018duj, toappear} that 4d unitarity implies the existence of an automorphism $\sigma$ of the VOA such that, in a subspace of fixed $h$ and $R$ and with $r=0$, the inner product defined by $\langle \sigma (O_i) O_j  \rangle$ should have a definite sign, namely 
\bea
\mathrm{sgn}\,\langle \sigma (O_i) O_j  \rangle = (-1)^{h -R}~.
\eea
  An important fact is that $\sigma$ is anti-linear. Moreover, $\sigma^2 = (-1)^{2R}$. 
As a simple example, we note that for the stress tensor $\sigma (T) = T, h = 2, R=1$ and hence $\langle  T T \rangle \sim c$  must be negative. Similarly we have $\sigma (W) = W, h=3, R=2$ and hence $\langle W W \rangle \sim c <0$.

To apply this in our current situation, we note that $\sigma (O_2) = \epsilon\, \bar O_2$ with $\epsilon = \pm 1$.   Indeed, $O_2$ and $\bar O_2$ have opposite imaginary OPE coefficients in the $W_3 O_2$ and $W_3 \bar O_2$ OPEs of (\ref{eq:OPEs}), and recalling that $\sigma (W_3) =  W_3$, for $ \sigma$ to act as an automorphism of the VOA (sending the $W_3 O_2$ OPE to the  $W_3 \bar O_2$ one)  we need  $\sigma (O_2) =  \epsilon\, \bar O_2$, for some arbitrary coefficient $\epsilon$. Using that $\sigma^2 = (-1)^{2R} = 1$, we then deduce that $\epsilon = \pm 1$. This holds for both $R=1$ and $R=2$. 

We now look at the $O_2O_2$ and $\bar O_2 \bar O_2$ OPEs in (\ref{eq:OPEs}). In order for $\sigma$ to act as an automorphism, we need to require that $ \epsilon =1$.  It follows that  the coefficient of the $\cO(z^4)$ piece of $O_2  (z) \bar O_2  (0)$  OPE is given by  $-{77\over 16}$. 

Finally we check unitarity. We need $\mathrm{sgn}\langle \sigma (O_2 ) O_2  \rangle = (-1)^{h-R}$ , which is $-1$ for the $R=1$ assignment and $+1$ for the $R=2$ assignment. Using $\sigma (O_2 ) = \epsilon \,\bar O_2 $, we conclude that $\langle \sigma (O_2 ) O_2  \rangle =  -77/16  < 0$. Thus we are forced to conclude that $R=1$, i.e. $O_2 $ and $\bar O_2$ correspond to additional spin-two conserved current multiplets. Curiously, it turns out that there exists a linear transformation from $T$, $O_2$, and $\bar O_2$ to $T_i$ with $i=1, 2, 3$ where the currents $T_i$ appear symmetrically. Each of them obeys a Virasoro algebra with $c = - 44/3$, but there are also cross terms. 

The presence of multiple spin-two currents is usually not forgiven, though as we have remarked in footnote \ref{indecomposibility:footnote} in Appendix \ref{superconformal:subsection}, we are not actually aware of an airtight argument that a fully interacting and indecomposable SCFT is forbidden to have additional spin-two currents on top of the stress tensor. It  would have been exciting to discover such an exotic SCFT. Unfortunately, as we have seen, our candidate theory fails the Coulomb geometry constraints.

\end{appendix}

\bibliographystyle{JHEP}
\bibliography{bib.bib}

\providecommand{\href}[2]{#2}\begingroup\raggedright\begin{thebibliography}{10}

\bibitem{Bhardwaj:2013qia}
L.~Bhardwaj and Y.~Tachikawa, \emph{{Classification of 4d N=2 gauge theories}},
  \href{http://dx.doi.org/10.1007/JHEP12(2013)100}{\emph{JHEP} {\bf 12} (2013)
  100}, [\href{http://arxiv.org/abs/1309.5160}{{\tt 1309.5160}}].

\bibitem{Argyres:2015ffa}
P.~Argyres, M.~Lotito, Y.~L\"u and M.~Martone, \emph{{Geometric constraints on
  the space of $ \mathcal{N} $ = 2 SCFTs. Part I: physical constraints on
  relevant deformations}},
  \href{http://dx.doi.org/10.1007/JHEP02(2018)001}{\emph{JHEP} {\bf 02} (2018)
  001}, [\href{http://arxiv.org/abs/1505.04814}{{\tt 1505.04814}}].

\bibitem{Argyres:2015gha}
P.~C. Argyres, M.~Lotito, Y.~L\"u and M.~Martone, \emph{{Geometric constraints
  on the space of $ \mathcal{N} $ = 2 SCFTs. Part II: construction of special
  K\"ahler geometries and RG flows}},
  \href{http://dx.doi.org/10.1007/JHEP02(2018)002}{\emph{JHEP} {\bf 02} (2018)
  002}, [\href{http://arxiv.org/abs/1601.00011}{{\tt 1601.00011}}].

\bibitem{Argyres:2016xmc}
P.~Argyres, M.~Lotito, Y.~L\"u and M.~Martone, \emph{{Geometric constraints on
  the space of $ \mathcal{N}$ = 2 SCFTs. Part III: enhanced Coulomb branches
  and central charges}},
  \href{http://dx.doi.org/10.1007/JHEP02(2018)003}{\emph{JHEP} {\bf 02} (2018)
  003}, [\href{http://arxiv.org/abs/1609.04404}{{\tt 1609.04404}}].

\bibitem{Argyres:2016xua}
P.~C. Argyres, M.~Lotito, Y.~L\"u and M.~Martone, \emph{{Expanding the
  landscape of $ \mathcal{N} $ = 2 rank 1 SCFTs}},
  \href{http://dx.doi.org/10.1007/JHEP05(2016)088}{\emph{JHEP} {\bf 05} (2016)
  088}, [\href{http://arxiv.org/abs/1602.02764}{{\tt 1602.02764}}].

\bibitem{Gaiotto:2009we}
D.~Gaiotto, \emph{{N=2 dualities}},
  \href{http://dx.doi.org/10.1007/JHEP08(2012)034}{\emph{JHEP} {\bf 08} (2012)
  034}, [\href{http://arxiv.org/abs/0904.2715}{{\tt 0904.2715}}].

\bibitem{Gaiotto:2009hg}
D.~Gaiotto, G.~W. Moore and A.~Neitzke, \emph{{Wall-crossing, Hitchin Systems,
  and the WKB Approximation}},  \href{http://arxiv.org/abs/0907.3987}{{\tt
  0907.3987}}.

\bibitem{Chacaltana:2010ks}
O.~Chacaltana and J.~Distler, \emph{{Tinkertoys for Gaiotto Duality}},
  \href{http://dx.doi.org/10.1007/JHEP11(2010)099}{\emph{JHEP} {\bf 11} (2010)
  099}, [\href{http://arxiv.org/abs/1008.5203}{{\tt 1008.5203}}].

\bibitem{Razamat:2016dpl}
S.~S. Razamat, C.~Vafa and G.~Zafrir, \emph{{4d $ \mathcal{N}=1 $ from 6d (1,
  0)}}, \href{http://dx.doi.org/10.1007/JHEP04(2017)064}{\emph{JHEP} {\bf 04}
  (2017) 064}, [\href{http://arxiv.org/abs/1610.09178}{{\tt 1610.09178}}].

\bibitem{Bah:2017gph}
I.~Bah, A.~Hanany, K.~Maruyoshi, S.~S. Razamat, Y.~Tachikawa and G.~Zafrir,
  \emph{{4d $ \mathcal{N}=1 $ from 6d $ \mathcal{N}=\left(1,0\right) $ on a
  torus with fluxes}},
  \href{http://dx.doi.org/10.1007/JHEP06(2017)022}{\emph{JHEP} {\bf 06} (2017)
  022}, [\href{http://arxiv.org/abs/1702.04740}{{\tt 1702.04740}}].

\bibitem{Kim:2017toz}
H.-C. Kim, S.~S. Razamat, C.~Vafa and G.~Zafrir, \emph{{E-String Theory on
  Riemann Surfaces}},
  \href{http://dx.doi.org/10.1002/prop.201700074}{\emph{Fortsch. Phys.} {\bf
  66} (2018) 1700074}, [\href{http://arxiv.org/abs/1709.02496}{{\tt
  1709.02496}}].

\bibitem{Kim:2018lfo}
H.-C. Kim, S.~S. Razamat, C.~Vafa and G.~Zafrir, \emph{{Compactifications of
  ADE conformal matter on a torus}},
  \href{http://dx.doi.org/10.1007/JHEP09(2018)110}{\emph{JHEP} {\bf 09} (2018)
  110}, [\href{http://arxiv.org/abs/1806.07620}{{\tt 1806.07620}}].

\bibitem{Razamat:2018gro}
S.~S. Razamat and G.~Zafrir, \emph{{Compactification of 6d minimal SCFTs on
  Riemann surfaces}},
  \href{http://dx.doi.org/10.1103/PhysRevD.98.066006}{\emph{Phys. Rev. D} {\bf
  98} (2018) 066006}, [\href{http://arxiv.org/abs/1806.09196}{{\tt
  1806.09196}}].

\bibitem{Ohmori:2018ona}
K.~Ohmori, Y.~Tachikawa and G.~Zafrir, \emph{{Compactifications of 6d $N = (1,
  0)$ SCFTs with non-trivial Stiefel-Whitney classes}},
  \href{http://dx.doi.org/10.1007/JHEP04(2019)006}{\emph{JHEP} {\bf 04} (2019)
  006}, [\href{http://arxiv.org/abs/1812.04637}{{\tt 1812.04637}}].

\bibitem{Katz:1996fh}
S.~H. Katz, A.~Klemm and C.~Vafa, \emph{{Geometric engineering of quantum field
  theories}},
  \href{http://dx.doi.org/10.1016/S0550-3213(97)00282-4}{\emph{Nucl. Phys. B}
  {\bf 497} (1997) 173--195}, [\href{http://arxiv.org/abs/hep-th/9609239}{{\tt
  hep-th/9609239}}].

\bibitem{Shapere:1999xr}
A.~D. Shapere and C.~Vafa, \emph{{BPS structure of Argyres-Douglas
  superconformal theories}},  \href{http://arxiv.org/abs/hep-th/9910182}{{\tt
  hep-th/9910182}}.

\bibitem{Xie:2015rpa}
D.~Xie and S.-T. Yau, \emph{{4d N=2 SCFT and singularity theory Part I:
  Classification}},  \href{http://arxiv.org/abs/1510.01324}{{\tt 1510.01324}}.

\bibitem{Wang:2016yha}
Y.~Wang, D.~Xie, S.~S.~T. Yau and S.-T. Yau, \emph{{$4d$ $\mathcal{N} = 2$ SCFT
  from complete intersection singularity}},
  \href{http://dx.doi.org/10.4310/ATMP.2017.v21.n3.a6}{\emph{Adv. Theor. Math.
  Phys.} {\bf 21} (2017) 801--855},
  [\href{http://arxiv.org/abs/1606.06306}{{\tt 1606.06306}}].

\bibitem{Chen:2016bzh}
B.~Chen, D.~Xie, S.-T. Yau, S.~S.~T. Yau and H.~Zuo, \emph{{4D $\mathcal{N} =
  2$ SCFT and singularity theory. Part II: complete intersection}},
  \href{http://dx.doi.org/10.4310/ATMP.2017.v21.n1.a2}{\emph{Adv. Theor. Math.
  Phys.} {\bf 21} (2017) 121--145},
  [\href{http://arxiv.org/abs/1604.07843}{{\tt 1604.07843}}].

\bibitem{Chen:2017wkw}
B.~Chen, D.~Xie, S.~S.~T. Yau, S.-T. Yau and H.~Zuo, \emph{{4d $\mathcal{N}=2$
  SCFT and singularity theory Part III: Rigid singularity}},
  \href{http://dx.doi.org/10.4310/ATMP.2018.v22.n8.a2}{\emph{Adv. Theor. Math.
  Phys.} {\bf 22} (2018) 1885--1905},
  [\href{http://arxiv.org/abs/1712.00464}{{\tt 1712.00464}}].

\bibitem{Closset:2020scj}
C.~Closset, S.~Schafer-Nameki and Y.-N. Wang, \emph{{Coulomb and Higgs Branches
  from Canonical Singularities: Part 0}},
  \href{http://dx.doi.org/10.1007/JHEP02(2021)003}{\emph{JHEP} {\bf 02} (2021)
  003}, [\href{http://arxiv.org/abs/2007.15600}{{\tt 2007.15600}}].

\bibitem{Beem:2013sza}
C.~Beem, M.~Lemos, P.~Liendo, W.~Peelaers, L.~Rastelli and B.~C. van Rees,
  \emph{{Infinite Chiral Symmetry in Four Dimensions}},
  \href{http://dx.doi.org/10.1007/s00220-014-2272-x}{\emph{Commun. Math. Phys.}
  {\bf 336} (2015) 1359--1433}, [\href{http://arxiv.org/abs/1312.5344}{{\tt
  1312.5344}}].

\bibitem{Beem:2017ooy}
C.~Beem and L.~Rastelli, \emph{{Vertex operator algebras, Higgs branches, and
  modular differential equations}},
  \href{http://dx.doi.org/10.1007/JHEP08(2018)114}{\emph{JHEP} {\bf 08} (2018)
  114}, [\href{http://arxiv.org/abs/1707.07679}{{\tt 1707.07679}}].

\bibitem{Bourget:2018ond}
A.~Bourget, A.~Pini and D.~Rodr\'\i{}guez-G\'omez, \emph{{Gauge theories from
  principally extended disconnected gauge groups}},
  \href{http://dx.doi.org/10.1016/j.nuclphysb.2019.02.004}{\emph{Nucl. Phys. B}
  {\bf 940} (2019) 351--376}, [\href{http://arxiv.org/abs/1804.01108}{{\tt
  1804.01108}}].

\bibitem{Argyres:2018wxu}
P.~C. Argyres and M.~Martone, \emph{{Coulomb branches with complex
  singularities}}, \href{http://dx.doi.org/10.1007/JHEP06(2018)045}{\emph{JHEP}
  {\bf 06} (2018) 045}, [\href{http://arxiv.org/abs/1804.03152}{{\tt
  1804.03152}}].

\bibitem{Seiberg:1994rs}
N.~Seiberg and E.~Witten, \emph{{Electric - magnetic duality, monopole
  condensation, and confinement in N=2 supersymmetric Yang-Mills theory}},
  \href{http://dx.doi.org/10.1016/0550-3213(94)90124-4}{\emph{Nucl. Phys. B}
  {\bf 426} (1994) 19--52}, [\href{http://arxiv.org/abs/hep-th/9407087}{{\tt
  hep-th/9407087}}].

\bibitem{Seiberg:1994aj}
N.~Seiberg and E.~Witten, \emph{{Monopoles, duality and chiral symmetry
  breaking in N=2 supersymmetric QCD}},
  \href{http://dx.doi.org/10.1016/0550-3213(94)90214-3}{\emph{Nucl. Phys. B}
  {\bf 431} (1994) 484--550}, [\href{http://arxiv.org/abs/hep-th/9408099}{{\tt
  hep-th/9408099}}].

\bibitem{Freed:1997dp}
D.~S. Freed, \emph{{Special Kahler manifolds}},
  \href{http://dx.doi.org/10.1007/s002200050604}{\emph{Commun. Math. Phys.}
  {\bf 203} (1999) 31--52}, [\href{http://arxiv.org/abs/hep-th/9712042}{{\tt
  hep-th/9712042}}].

\bibitem{Argyres:2018urp}
P.~C. Argyres and M.~Martone, \emph{{Scaling dimensions of Coulomb branch
  operators of 4d N=2 superconformal field theories}},
  \href{http://arxiv.org/abs/1801.06554}{{\tt 1801.06554}}.

\bibitem{Caorsi:2018zsq}
M.~Caorsi and S.~Cecotti, \emph{{Geometric classification of 4d $\mathcal{N}=2$
  SCFTs}}, \href{http://dx.doi.org/10.1007/JHEP07(2018)138}{\emph{JHEP} {\bf
  07} (2018) 138}, [\href{http://arxiv.org/abs/1801.04542}{{\tt 1801.04542}}].

\bibitem{Argyres:2020wmq}
P.~C. Argyres and M.~Martone, \emph{{Towards a classification of rank r$
  \mathcal{N} $ = 2 SCFTs. Part II. Special Kahler stratification of the
  Coulomb branch}},
  \href{http://dx.doi.org/10.1007/JHEP12(2020)022}{\emph{JHEP} {\bf 12} (2020)
  022}, [\href{http://arxiv.org/abs/2007.00012}{{\tt 2007.00012}}].

\bibitem{Martone:2020nsy}
M.~Martone, \emph{{Towards the classification of rank-r$ \mathcal{N} $ = 2
  SCFTs. Part I. Twisted partition function and central charge formulae}},
  \href{http://dx.doi.org/10.1007/JHEP12(2020)021}{\emph{JHEP} {\bf 12} (2020)
  021}, [\href{http://arxiv.org/abs/2006.16255}{{\tt 2006.16255}}].

\bibitem{Cecotti:2021ouq}
S.~Cecotti, M.~Del~Zotto, M.~Martone and R.~Moscrop, \emph{{The Characteristic
  Dimension of Four-dimensional $\mathcal{N} = 2$ SCFTs}},
  \href{http://arxiv.org/abs/2108.10884}{{\tt 2108.10884}}.

\bibitem{Cordova:2015nma}
C.~Cordova and S.-H. Shao, \emph{{Schur Indices, BPS Particles, and
  Argyres-Douglas Theories}},
  \href{http://dx.doi.org/10.1007/JHEP01(2016)040}{\emph{JHEP} {\bf 01} (2016)
  040}, [\href{http://arxiv.org/abs/1506.00265}{{\tt 1506.00265}}].

\bibitem{Martone:2021ixp}
M.~Martone, \emph{{Testing our understanding of SCFTs: a catalogue of rank-2
  $\mathcal{N}$=2 theories in four dimensions}},
  \href{http://arxiv.org/abs/2102.02443}{{\tt 2102.02443}}.

\bibitem{Cecotti:2013lda}
S.~Cecotti, M.~Del~Zotto and S.~Giacomelli, \emph{{More on the N=2
  superconformal systems of type $D_p(G)$}},
  \href{http://dx.doi.org/10.1007/JHEP04(2013)153}{\emph{JHEP} {\bf 04} (2013)
  153}, [\href{http://arxiv.org/abs/1303.3149}{{\tt 1303.3149}}].

\bibitem{Arakawa:2016hkg}
T.~Arakawa and K.~Kawasetsu, \emph{{Quasi-lisse vertex algebras and modular
  linear differential equations}},  \href{http://arxiv.org/abs/1610.05865}{{\tt
  1610.05865}}.

\bibitem{brieskorn1970singular}
E.~Brieskorn, \emph{Singular elements of semi-simple algebraic groups},  in
  \emph{Actes du Congres International des Math{\'e}maticiens (Nice, 1970)},
  vol.~2, pp.~279--284, 1970.

\bibitem{slodowy1980simple}
P.~Slodowy, \emph{Simple singularities},  in \emph{Simple Singularities and
  Simple Algebraic Groups}, pp.~70--102.
\newblock Springer, 1980.

\bibitem{beauville1999symplectic}
A.~Beauville, \emph{Symplectic singularities}, {\emph{arXiv preprint
  math/9903070} (1999) }.

\bibitem{Pan:2021mrw}
Y.~Pan and W.~Peelaers, \emph{{The exact Schur index in closed form}},
  \href{http://arxiv.org/abs/2112.09705}{{\tt 2112.09705}}.

\bibitem{Beem:2021zvt}
C.~Beem, S.~S. Razamat and P.~Singh, \emph{{Schur Indices of Class
  $\mathcal{S}$ and Quasimodular Forms}},
  \href{http://arxiv.org/abs/2112.10715}{{\tt 2112.10715}}.

\bibitem{Bae:2020xzl}
J.-B. Bae, Z.~Duan, K.~Lee, S.~Lee and M.~Sarkis, \emph{{Fermionic rational
  conformal field theories and modular linear differential equations}},
  \href{http://dx.doi.org/10.1093/ptep/ptab033}{\emph{PTEP} {\bf 2021} (2021)
  08B104}, [\href{http://arxiv.org/abs/2010.12392}{{\tt 2010.12392}}].

\bibitem{Bae:2021mej}
J.-B. Bae, Z.~Duan, K.~Lee, S.~Lee and M.~Sarkis, \emph{{Bootstrapping
  Fermionic Rational CFTs with Three Characters}},
  \href{http://arxiv.org/abs/2108.01647}{{\tt 2108.01647}}.

\bibitem{Mathur:1988rx}
S.~D. Mathur, S.~Mukhi and A.~Sen, \emph{{Differential Equations for
  Correlators and Characters in Arbitrary Rational Conformal Field Theories}},
  \href{http://dx.doi.org/10.1016/0550-3213(89)90022-9}{\emph{Nucl. Phys. B}
  {\bf 312} (1989) 15--57}.

\bibitem{Mathur:1988na}
S.~D. Mathur, S.~Mukhi and A.~Sen, \emph{{On the Classification of Rational
  Conformal Field Theories}},
  \href{http://dx.doi.org/10.1016/0370-2693(88)91765-0}{\emph{Phys. Lett. B}
  {\bf 213} (1988) 303--308}.

\bibitem{Mathur:1988gt}
S.~D. Mathur, S.~Mukhi and A.~Sen, \emph{{Reconstruction of Conformal Field
  Theories From Modular Geometry on the Torus}},
  \href{http://dx.doi.org/10.1016/0550-3213(89)90615-9}{\emph{Nucl. Phys. B}
  {\bf 318} (1989) 483--540}.

\bibitem{Gaberdiel:2008pr}
M.~R. Gaberdiel and C.~A. Keller, \emph{{Modular differential equations and
  null vectors}},
  \href{http://dx.doi.org/10.1088/1126-6708/2008/09/079}{\emph{JHEP} {\bf 09}
  (2008) 079}, [\href{http://arxiv.org/abs/0804.0489}{{\tt 0804.0489}}].

\bibitem{Hampapura:2015cea}
H.~R. Hampapura and S.~Mukhi, \emph{{On 2d Conformal Field Theories with Two
  Characters}}, \href{http://dx.doi.org/10.1007/JHEP01(2016)005}{\emph{JHEP}
  {\bf 01} (2016) 005}, [\href{http://arxiv.org/abs/1510.04478}{{\tt
  1510.04478}}].

\bibitem{Gaberdiel:2016zke}
M.~R. Gaberdiel, H.~R. Hampapura and S.~Mukhi, \emph{{Cosets of Meromorphic
  CFTs and Modular Differential Equations}},
  \href{http://dx.doi.org/10.1007/JHEP04(2016)156}{\emph{JHEP} {\bf 04} (2016)
  156}, [\href{http://arxiv.org/abs/1602.01022}{{\tt 1602.01022}}].

\bibitem{Hampapura:2016mmz}
H.~R. Hampapura and S.~Mukhi, \emph{{Two-dimensional RCFT\textquoteright{}s
  without Kac-Moody symmetry}},
  \href{http://dx.doi.org/10.1007/JHEP07(2016)138}{\emph{JHEP} {\bf 07} (2016)
  138}, [\href{http://arxiv.org/abs/1605.03314}{{\tt 1605.03314}}].

\bibitem{Mukhi:2017ugw}
S.~Mukhi and G.~Muralidhara, \emph{{Universal RCFT Correlators from the
  Holomorphic Bootstrap}},
  \href{http://dx.doi.org/10.1007/JHEP02(2018)028}{\emph{JHEP} {\bf 02} (2018)
  028}, [\href{http://arxiv.org/abs/1708.06772}{{\tt 1708.06772}}].

\bibitem{Chandra:2018pjq}
A.~R. Chandra and S.~Mukhi, \emph{{Towards a Classification of Two-Character
  Rational Conformal Field Theories}},
  \href{http://dx.doi.org/10.1007/JHEP04(2019)153}{\emph{JHEP} {\bf 04} (2019)
  153}, [\href{http://arxiv.org/abs/1810.09472}{{\tt 1810.09472}}].

\bibitem{Mukhi:2019xjy}
S.~Mukhi, \emph{{Classification of RCFT from Holomorphic Modular Bootstrap: A
  Status Report}},  in \emph{{Pollica Summer Workshop 2019}: {Mathematical and
  Geometric Tools for Conformal Field Theories}}, 10, 2019.
\newblock \href{http://arxiv.org/abs/1910.02973}{{\tt 1910.02973}}.

\bibitem{Mukhi:2020gnj}
S.~Mukhi, R.~Poddar and P.~Singh, \emph{{Rational CFT with three characters:
  the quasi-character approach}},
  \href{http://dx.doi.org/10.1007/JHEP05(2020)003}{\emph{JHEP} {\bf 05} (2020)
  003}, [\href{http://arxiv.org/abs/2002.01949}{{\tt 2002.01949}}].

\bibitem{Das:2020wsi}
A.~Das, C.~N. Gowdigere and J.~Santara, \emph{{Wronskian Indices and Rational
  Conformal Field Theories}},
  \href{http://dx.doi.org/10.1007/JHEP04(2021)294}{\emph{JHEP} {\bf 04} (2021)
  294}, [\href{http://arxiv.org/abs/2012.14939}{{\tt 2012.14939}}].

\bibitem{Mason:2021xfs}
G.~Mason, K.~Nagatomo and Y.~Sakai, \emph{{Vertex operator algebras of rank
  $2$: The Mathur\textendash{}Mukhi\textendash{}Sen theorem revisited}},
  \href{http://dx.doi.org/10.4310/CNTP.2021.v15.n1.a2}{\emph{Commun. Num.
  Theor. Phys.} {\bf 15} (2021) 59--90}.

\bibitem{Kaidi:2020ecu}
J.~Kaidi and E.~Perlmutter, \emph{{Discreteness and integrality in Conformal
  Field Theory}}, \href{http://dx.doi.org/10.1007/JHEP02(2021)064}{\emph{JHEP}
  {\bf 02} (2021) 064}, [\href{http://arxiv.org/abs/2008.02190}{{\tt
  2008.02190}}].

\bibitem{Kaidi:2021ent}
J.~Kaidi, Y.-H. Lin and J.~Parra-Martinez, \emph{{Holomorphic modular bootstrap
  revisited}}, \href{http://dx.doi.org/10.1007/JHEP12(2021)151}{\emph{JHEP}
  {\bf 12} (2021) 151}, [\href{http://arxiv.org/abs/2107.13557}{{\tt
  2107.13557}}].

\bibitem{Das:2021uvd}
A.~Das, C.~N. Gowdigere and J.~Santara, \emph{{Classifying three-character
  RCFTs with Wronskian index equalling 0 or 2}},
  \href{http://dx.doi.org/10.1007/JHEP11(2021)195}{\emph{JHEP} {\bf 11} (2021)
  195}, [\href{http://arxiv.org/abs/2108.01060}{{\tt 2108.01060}}].

\bibitem{Bae:2021jkc}
J.-B. Bae, Z.~Duan and S.~Lee, \emph{{Can the energy bound $E \geq 0$ imply
  supersymmetry?}},  \href{http://arxiv.org/abs/2112.14130}{{\tt 2112.14130}}.

\bibitem{Cecotti:2015lab}
S.~Cecotti, J.~Song, C.~Vafa and W.~Yan, \emph{{Superconformal Index, BPS
  Monodromy and Chiral Algebras}},
  \href{http://dx.doi.org/10.1007/JHEP11(2017)013}{\emph{JHEP} {\bf 11} (2017)
  013}, [\href{http://arxiv.org/abs/1511.01516}{{\tt 1511.01516}}].

\bibitem{DiPietro:2014bca}
L.~Di~Pietro and Z.~Komargodski, \emph{{Cardy formulae for SUSY theories in $d
  =$ 4 and $d =$ 6}},
  \href{http://dx.doi.org/10.1007/JHEP12(2014)031}{\emph{JHEP} {\bf 12} (2014)
  031}, [\href{http://arxiv.org/abs/1407.6061}{{\tt 1407.6061}}].

\bibitem{ArabiArdehali:2015ybk}
A.~Arabi~Ardehali, \emph{{High-temperature asymptotics of supersymmetric
  partition functions}},
  \href{http://dx.doi.org/10.1007/JHEP07(2016)025}{\emph{JHEP} {\bf 07} (2016)
  025}, [\href{http://arxiv.org/abs/1512.03376}{{\tt 1512.03376}}].

\bibitem{Hofman:2008ar}
D.~M. Hofman and J.~Maldacena, \emph{{Conformal collider physics: Energy and
  charge correlations}},
  \href{http://dx.doi.org/10.1088/1126-6708/2008/05/012}{\emph{JHEP} {\bf 05}
  (2008) 012}, [\href{http://arxiv.org/abs/0803.1467}{{\tt 0803.1467}}].

\bibitem{Argyres:2022yet}
P.~Argyres, S.~Cecotti, M.~Del~Zotto, M.~Martone and R.~Moscrop, \emph{{to
  appear}}, .

\bibitem{Shapere:2008zf}
A.~D. Shapere and Y.~Tachikawa, \emph{{Central charges of N=2 superconformal
  field theories in four dimensions}},
  \href{http://dx.doi.org/10.1088/1126-6708/2008/09/109}{\emph{JHEP} {\bf 09}
  (2008) 109}, [\href{http://arxiv.org/abs/0804.1957}{{\tt 0804.1957}}].

\bibitem{Argyres:2018zay}
P.~C. Argyres, C.~Long and M.~Martone, \emph{{The Singularity Structure of
  Scale-Invariant Rank-2 Coulomb Branches}},
  \href{http://dx.doi.org/10.1007/JHEP05(2018)086}{\emph{JHEP} {\bf 05} (2018)
  086}, [\href{http://arxiv.org/abs/1801.01122}{{\tt 1801.01122}}].

\bibitem{Argyres:2016yzz}
P.~C. Argyres and M.~Martone, \emph{{4d $ \mathcal{N} $ =2 theories with
  disconnected gauge groups}},
  \href{http://dx.doi.org/10.1007/JHEP03(2017)145}{\emph{JHEP} {\bf 03} (2017)
  145}, [\href{http://arxiv.org/abs/1611.08602}{{\tt 1611.08602}}].

\bibitem{Beem:2018duj}
C.~Beem, \emph{{Flavor Symmetries and Unitarity Bounds in ${\mathcal N}=2$
  Superconformal Field Theories}},
  \href{http://dx.doi.org/10.1103/PhysRevLett.122.241603}{\emph{Phys. Rev.
  Lett.} {\bf 122} (2019) 241603}, [\href{http://arxiv.org/abs/1812.06099}{{\tt
  1812.06099}}].

\bibitem{Duncan:2014eha}
J.~F.~R. Duncan and S.~Mack-Crane, \emph{{The Moonshine Module for
  Conway\textquoteright{}s Group}},
  \href{http://dx.doi.org/10.1017/fms.2015.7}{\emph{SIGMA} {\bf 3} (2015) e10},
  [\href{http://arxiv.org/abs/1409.3829}{{\tt 1409.3829}}].

\bibitem{Argyres:2017tmj}
P.~C. Argyres, Y.~L\"u and M.~Martone, \emph{{Seiberg-Witten geometries for
  Coulomb branch chiral rings which are not freely generated}},
  \href{http://dx.doi.org/10.1007/JHEP06(2017)144}{\emph{JHEP} {\bf 06} (2017)
  144}, [\href{http://arxiv.org/abs/1704.05110}{{\tt 1704.05110}}].

\bibitem{Shephard:1954}
G.~Shephard and J.~Todd, \emph{{{Finite unitary reflection groups}}},
  {\emph{Canadian J. Math.} {\bf 6} (1954) 274}.

\bibitem{Chevalley:1955}
C.~Chevalley, \emph{{{Invariants of finite groups generated by reflections}}},
  {\emph{Amer. J. Math.} {\bf 77} (1955) 778--782}.

\bibitem{Fujiki:1988}
A.~Fujiki, \emph{{Finite automorphism groups of complex tori of dimension 2}},
  {\emph{Publ. RIMS Kyoto Univ.} {\bf 24} (1988) 1--97}.

\bibitem{Cecotti:2021yet}
S.~Cecotti, M.~Del~Zotto, M.~Martone and R.~Moscrop, \emph{{to appear}}, .

\bibitem{lehrer2009unitary}
G.~I. Lehrer and D.~E. Taylor, \emph{Unitary reflection groups}, vol.~20.
\newblock Cambridge University Press, 2009.

\bibitem{Buican:2015ina}
M.~Buican and T.~Nishinaka, \emph{{On the superconformal index of
  Argyres\textendash{}Douglas theories}},
  \href{http://dx.doi.org/10.1088/1751-8113/49/1/015401}{\emph{J. Phys. A} {\bf
  49} (2016) 015401}, [\href{http://arxiv.org/abs/1505.05884}{{\tt
  1505.05884}}].

\bibitem{Giacomelli:2020jel}
S.~Giacomelli, C.~Meneghelli and W.~Peelaers, \emph{{New $ \mathcal{N} $ = 2
  superconformal field theories from $ \mathcal{S} $-folds}},
  \href{http://dx.doi.org/10.1007/JHEP01(2021)022}{\emph{JHEP} {\bf 01} (2021)
  022}, [\href{http://arxiv.org/abs/2007.00647}{{\tt 2007.00647}}].

\bibitem{CCLMW2021}
C.~Beem, M.~Martone, C.~Meneghelli, W.~Peelaers and L.~Rastelli, \emph{{A
  bottom up approach for $\cN=2$ SCFTs: rank-1, to appear}}, .

\bibitem{Beem:2019tfp}
C.~Beem, C.~Meneghelli and L.~Rastelli, \emph{{Free Field Realizations from the
  Higgs Branch}}, \href{http://dx.doi.org/10.1007/JHEP09(2019)058}{\emph{JHEP}
  {\bf 09} (2019) 058}, [\href{http://arxiv.org/abs/1903.07624}{{\tt
  1903.07624}}].

\bibitem{Beem:2019snk}
C.~Beem, C.~Meneghelli, W.~Peelaers and L.~Rastelli, \emph{{VOAs and rank-two
  instanton SCFTs}},
  \href{http://dx.doi.org/10.1007/s00220-020-03746-9}{\emph{Commun. Math.
  Phys.} {\bf 377} (2020) 2553--2578},
  [\href{http://arxiv.org/abs/1907.08629}{{\tt 1907.08629}}].

\bibitem{Dolan:2002zh}
F.~A. Dolan and H.~Osborn, \emph{{On short and semi-short representations for
  four-dimensional superconformal symmetry}},
  \href{http://dx.doi.org/10.1016/S0003-4916(03)00074-5}{\emph{Annals Phys.}
  {\bf 307} (2003) 41--89}, [\href{http://arxiv.org/abs/hep-th/0209056}{{\tt
  hep-th/0209056}}].

\bibitem{Donagi:1995cf}
R.~Donagi and E.~Witten, \emph{{Supersymmetric Yang-Mills theory and integrable
  systems}}, \href{http://dx.doi.org/10.1016/0550-3213(95)00609-5}{\emph{Nucl.
  Phys. B} {\bf 460} (1996) 299--334},
  [\href{http://arxiv.org/abs/hep-th/9510101}{{\tt hep-th/9510101}}].

\bibitem{Beem:2014zpa}
C.~Beem, M.~Lemos, P.~Liendo, L.~Rastelli and B.~C. van Rees, \emph{{The $
  \mathcal{N}=2 $ superconformal bootstrap}},
  \href{http://dx.doi.org/10.1007/JHEP03(2016)183}{\emph{JHEP} {\bf 03} (2016)
  183}, [\href{http://arxiv.org/abs/1412.7541}{{\tt 1412.7541}}].

\bibitem{Perlmutter:2020buo}
E.~Perlmutter, L.~Rastelli, C.~Vafa and I.~Valenzuela, \emph{{A CFT distance
  conjecture}}, \href{http://dx.doi.org/10.1007/JHEP10(2021)070}{\emph{JHEP}
  {\bf 10} (2021) 070}, [\href{http://arxiv.org/abs/2011.10040}{{\tt
  2011.10040}}].

\bibitem{Bilal:1991eu}
A.~Bilal, \emph{{Introduction to W algebras}},  in \emph{{Spring School on
  String Theory and Quantum Gravity (to be followed by Workshop)}}, 4, 1991.

\bibitem{Iles:2014gra}
N.~J. Iles and G.~M.~T. Watts, \emph{{Modular properties of characters of the
  W$_{3}$ algebra}},
  \href{http://dx.doi.org/10.1007/JHEP01(2016)089}{\emph{JHEP} {\bf 01} (2016)
  089}, [\href{http://arxiv.org/abs/1411.4039}{{\tt 1411.4039}}].

\bibitem{Iles:2013jha}
N.~J. Iles and G.~M.~T. Watts, \emph{{Characters of the $W_3$ algebra}},
  \href{http://dx.doi.org/10.1007/JHEP02(2014)009}{\emph{JHEP} {\bf 02} (2014)
  009}, [\href{http://arxiv.org/abs/1307.3771}{{\tt 1307.3771}}].

\bibitem{Beltaos:2010ka}
E.~Beltaos and T.~Gannon, \emph{{The $W_{N}$ minimal model classification}},
  \href{http://dx.doi.org/10.1007/s00220-012-1473-4}{\emph{Commun. Math. Phys.}
  {\bf 312} (2012) 337--360}, [\href{http://arxiv.org/abs/1004.1205}{{\tt
  1004.1205}}].

\bibitem{toappear}
C.~Beem and L.~Rastelli, \emph{{Unpublished}}, .

\end{thebibliography}\endgroup
\end{document}